\documentclass[aps,prl,twocolumn,superscriptaddress]{revtex4-2}

\usepackage[]{standalone}
\usepackage{import}
\usepackage{soul}
\usepackage{amsthm}
\usepackage{amsmath}
\usepackage{amssymb}
\usepackage{graphicx}
\usepackage{epstopdf}
\usepackage{multirow}
\usepackage{url}
\usepackage{float}
\usepackage{bm}
\usepackage{ifthen}
\usepackage[usenames,dvipsnames]{color}
\usepackage{mathrsfs}
\usepackage[colorlinks=false,citecolor=blue,urlcolor=black]{hyperref}
\usepackage{orcidlink}

\usepackage[absolute]{textpos}
\usepackage{xcolor}

\hypersetup{
 bookmarksopen=true,
 bookmarksopenlevel=1,
 colorlinks=true,
 linkcolor=blue,
 anchorcolor=blue,
 citecolor=blue,
 filecolor=blue,
 urlcolor=blue,
 pdfpagemode=UseOutlines,
 pdfstartview={XYZ null null 1},
}
\usepackage{placeins}
\usepackage{physics}
\usepackage[
 capitalise,
]{cleveref}

\usepackage{pgfplots}
\pgfplotsset{compat=newest}
\usepackage{tikz}

\newcommand{\coloneqq}{:=}

\theoremstyle{plain}

\theoremstyle{definition}

\newif \ifcmnt
\cmnttrue
\ifcmnt
    \providecommand{\aucmnt}[1]{#1}

 \else
    \providecommand{\aucmnt}[1]{}
    
\fi

\graphicspath{{.}{./figs/}}

\begin{document}


\title{Real-time heralded non-Gaussian teleportation resource-state generator} 


\author{Joseph C. Chapman\,\orcidlink{0000-0002-3346-0914}}
\email{chapmanjc@ornl.gov}
\affiliation{Quantum Information Science Section, Oak Ridge National Laboratory, Oak Ridge, TN, USA}

\author{Yanbao Zhang\,\orcidlink{0000-0002-4553-0561}}
\affiliation{Quantum Information Science Section, Oak Ridge National Laboratory, Oak Ridge, TN, USA}

\author{Joseph M. Lukens\,\orcidlink{0000-0001-9650-4462}}
\affiliation{Quantum Information Science Section, Oak Ridge National Laboratory, Oak Ridge, TN, USA}
\affiliation{Elmore Family School of Electrical and Computer Engineering and Purdue Quantum Science and Engineering Institute, Purdue University, West Lafayette, IN, USA}

\author{Alberto M. Marino\,\orcidlink{0000-0001-5377-1122}}
\affiliation{Quantum Information Science Section, Oak Ridge National Laboratory, Oak Ridge, TN, USA}

\author{Eugene Dumitrescu\,\orcidlink{0000-0001-5851-9567}}
\affiliation{Quantum Information Science Section, Oak Ridge National Laboratory, Oak Ridge, TN, USA}

\author{Yan Wang\,\orcidlink{0000-0002-6545-6434}}
\affiliation{Quantum Information Science Section, Oak Ridge National Laboratory, Oak Ridge, TN, USA}

\author{Nicholas A. Peters\,\orcidlink{0000-0002-7215-9630}}
\affiliation{Quantum Information Science Section, Oak Ridge National Laboratory, Oak Ridge, TN, USA}



\begin{abstract}
Quantum teleportation is a fundamental quantum communications primitive that requires an entangled resource state.  In the continuous-variable regime, non-Gaussian entangled resources have been shown theoretically to improve teleportation fidelity compared to Gaussian squeezed vacuum. We experimentally demonstrate a heralded two-mode resource state for non-Gaussian teleportation capable of real-time use. We characterize this state with two-mode homodyne tomography showing it has fidelity $F=0.973\pm 0.005$ with the expected resource state. Real-time use is enabled by a photon-subtraction orchestrator system performing live coincidence detection and outputting low-jitter and low-latency heralding signals. Live collection of real-time quadrature measurements of photon-subtracted states is enabled by the development of a synchronized homodyne detection server, where the orchestrator system queries the server to collect the real-time quadrature samples corresponding to the heralded state. 
These results demonstrate significant advancement in enabling the use of heralded non-Gaussian states in quantum networking protocols, especially in the context of quantum repeaters,  non-Gaussian quantum sensing, and measurement-based quantum computing.
\end{abstract}


\maketitle

\textit{Introduction}---Quantum computing and quantum sensing systems continue to show growing advantages over  classical methods~\cite{Morvan_etal_2024,Kim_Eddins_etal_2023,PhysRevLett.134.090601,PhysRevLett.123.231107,sciadv.adw9757,acsphotonics4c00256}. To provide further advantages, the field of quantum computing is currently focused on performance improvements and system scaling. For example, quantum computing platforms will benefit from the use of inter-node optical quantum networking~\cite{science1231298,BrownKimMonroe2016,IonQ2025Q3,XanaduNature2025a,Psiquantum} over short and long distances. Based on current teleportation fidelities~\cite{HuGuoLiuLiGuo2023}, these links will require quantum-repeater functionality, regardless of their length, to ensure sufficiently high-quality connections. Quantum repeaters can broadly be classified in two ways: ``one-way'' quantum repeaters use quantum error-correction on large entangled resource states, whereas, ``two-way'' quantum repeaters use heralded entanglement distribution and purification (also known as distillation) or error-correction~\cite{Muralidharan_etal_2016}. After a ``two-way'' quantum-repeater system generates purified shared entanglement, quantum teleportation then uses the heralded purified quantum entanglement for quantum state transfer or quantum entanglement swapping between user nodes, e.g., disparate quantum computers. 

In the context of continuous variables (CV), quantum teleportation is based on the Gaussian two-mode squeezed-vacuum entangled state. With infinite squeezing and no transmission loss, 100\% teleported state fidelity is theorized~\cite{PhysRevLett.80.869}. In reality, however, finite squeezing and losses limit the fidelity significantly. Non-Gaussian teleportation has been introduced~\cite{PhysRevA.61.032302} and analyzed~\cite{PhysRevA.65.062306,PhysRevA.76.022301,PhysRevA.81.012333,PhysRevA.91.063832,PhysRevA.103.043701,PhysRevA.107.012418,https://doi.org/10.1002/qute.202300344} providing improved teleported state fidelity, even in realistic conditions, when using a resource state generated from coincident photon subtraction of the two-mode squeezed vacuum initial state, where the resultant resource state may or may not be quantum non-Gaussian~\cite{PRXQuantum.2.030204} but has increased entanglement. This improvement can be understood as first applying a type of entanglement distillation to the initial state, to create a heralded resource state with greater entanglement and less vacuum contribution~\cite{PhysRevA.65.062306, EntDist_Furusawa_2010}. Thus, this resource state was effectively created before in the context of CV entanglement distillation~\cite{EntDist_Furusawa_2010,PhysRevLett.112.070402,PhysRevLett.129.273604} but without real-time operation, requiring lengthy post-processing, and at wavelengths incompatible with fiber networks.

In this Letter, we demonstrate a real-time heralded resource-state generator via two-mode photon subtraction on fiber-coupled frequency non-degenerate two-mode squeezed vacuum in the optical C-band (1530-1565 nm). This state is compatible with wavelength-division-multiplexed deployed fiber networks and high-quality coexistence with classical signals~\cite{ChapmanDeployedTMSV2023, breum2025NGdep}. To our knowledge, this is the first demonstration of two-mode photon subtraction on frequency non-degenerate two-mode squeezed vacuum.  Real-time operation is demonstrated using relatively low-cost hardware in two ways: (1) a newly developed photon-subtraction system orchestrator (PSO) implements live multi-mode multi-detector coincidence-analysis circuitry---that also creates a low-latency low-jitter ($91\pm1$~ns) heralding output signal (with a maximum rate of about $150$~MHz) enabling a variety of protocols and applications with this resource state and (2) the homodyne detection implements real-time quadrature measurements of photon-subtracted wave packets~\cite{PhysRevLett.116.233602}. The quadrature measurement results are then directed to a newly developed homodyne detection server (HDS), synchronized to the PSO (with only about 500~ps jitter), that enables buffering (up to 691~ms) and live transmission of 100-MHz samples requested by the PSO that correspond to detected heralded resource states. This coordination enables distributed two-mode homodyne tomography of heralded photon-subtracted resource states, which is used for state verification. The PSO/HDS combined system can reliably continuously process events up to about 250k distinct photon-subtraction events per second (including the corresponding homodyne tomography samples). The continuous processing capability has been tested for a continuous data collection run of as much as seven hours.

The generation of heralded entanglement, purified by the photon subtraction operation before use, demonstrates major components of two-way quantum repeaters enabled by our real-time operation methods. Additionally, between the PSO and HDS, this combined measurement system demonstrates several key enabling technologies for real-time non-Gaussian measurement-based quantum processing, providing an alternative in some respects to Refs.~\cite{XanaduNature2025a, XanaduNature2025b}, with our focus on quantum networking protocols.

\begin{figure}
    \centerline{\includegraphics[width=1\columnwidth]{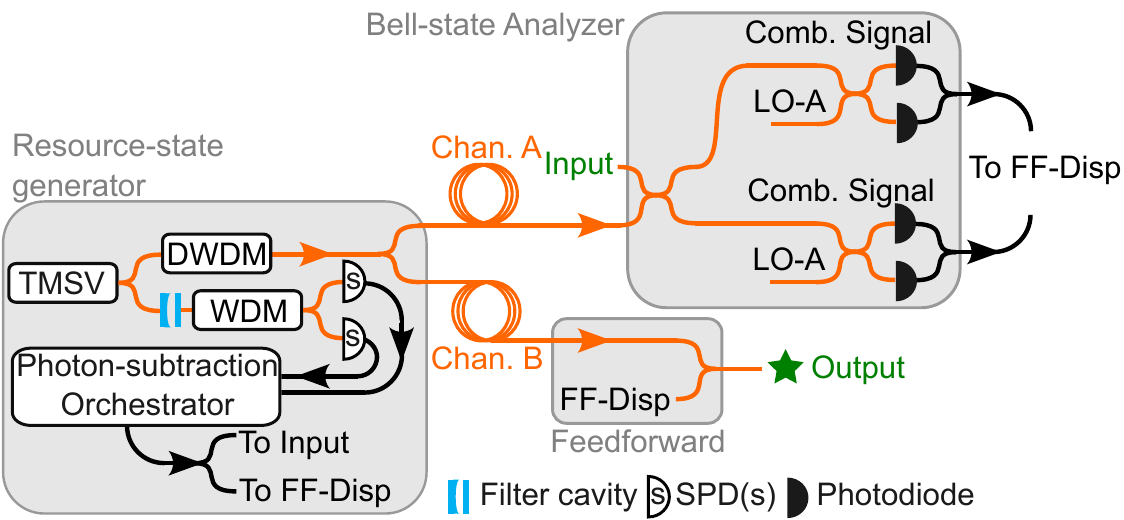}}
    \caption{Scheme for Non-Gaussian teleportation using our heralded resource state. The input state (green) is teleported after using this resource state to the green star using Bell-state analyzer results for feedforward correction. Definitions (in alphabetical order). DWDM: dense-WDM. FF-Disp: feedforward displacement. LO: local oscillator. SPD: single-photon detector. TMSV: two-mode squeezed vacuum. WDM: wavelength division multiplexer.}
    \label{fig:NGTelsimpsetup}
\end{figure}

To place our results in context, we start by explaining a  protocol as a specific example where our resource generator is ideally suited (see Fig.~\ref{fig:NGTelsimpsetup}). Non-Gaussian teleportation starts by generating two-mode squeezed vacuum which is then purified by photon-subtraction. The two modes of the resource state are then transmitted to their respective destinations. Mode A is mixed with the input state to be teleported (which received a heralding signal from the PSO to synchronize the mode-matched interaction). The combined output is directed to the Bell-state measurement system consisting of a dual-homodyne detector whose outputs are transmitted for feedforward (and can be sampled by a local HDS). After transmission, Mode B is delayed (via optical delay or quantum memory) to allow for feedforward of Bell-state measurement results from Mode A to perform the unitary operation to complete the teleportation. 

After this feedforward operation, the protocol is complete and the teleported state is available for use. This could entail reception into a quantum computer, e.g., if the input state was the output of some quantum sensor buffered by a quantum memory. Alternatively, if the input state was entangled with another mode, this could be one step in a larger two-way quantum repeater protocol using entanglement swapping. Finally, to characterize the heralded resource state, each mode can be sent to a homodyne detector and the HDS facilitates joint two-mode tomography.

\textit{Experiment}---In the experiment described herein, we demonstrate the left half of Fig.~\ref{fig:NGTelsimpsetup} and on the right half, instead of the Bell-state measurement and feedforward, we characterize both modes with homodyne tomography to fully characterize the heralded resource state. The simplified experimental setup of our actual system is shown in Fig.~\ref{fig:simpexpsetup}. A complete diagram and description are presented in the Supplemental Material~\cite{SM1}. Moreover, a complete diagram of a proposed full non-Gaussian teleportation system using our methods is also presented in the Supplemental Material~\cite{SM4}. 

The overall flow of our demonstrated system starts with the generation of frequency non-degenerate two-mode squeezed vacuum (TMSV) with some of the light from each mode reflected into another mode for photon subtraction via single-photon detection. The remaining transmitted TMSV is split such that each mode goes into its own fiber using a wavelength demultiplexing filter. Each mode is then directed towards a homodyne detector with a variable local-oscillator phase for two-mode homodyne tomography. Homodyne samples heralded to correspond to photon-subtraction events are collected and used to estimate the two-mode photon-subtracted state using tomographic analysis. An $(n,m)$ photon subtraction from modes $(A=1,B=2)$ on TMSV results in the unnormalized state:
\begin{align}
    \ket{\psi_{n,m}}= \sum_{k=\max(n,m)}^{\infty} c_{k,n,m} \ket{k-n}_1\otimes \ket{k-m}_2. 
\label{eq:PSTMS}
\end{align}
Using $R_i$ to denote the  reflectivity of the beamsplitter acting on mode $i$, $B_{k,n}(R) = \sqrt{\binom{k}{n}(1-R)^{k-n} R^{n}}$ as binomial factors, and $t_r=\tanh{r}$ with squeezing parameter $r$, the coefficients can be written as $c_{k,n,m} = (-t_r)^k B_{k,n}(R_1) B_{k,m}(R_2)$. A complete derivation is available in the Supplemental Material~\cite{SM7}.
\begin{figure}
    \centerline{\includegraphics[width=1\columnwidth]{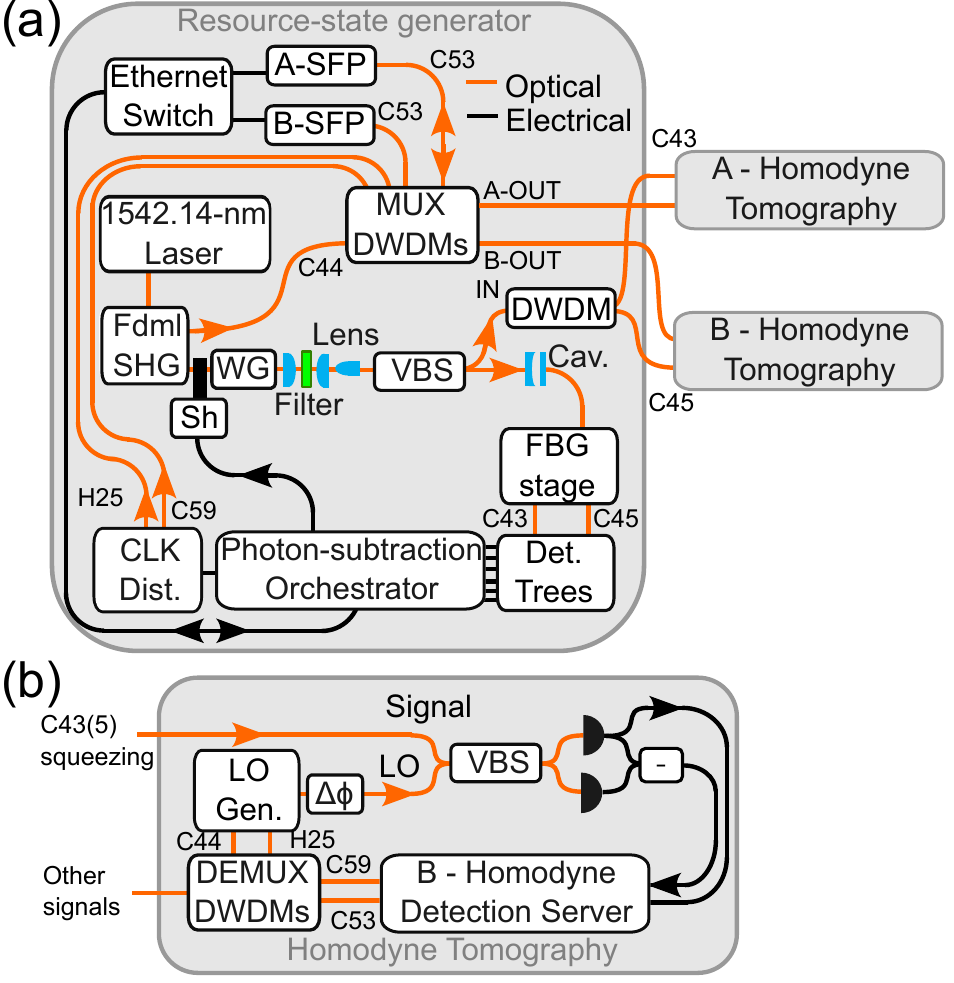}}
    \caption{Simplified experimental setup. (a) Detailed resource state generator showing connections to homodyne tomography subsystems. (b) Detailed homodyne tomography subsystem. Definitions (in alphabetical order). CLK: clock. $\Delta\phi$: phase shifter(s). DEMUX: demultiplex. DWDM: dense-wavelength division multiplexer. FBG: fiber Bragg grating. Fdml: fundamental. LO: local oscillator. MUX: multiplex. SFP: small-form-factor pluggable transceiver. Sh: Shutter. SHG: second-harmonic generation. VBS: variable beamsplitter. WG: waveguide. 100-GHz channel center wavelength and description for H25: Sideband reference (25-GHz modulation on carrier) at 1556.96~nm. C43: Resource-state mode A at 1542.95~nm. C44: Phase reference at 1542.14~nm. C45: Resource-state mode B at 1541.35~nm. C53: bi-directional fiber transceiver at 1535.04~nm. C59: 10-MHz clock reference modulation on carrier at 1530.33 nm}
    \label{fig:simpexpsetup}
\end{figure}

To generate TMSV, we use a  doubled continuous-wave fiber laser emitting two beams; the fundamental wavelength at $\lambda_f=1542.14$~nm, aligned with International Telecommunications Union (ITU) grid 100-GHz Channel C44, and the second-harmonic at  771.07~nm are filtered and independently fiber coupled. The second-harmonic is used to pump a fiber-coupled type-0 lithium niobate waveguide (WG). For photon subtraction, the squeezed vacuum is then directed to a variable fiber beamsplitter (VBS) with reflectivity $R_S=14$\% to balance the trade-off between heralding rate and state quality.

To enable two-mode photon-subtraction of broadband two-mode squeezing, precise filtering of each mode is required so the modes directed to the single-photon detectors are the same, as detected by the homodyne detectors. This filtering is mainly done with a  9-MHz half-width-at-half-maximum 25-GHz free-spectral range fiber-coupled locked optical cavity. The targeted two-mode squeezing is at $\pm$100 GHz from the $\lambda_f$ (which aligns with ITU grid channels C43 and  C45 for modes A and B, respectively). After the cavity, 25-GHz fiber Bragg gratings are used to select the two desired resonances from the cavity transmission spectrum. Our methods enable continuous cavity locking with negligible added photon-subtraction noise from any stabilization system (e.g., phase or cavity). Each resonance (one at C43 and the other at C45) is then sent to a 3-output beamsplitter tree connected to single-photon detectors to enable multi-photon subtraction. 

The detector outputs are collected by a newly developed PSO for processing. Instead of the conventional method used in photon-subtraction experiments, where the single-photon detection directly triggers a sampling oscilloscope to save a signal trace for later processing, we have implemented a more practical and capable method using pipelined digital logic that enables real-time use of the resource state so it can be leveraged for a variety of protocols and applications. The PSO's pipelined multi-mode multi-detector coincidence-analysis circuitry outputs a low-latency low-jitter ($91\pm1$~ns) heralding signal (with a maximum high-rate of about $150$~MHz) for application use. 

On the other hand, to enable distributed multi-mode homodyne tomography on photon-subtracted states, these detection events are also time-stamped and saved in a buffer. The PSO uses these timestamped events to collect corresponding synchronously buffered homodyne samples from the HDS (described below) for tomographic analysis.  These homodyne samples are then automatically triaged into separate buffers then saved based on the corresponding photon-subtraction type. This system can run for hours, or more, continuously capturing data for whatever detection signatures are collected.

The transmitted modes are sent to a dense-wavelength division multiplexer (DWDM) which is configured to demultiplex the squeezed modes. For this demonstration, each mode is then directed towards a separate homodyne detector for relative phase-stabilized two-mode homodyne tomography (see Supplemental Materials~\cite{SM8}). The homodyne detection and local oscillator (LO) generation largely follow the methods of Ref.~\cite{ChapmanDeployedTMSV2023} with several improvements. The real-time quadrature samples~\cite{PhysRevLett.116.233602,RTPSquadOE2017} from the homodyne detector are directed to the newly developed HDS. The HDS's hardware samples the homodyne detector output at 100~MHz into a buffer of about 70~Msamples (equating to about 700~ms) for the software to service client sample requests.

To ensure synchronization and communication between the PSO and each HDS, as well as to generate coherent LOs, we transmitted several reference and control signals on side wavelength channels in addition to the resource state modes. We achieve a synchronization between the PSO and each HDS with only about 500~ps jitter. All of these signals and the resource state modes are contained within the optical C-band enabling multiplexing for fiber transmission~\cite{ChapmanDeployedTMSV2023}. 

\begin{figure*}
    \centerline{\includegraphics[width=1\textwidth]{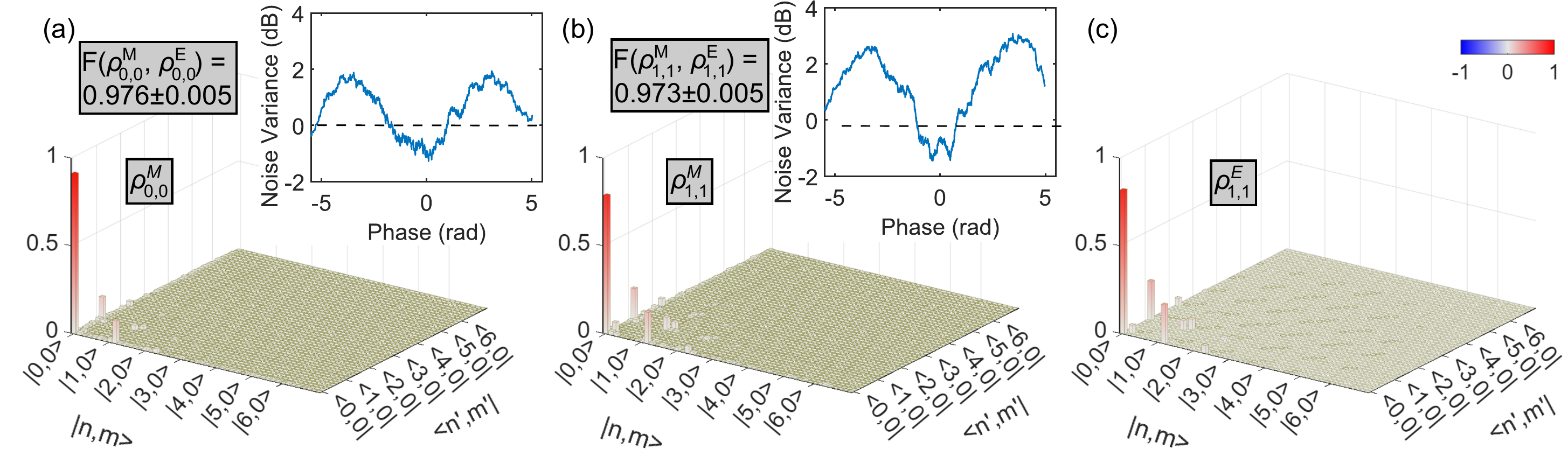}}
    \caption{Density matrices (absolute value) from tomographic reconstruction of measured data with inset showing rolling variance (window size 500 samples) of measured combined quadrature samples sorted by phase  for (a) zero photons subtracted ($\rho^M_{0,0}$) (b) 1 photon subtracted per mode ($\rho^M_{1,1}$), as well as (c) expected theoretical state for 1 photon subtracted per mode ($\rho^E_{1,1}$). For these measurement reconstructions, the photon-number cut-off per mode is 6. The full density matrices are shown in the Supplemental Material.}
    \label{fig:DMs6969}
\end{figure*}
\textit{Results}---To characterize our two-mode photon-subtracted resource state, we need to know which homodyne detector samples correspond to the photon-subtracted temporal modes. This requires calibrating the requisite delays between the time-stamped photon-subtraction detection events to the time-stamped homodyne detection samples at the servers. We do this calibration using cross-correlation (CC) analysis of pulsed thermal states detection (see Supplemental Material~\cite{SM5} for delay calibration details). With an approximate delay for each mode, we acquire tomographic data with 10000 samples (at a rate of about 20 samples per second) each for a variety of delay combinations around the thermal-state peak-CC delay. 

In Fig.~\ref{fig:DMs6969}, we show the measured density matrices, for zero photons subtracted $\rho^M_{0,0}$ and one photon subtracted per mode $\rho^M_{1,1}$, at the optimal delay combination D(0,1) based on the state fidelity and squeezing amount. Here D($j,j$) is the delay where the sample for each mode is shifted $j$ time bin(s) with respect to the approximate delay combination, i.e., thermal-state peak-CC delays. There is a clear reduction of the vacuum contribution as well as a rise in the one-photon pair term in the one-photon-subtracted measured state [Fig.~\ref{fig:DMs6969}(b)] compared to the zero-photon subtracted measured state [Fig.~\ref{fig:DMs6969}(a)]. This is a clear indication that the measured photon-subtraction state does indeed exhibit the signs of the expected photon subtraction. Moreover, comparing the entanglement of each state using the log negativity~\cite{PhysRevA.65.032314} $E_\mathcal{N}(\rho)$, we see a slight rise in entanglement from $E_\mathcal{N}(\rho^M_{0,0})=0.49\pm 0.03$ to $E_\mathcal{N}(\rho^M_{1,1})=0.52\pm 0.03$ (limited in our implementation by some thermal noise due to mode mismatch). This is also evidenced by the insets of Fig.~\ref{fig:DMs6969} showing increased two-mode squeezing and anti-squeezing for $\rho^M_{1,1}$.

It should be noted, the log negativity is disproportionately affected by noise in the smaller eigenvalues of the tomographically reconstructed density matrices (which are influenced by sample size and photon-number cut-off) leading to somewhat inflated values of $E_\mathcal{N}(\rho)$ for our sample size of 10000~\cite{EntDist_Furusawa_2010} and leading to the comparatively large error bar which is derived from the standard deviation of 50 tomographies of $\rho^M_{0,0}$. But to enable a relatively fair comparison without a significantly larger sample size, we use the same sample size for the $\rho^M_{0,0}$ and $\rho^M_{1,1}$ tomographies used to calculate $E_\mathcal{N}(\rho)$. On the other hand, we also analyzed another entanglement metric, most negative eigenvalue of the partial transpose (a lower bound to the log negativity)~\cite{Eisert01011999}, and found it to not have the aforementioned sensitivity to sample size and photon-number cut-off (see Supplemental Material for complete analysis). $|\min(\lambda^M_{0,0})|=0.11\pm0.01$ and $|\min(\lambda^M_{1,1})|=0.12\pm0.01$ which shows a similarly significant increase in entanglement, by one standard deviation.

Moreover, at a less optimal delay (less squeezing and lower fidelity) where it is easier to make a larger improvement, D(0,-1), we observed a larger, more statistically significant, entanglement increase from $E_\mathcal{N}(\rho^M_{0,0})=0.35\pm 0.03$ to $E_\mathcal{N}(\rho^M_{1,1})=0.45\pm 0.03$ which is an increase greater than three standard deviations. Here the sub-optimal $E_\mathcal{N}(\rho^M_{0,0})$ shows an greater improvement with photon subtraction (but still lower) than the higher $E_\mathcal{N}(\rho^M_{0,0})$ at the optimal delay which is understandable when the optimal starting point is nearer to the maximum $E_\mathcal{N}$ capable of this system (even after photon subtraction). So photon subtraction provides an improvement but the improvement can be reduced when the system is already near the best the system can produce like a saturation effect. This reduced improvement at higher initial performance is also seen in analyses of teleportation fidelity improvement using this photon-subtracted resource state (See Ref.~\cite{PhysRevA.65.062306,PhysRevA.76.022301,PhysRevA.81.012333,PhysRevA.91.063832,PhysRevA.103.043701,PhysRevA.107.012418,https://doi.org/10.1002/qute.202300344} and Supplemental Material) where the fidelity improvement due to a photon-subtracted resource state is greater in some lower fidelity range than at higher fidelities.
Moreover, this entanglement increase is a form of entanglement distillation~\cite{PhysRevA.65.062306}. Additionally, one can consider the quantum non-Gaussian character of the resource state which we do as an aside in the Supplemental Material for the interested reader.

In addition, for comparison to $\rho^M_{1,1}$ [Fig.~\ref{fig:DMs6969}(b)],  Fig.~\ref{fig:DMs6969}(c) shows the theoretically expected density matrix $\rho^E_{1,1}$ for the one-photon-subtracted-per-mode state using squeezing parameter $r=0.3$ and total transmission after photon subtraction $\eta_A=0.55$ and $\eta_B=0.5$ for photon-subtraction beamsplitter reflectivity $R_1=R_2=R_S=0.14$. These parameters are derived from analysis on independent measurements and not from the tomography data. The fidelity between the measured and expected one-photon-subtracted-per-mode state is $F(\rho^M_{1,1},\rho^E_{1,1})=0.973\pm 0.005$. The error bar for all fidelity calculations corresponds to the standard deviation of $F(\rho^M_{0,0},\rho^E_{0,0})$ for 50 tomographies.

\begin{figure}
    \centerline{\includegraphics[width=1\columnwidth]{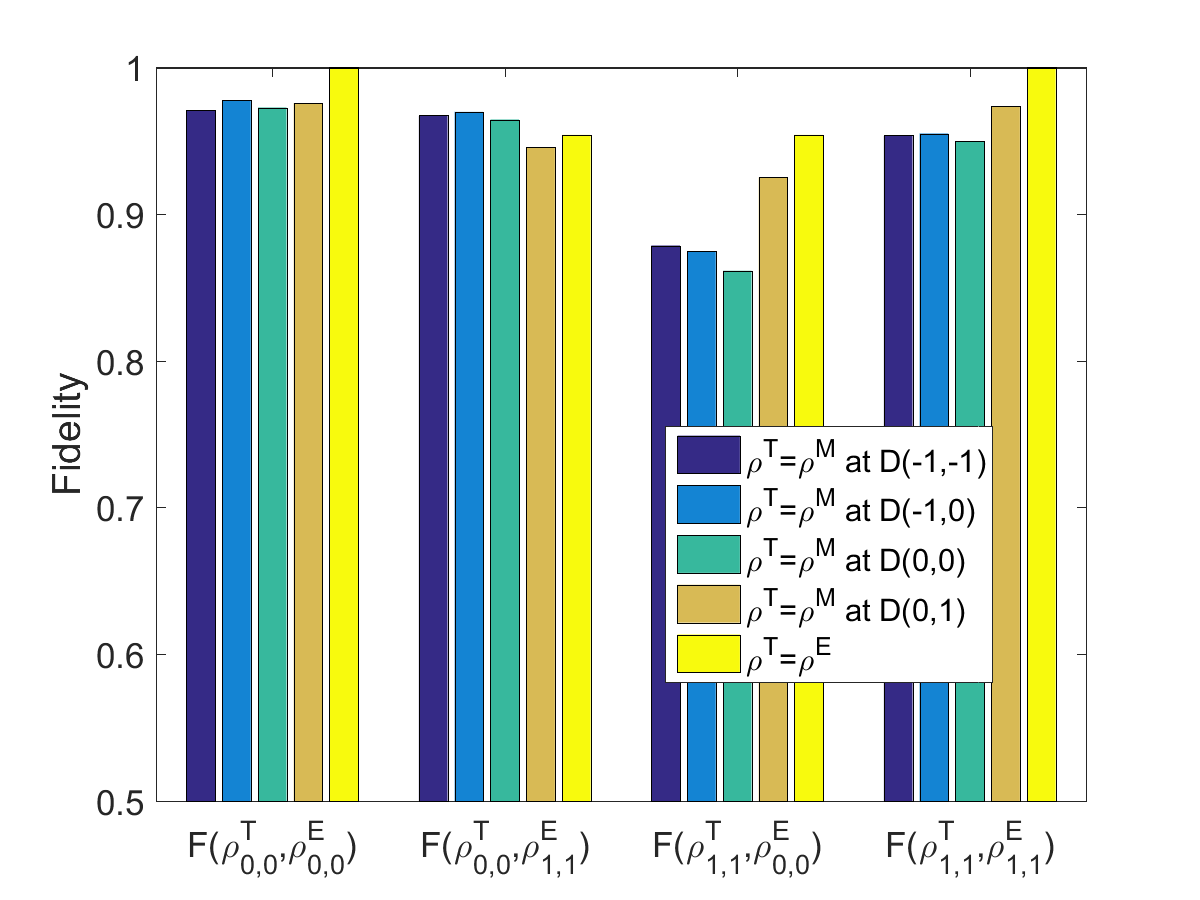}}
    \caption{Fidelity comparison between the the theoretically expected states $\rho_E$ and several test states $\rho^T$. Yellow bars provide theoretical fidelities.}
    \label{fig:FidsDT}
\end{figure}

Due to the large overlap between various similar continuous-variable states, we also show a more complete fidelity comparison for $\rho^M_{1,1}$, $\rho^E_{1,1}$, $\rho^M_{0,0}$, and $\rho^E_{0,0}$ in Fig.~\ref{fig:FidsDT} where $\rho^E_{0,0}$ has $\eta_A=0.55$ and $\eta_B=0.5$. 
In Fig.~\ref{fig:FidsDT}, it is clear that the fidelity of $\rho^M$ for D(0,1) most closely matches the theoretical expectation (yellow bar) for all 4 cases. For all these near optimal delays (see Fig.~\ref{fig:FidsDT} legend), there is a clear trend where $\rho^M_{0,0}$ and $\rho^M_{1,1}$ each have their highest fidelity with their corresponding expected theoretical state. In particular, D(0,1) is the delay combination where $F(\rho^M_{1,1},\rho^E_{1,1})>F(\rho^M_{0,0},\rho^E_{1,1})$ by more than four standard deviations, showing very clearly the $\rho^M_{1,1}$ resource state at the optimal delay most closely matches the one-photon-subtracted-per-mode target state. 

This resource state is directly usable in non-Gaussian teleportation. Using $\rho^M_{1,1}$ ($\rho^E_{1,1}$), we calculate a coherent-state teleportation fidelity of $0.5635$ ($0.5952$). Using $\rho^M_{0,0}$ (and $\rho^E_{0,0}$ without the photon-subtraction beamsplitter) we calculate a coherent-state teleportation fidelity of $0.5539$($0.5670$). This analysis shows our measured photon-subtracted resource state does provide improvement over the measured non-photon-subtracted version and about breaks even compared to the expected state without the beamsplitter. The supporting derivations and full analysis are provided in the Supplemental Material. With improved transmission and photon-subtraction-induced entanglement increase, this teleportation fidelity improvement could rise to about 0.1~\cite{PhysRevA.107.012418}. 
\textit{Conclusion}---To generate the presented resource state, we developed a source of two-mode photon subtraction on frequency non-degenerate two-mode squeezed vacuum. This development was enabled by frequency non-degenerate two-mode cavity filtering indirectly locked to the frequency modes of the LOs and squeezed-light modes, with negligible added photon-subtraction background counts, which has not been previously demonstrated.

Additionally, leveraging two-mode generalized photon-subtraction~\cite{PhysRevA.103.013710,PhysRevA.88.043818,PhysRevA.67.062320}, the non-Gaussian teleportation rate could be increased dramatically. 
Moreover, this type of resource could enable improved quantum sensing leveraging the greater squeezing and entanglement. 
Finally, with some further development, operation of the PSO and HDS in closer coordination enables (1) breeding of non-Gaussian states~\cite{XanaduNature2025a} and (2) feedforward for non-Gaussian measurement-based quantum computing. 

In summary, we have demonstrated a heralded resource state generator capable of real-time use for a variety of applications with a focus on non-Gaussian teleportation. This is enabled by numerous advancements in the squeezed-light source, photon-subtraction system, as well as the homodyne detection, and newly developed data acquisition systems. These methods are useful not only towards non-Gaussian teleportation but also for other impactful protocols and applications of quantum technology.

\textit{Acknowledgements}---We acknowledge Muneer Alshowkan and Nageswara Rao for linux and conventional networking assistance. We acknowledge Christopher M. Seck for assistance with cavity locking. We acknowledge Trevor Michelson and Blake Van Hoy for cavity vibration testing assistance. We acknowledge Jack Postlewaite for assistance with experimental setup. This work was performed at Oak Ridge National Laboratory, operated by UT-Battelle for the U.S.\ Department of Energy under contract No.\ DE-AC05-00OR22725. Funding was provided by the U.S. Department of Energy, Office of Science, Office of Advanced Scientific Computing Research, through
the Transparent Optical Quantum Networks for Distributed Science Program,
the Early Career Research Program,
the Quantum Internet to Accelerate Scientific Discovery program, and
the Accelerated Research in Quantum Computing program 
(Field Work Proposals ERKJ355, ERKJ353, ERKJ420, ERKJ381, ERKJ445).

 J.C.C. designed and constructed the experimental setup. J.C.C. developed photon-subtraction system orchestrator and homodyne detection servers. J.C.C. collected all measurements.  J.C.C. devised measurements and analysis for experiment. J.C.C. and Y.Z. analyzed the data and ran simulations. Y.Z., J.M.L. and J.C.C. developed the two-mode tomography analysis used and implemented it in software. A.M.M. assisted in designing the phase and cavity stabilization systems. Y.Z., E.D., and Y.W. developed the quantum state simulations for the fidelity calculations. N.A.P supervised project and provided direction for experimental design, measurements, and analysis. All authors contributed to manuscript preparation.

\let\oldaddcontentsline\addcontentsline
\renewcommand{\addcontentsline}[3]{}
%

\let\addcontentsline\oldaddcontentsline

\clearpage \newpage
\pagenumbering{arabic}
\renewcommand{\thetable}{S\arabic{table}}
\renewcommand{\theequation}{S\arabic{equation}}
\renewcommand{\thepage}{S\arabic{page}}
\setcounter{table}{0}
\setcounter{equation}{0}

\newcounter{offset}
\setcounter{offset}{\value{figure}}
\renewcommand{\thefigure}{S\the\numexpr\value{figure}-\value{offset}\relax}

\setcounter{section}{0}
\setcounter{secnumdepth}{4}
\setcounter{tocdepth}{4}
\makeatletter
\def\l@paragraph#1#2{%
  \begingroup
  \def\@dotsep{10000}%
  \let\oldnumberline\numberline
  \def\numberline##1{\oldnumberline{\makebox[0.6em][r]{##1}.}}%
  \@dottedtocline{4}{4.7em}{1.3em}{#1}{#2}%
  \endgroup
}
\renewcommand{\paragraph}{%
  \@startsection{paragraph}{4}{\z@}%
  {3.25ex \@plus 1ex \@minus .2ex}%
  {1.5ex \@plus .2ex}%
  {\normalfont\normalsize\itshape\noindent}%
}
\makeatother

\allowdisplaybreaks

\onecolumngrid
\begin{center}
\bf\large Supplemental Material for\\``Real-time heralded non-Gaussian teleportation resource-state generator''
\end{center}
\tableofcontents
\section{Quantum non-Gaussian testing}\label{sec:nonGauss_test}
Given the name, non-Gaussian teleportation, it is valuable to consider if the resource state used therein has quantum non-Gaussian characteristics. These characteristics can be seen by Wigner negativity and other tests~\cite{PRXQuantum.2.030204} that are especially useful for states without Wigner negativity but that are still quantum non-Gaussian. Specifically for two-mode states, besides Wigner negativity, we find three methods applicable: (1) Stellar Rank~\cite{PRXQuantum.2.020333,Chabaud2025}, (2) Wigner function amplitude at the origin~\cite{PhysRevA.87.062104}, and (3) the non-Gaussianity~\cite{PhysRevA.76.042327}. Given a density matrix, method (1) is easiest since it just requires the calculation of the fidelity with a Fock state to say it has at least the rank of the Fock state. It is thus a discrete metric and not one that can be used for a more continuous quantification of the non-Gaussianity. Method (1) also can only provide a lower bound on the stellar rank. Methods (2) and (3) provide a more continuous quantification. Method (2) is fairly straightforward to calculate if the Wigner function is already calculated, but this method is known to not work accurately for all quantum non-Gaussian states~\cite{PRXQuantum.2.030204}. Method (3) provides the most rigorous and quantitative measure of the quantum non-Gaussian character of a state, but is also by far the most difficult to calculate, at least in the multi-mode case due to the calculation of the reference state with the same mean and covariance as the input.

Our tomography method natively outputs the density matrix in the Fock basis, so we will focus on method (1) for which we calculate the fidelity $F$ with respect to $\ket{1}\otimes\ket{1}$ (see Ref.~\cite{PRXQuantum.2.020333} Appendix F and related Erratum~\cite{Chabaud2025}). If $F(\rho,\ket{1}\otimes\ket{1})\leq0.25$, the state has stellar rank 0+ and is likely not quantum non-Gaussian. If $0.25<F(\rho,\ket{1}\otimes\ket{1})<0.532$, the state is quantum non-Gaussian with stellar rank 1+. If $F(\rho,\ket{1}\otimes\ket{1})>0.532$, the state is quantum non-Gaussian with rank 2+. 

Numerically, we find that the photon-subtracted TMSV state has stellar rank $>0$ only for low loss ($\eta>0.85$, agreeing with the middle panel in \cref{fig:stellar_rank_fid}) and certain ranges of combinations of the squeezing parameter and photon-subtraction beamsplitter reflectivity. To approximately maximize $F(\rho,\ket{1}\otimes\ket{1})$ here are some different parameter combinations we found numerically, e.g., $r=0.5$ and $R_S=0.01$, $r=0.6$ and $R_S=0.1$, or $r=0.7$ and $R_S=0.2$. Adjusting the parameters away from these optimal pairings will reduce the fidelity; for example, given $R_S=0.01$, the stellar rank is still 1+ from $r=0.5$ until down to about $r=0.31$. By observation, we find that the optimal $r$ increases as does $R_S$ by a similar amount to preserve the optimal $F(\rho,\ket{1}\otimes\ket{1})$ with Stellar rank of 1+ up to $r=0.8$ and $R_S=0.3$; at $r=0.9$ and $R_S=0.4$ the fidelity starts to drop but the Stellar rank is still 1+ for low enough loss (which gets more stringent) until $r=1.1$ and $R_S=0.6$. $r=1.2$ and $R_S=0.7$ have stellar rank of 0+ even with $\eta=1$.

Due to the losses in our experiment, our measured states have stellar rank 0+ and would still be rank 0+ even if our experimental $r$ would be higher. Non-Gaussian teleportation is so named not because the resource state is necessarily quantum non-Gaussian, though it can be, but because a non-Gaussian operation (photon-subtraction) is used to deGaussify the entangled resource state to improve the teleportation fidelity.

\section{Two-mode tomography methods}
\label{sect:mle}

Quantum state tomography seeks to reconstruct the full density matrix of a
quantum system from measurement data. For continuous-variable (CV) optical
systems, such as entangled two-mode states, this reconstruction is typically
based on homodyne detection, which measures field quadratures at various
local oscillator phases. By collecting sufficient quadrature samples
\((x_1, x_2)\) over a range of phase settings \((\theta_1, \theta_2)\),
one can infer the two-mode density matrix \(\rho_{1,2}\) that best describes
the experimental data. Because CV systems formally occupy an infinite-dimensional Hilbert space, the reconstruction is carried out in a truncated photon-number basis up to a cut-off \(n_c\) per mode, chosen to capture the relevant photon-number support of the measured state.  In what follows, we first describe the state and measurement representations in the truncated photon-number basis, and then present the tomography methods used for state reconstruction.


For two optical modes \(1\) and \(2\), each truncated to a maximum photon number
\(n_c\), the joint Hilbert space is spanned by the photon-number basis
\(\{\,\ket{n, m}\,\}_{n,m=0}^{n_c}\),
with total dimension \(D = (n_c+1)^2\).
A general density matrix \(\rho\) in this basis can be written as
\begin{align}
    \rho
 &= \sum_{n,n'=0}^{n_c} \sum_{m,m'=0}^{n_c}
    \rho_{ n m,\, n' m'} \op{n,m}{n',m'},
\end{align}
where each coefficient \(\rho_{n m,\, n' m'}\) is a complex number satisfying
\begin{align*}
    \rho &= \rho^\dagger, \qquad
    \Tr(\rho) = 1, \qquad
    \rho \ge 0.
\end{align*}


For each optical mode, the homodyne measurement corresponds to a projective
measurement onto the eigenstates of the quadrature operator
\begin{align*}
    \hat{x}_\theta 
 &= \frac{1}{\sqrt{2}}\left(\hat{a} e^{-i\theta} + \hat{a}^\dagger e^{i\theta}\right),
\end{align*}
where $\theta$ is the local oscillator phase and $\hat{a}$ ($\hat{a}^\dagger$)
denote the annihilation (creation) operators.
The corresponding quadrature eigenstate $\ket{x_\theta}$ satisfies
\begin{align*}
\hat{x}_\theta \ket{x_\theta} = x_\theta \ket{x_\theta}.
\end{align*}
Here, the eigenvalue $x_\theta$ must be real because $\hat{x}_\theta$ is Hermitian. At $\theta = 0$, denote the position operator $\hat{x}_0 = \hat{x}$ and the eigenvalue $x_0 = x$. Using commutation algebra, we can write the quadrature operator $\hat{x}_\theta = e^{i\theta \hat{n}} \hat{x} e^{-i\theta \hat{n}}$, where $\hat{n} = \hat{a}^\dagger \hat{a}$, as the Heisenberg picture operator of the position operator $\hat{x}$ under the simple harmonic oscillator Hamiltonian. Then, it is easy see the eigenstate of $\ket{\hat{x}_\theta} = e^{i\theta\hat{n}} \ket{x}$, with the eigenvalue $x_\theta = x$ (because $\hat{x}_\theta  e^{i\theta\hat{n}} \ket{x} = e^{i\theta\hat{n}} \hat{x} \ket{x} = x \ket{\hat{x}_\theta}$). So the eigenvalue is independent of the quadrature phase.

In the photon-number basis $\{\ket{n}\}_{n=0}^{n_c}$,
the wavefunction of this state is given by
\begin{align*}
    \ip{n}{x_\theta}
  = \mel{n}{e^{i\theta \hat{n}}}{x}
 &= \frac{1}{\pi^{1/4} \sqrt{2^n n!}}
    e^{-i n \theta}
    H_n(x_\theta) e^{-x_\theta^2/2},
\end{align*}
where $H_n(x)$ is the Hermite polynomial of order $n$ and the convention that $\hbar=1$ is used [see Eqs.~(26) and (44) of Ref.~\cite{RevModPhys.81.299}]. In the last step, we used the harmonic oscillator wavefunction $\ip{n}{x} = \psi_n(x)$ and plugged in back $x = x_\theta$. Hence, the local homodyne measurement operator corresponding to measuring quadrature value $x_\theta$ is
\begin{align}
    \Pi(x_\theta; \theta)
 &= \op{x_\theta} \notag\\
 &= \sum_{m,n=0}^{n_c}
    \bra{m}\ket{x_\theta}\bra{x_\theta}\ket{n}
    \op{m}{n} \notag\\
 &= \frac{e^{-x_\theta^2}}{\sqrt{\pi}}
    \sum_{m,n=0}^{n_c}
    \frac{H_m(x_\theta) H_n(x_\theta)}{\sqrt{2^{(m+n)} m! n!}}
    e^{-i (m-n)\theta}
    \op{m}{n}.
\end{align}

For a two-mode homodyne measurement on modes \(1\) and \(2\)
with respective phases \(\theta_1\) and \(\theta_2\),
the joint measurement operator for outcomes \(x_1\) and \(x_2\) is simply the tensor product:
\begin{equation}
\Pi(x_1,x_2;\theta_1,\theta_2)
  = \Pi_1(x_1;\theta_1)\,\otimes\,\Pi_2(x_2;\theta_2).
\end{equation}
The corresponding joint probability density for observing quadrature
values $(x_1, x_2)$ given local oscillator phases
$(\theta_1, \theta_2)$ is
\begin{equation}
p(x_1, x_2 | \theta_1, \theta_2)
  = \operatorname{Tr}\!\left[
      \rho\,
      \Pi(x_1,x_2;\theta_1,\theta_2)
    \right].
\label{eq:homodyne_prob}
\end{equation}
This probability distribution provides the fundamental link between
experimental homodyne data and the reconstructed density matrix
$\rho$ in the photon-number basis, and forms the basis of
maximum-likelihood quantum state tomography.

Reconstructing the quantum state of a two-mode CV system requires estimating the density matrix $\rho$ that best explains a collection of homodyne measurement data. Each mode's quadrature is measured at a range of local oscillator phases, yielding samples of the joint quadrature distribution $p(x_1, x_2 | \theta_1, \theta_2)$. 
To reconstruct the state, the maximum-likelihood method is employed. Specifically, the \emph{R$\rho$R} iterative algorithm~\cite{Lvovsky2004} is used to maximize the log-likelihood functional
\begin{equation}
\mathcal{L}(\rho) = \sum_{x_1, x_2, \theta_1, \theta_2} f(x_1,x_2|\theta_1,\theta_2) \ln \!\big( \operatorname{Tr}[\Pi(x_1,x_2;\theta_1,\theta_2) \rho] \big),
\end{equation}
where $f(x_1,x_2|\theta_1,\theta_2)$ are the observed frequencies of 
quadrature values \((x_1,x_2)\) given local phases \((\theta_1,\theta_2)\). At each iteration, the density matrix is updated as
\begin{equation}
\rho_{k+1} = \frac{R(\rho_k)\, \rho_k\, R(\rho_k)}{\operatorname{Tr}[R(\rho_k)\, \rho_k\, R(\rho_k)]},
\end{equation}
where
\begin{align}
R(\rho_k)
  &= \sum_{x_1, x_2, \theta_1, \theta_2}
     \frac{f(x_1,x_2|\theta_1,\theta_2)}
          {p_k(x_1,x_2|\theta_1,\theta_2)}\,
     \Pi(x_1,x_2;\theta_1,\theta_2),
\label{eq:R_operator}
\end{align}
and
\begin{equation}
p_k(x_1,x_2|\theta_1,\theta_2)
  = \operatorname{Tr}\!\left[
      \Pi(x_1,x_2;\theta_1,\theta_2)\,\rho_k
    \right]
\label{eq:predicted_probs}
\end{equation}
are the probabilities predicted by the current estimate $\rho_k$.
Although this iterative rule does not guarantee monotonic increase of the
likelihood~\cite{PhysRevA.75.042108}, the stopping criterion formulated in
Ref.~\cite{Glancy_2012} provides an upper bound on the possible
remaining likelihood improvement. We adopt this criterion to terminate
the iterations once further updates yield negligible change in the
log-likelihood value.

In practice, this reconstruction method efficiently handles large homodyne datasets. The photon-number truncation defines the reconstruction subspace, while the homodyne quadrature data ensure informational completeness. In our implementation, we set \(n_c=6\) for the maximum-likelihood estimation in view of the justification presented in the next paragraph.

\begin{figure}[htbp]
    \centering
    \includegraphics[width=0.8\linewidth]{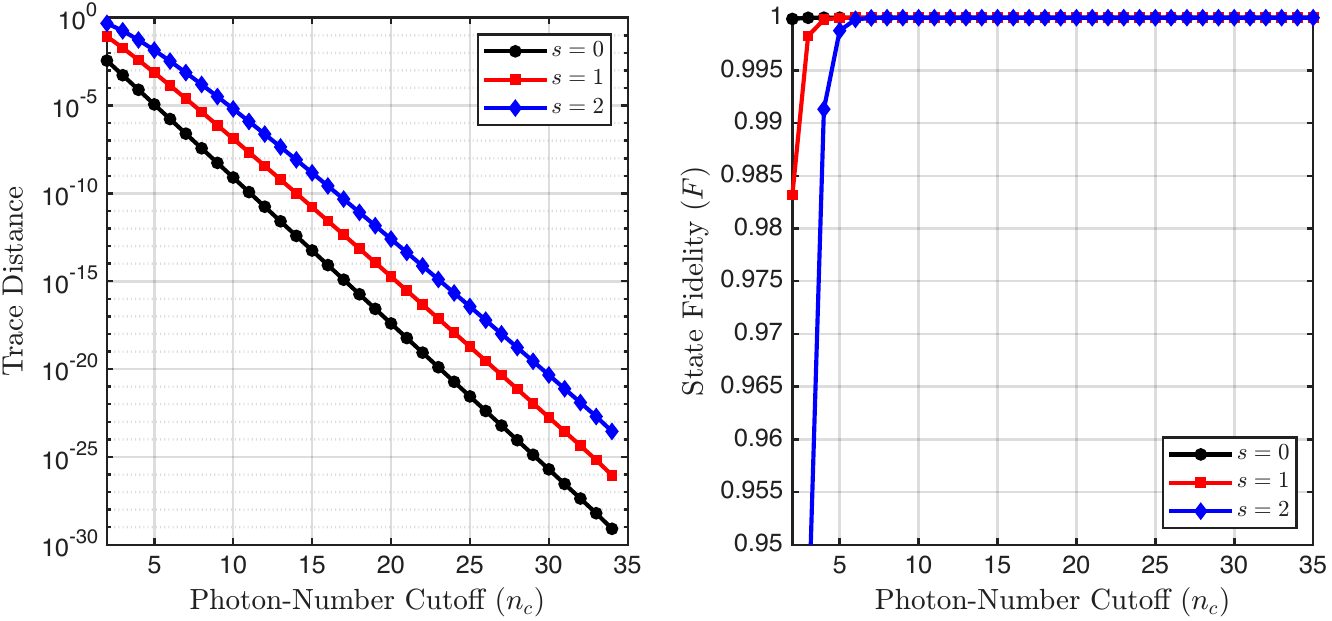} 
    \caption{Theoretical convergence of the truncated state as a function of the photon-number cut-off \(n_c\). The left panel displays the trace distance (on a logarithmic scale) and the right panel displays the quantum state fidelity (on a linear scale) between the state truncated at \(n_c\) and a high-dimensional reference state evaluated at \(n_{c,\max} = 35\). The convergence is evaluated for zero, one, and two subtracted photons per mode (\(s = 0, 1, 2\)). The state is generated utilizing an experimentally realistic squeezing parameter of \(r=0.3\), transmission efficiencies of \(\eta_1 = 0.55\) and \(\eta_2 = 0.5\), and subtraction beam-splitter reflectivities of \(R_1 = R_2 = 0.14\). }
    \label{fig:cut-off_robustness}
\end{figure}

To establish a physically robust and computationally stable cut-off, we analyzed the theoretical truncation error of the expected state.  Specifically, we generated the density matrices for various photon-subtracted TMSV states across a range of truncation values \(n_c\), utilizing an experimentally accessible squeezing parameter \(r=0.3\) and accounting for experimental transmission efficiencies \(\eta_1 = 0.55\) and \(\eta_2= 0.5\). We then computed the trace distance and state fidelity between these truncated states and a high-dimensional reference state,  which is generated with the same experimental parameters but evaluated at a sufficiently high cut-off of \(n_{c, \max} = 35\). Given the parameters used for state generation, this reference cut-off is high enough to ensure that the tail of the photon-number distribution is completely captured. As shown in Fig.~\ref{fig:cut-off_robustness}, the truncated state rapidly converges. At a cut-off of \(n_c = 6\) and when the number of photons subtracted from each mode \(s\) is no more than \(1\), the trace distance drops below \(1.4 \times 10^{-4}\), meaning that the maximum probability of distinguishing
between the truncated state and the reference state by any physical measurement is less than \(1.4 \times 10^{-4}\). Concurrently, the state fidelity reaches near-perfect unity; for example, when \(s=1\) and \(n_c=6\), the state fidelity is \(F=1-1.5\times 10^{-6}\).
This definitively illustrates that a truncation at \(n_c = 6\) is sufficient to capture the physical properties of the ground-truth state with \(s\leq 1\), providing a rigorous foundation for restricting the maximum-likelihood reconstruction to the subspace of up to \(n_c = 6\) photons per mode. 

\begin{figure}[htbp]
    \centering
    \includegraphics[width=0.8\linewidth]{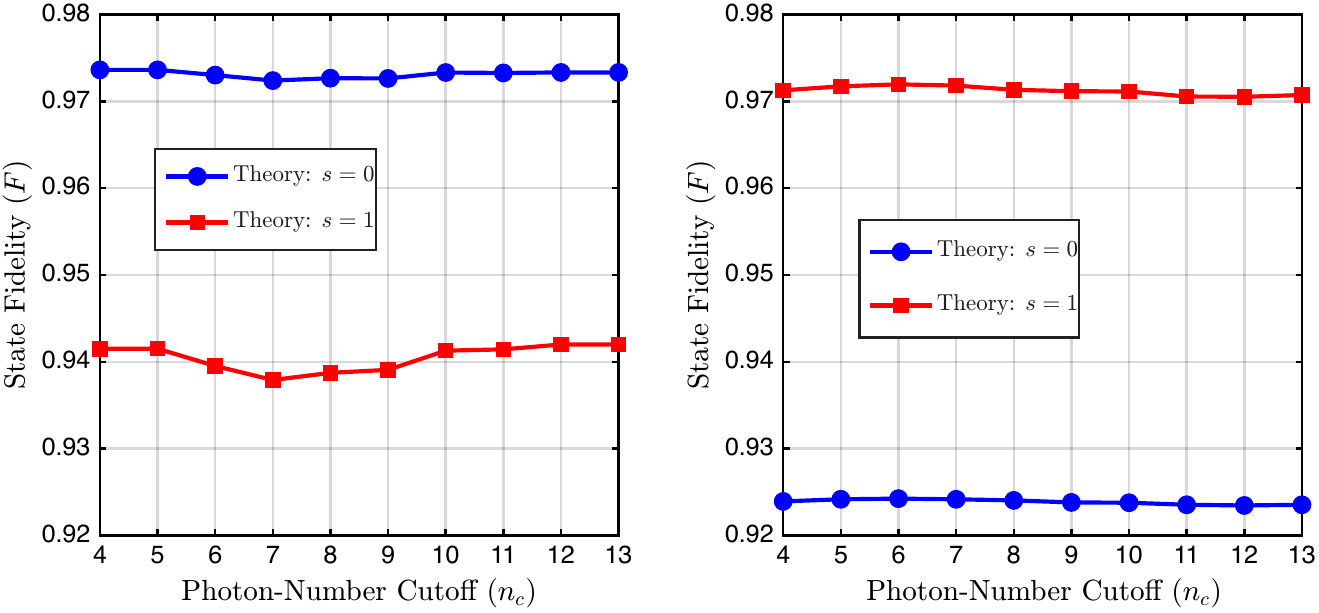} 
    \caption{Empirical convergence of the state fidelity for the reconstructed density matrices. The left panel shows the fidelities of states reconstructed using $(0,0)$-photon subtracted experimental data, compared against high-dimensional theoretical reference states evaluated at $n_{c,\max} = 35$ with $s$ photons subtracted from each mode (blue circles for $s=0$ and red squares for $s=1$). The right panel shows the corresponding fidelities for states reconstructed using $(1,1)$-photon subtracted experimental data. In both cases, the reconstructed states correctly align with their respective theoretical targets, and the state fidelity tightly saturates at a photon-number cut-off of $n_c = 6$. The negligible variation in fidelity beyond $n_c = 6$ empirically confirms that this cut-off is sufficient to capture the physical structure of the ground-truth state without introducing truncation artifacts.}
    \label{fig:mle_fidelity}
\end{figure}

The density matrix $\rho_{\mathrm{MLE}}$ obtained by maximum-likelihood estimation (MLE) faithfully reproduces both marginal and joint quadrature distributions and allows one to compute derived quantities such as the Wigner function, fidelity, purity, and entanglement measures of the two-mode CV state. When we reconstructed the $(0,0)$-photon subtracted and $(1,1)$-photon subtracted density matrices, we varied the photon-number cut-off $n_c$ up to $n_c = 13$ to evaluate the empirical robustness of the reconstruction. (We truncated our empirical sweep at $n_c = 13$ because the memory and runtime scaling of MLE becomes computationally intractable for larger cut-offs on a personal computer.) For this, we computed the state fidelity between the reconstructed states by MLE and their corresponding high-dimensional theoretical reference states evaluated at $n_{c,\max} = 35$. As shown in Fig.~\ref{fig:mle_fidelity}, the experimental data demonstrates that the state fidelity saturates at $n_c = 6$; increasing the cut-off beyond this threshold just yields negligible variations, with the fidelity fluctuating by less than $\Delta F = 0.005$ across the entire evaluated range. 

We note that utilizing excessively large cut-offs provides the MLE method with surplus free parameters, causing it to systematically overfit finite statistical noise in the experimental data. This noise fitting populates high photon-number states, which barely alters the state fidelity as shown in Fig.~\ref{fig:mle_fidelity}. Furthermore, we numerically confirmed that the expected teleportation fidelity, using the reconstructed density matrix $\rho_{\mathrm{MLE}}$ as the resource state, is insensitive to the chosen cut-off. 
 However, we observed that the overfitting causes non-linear, tail-sensitive metrics such as the logarithmic  negativity to artificially increase with $n_c$, falsely inflating the measured entanglement. In contrast,  
 unlike logarithmic negativity, which artificially inflates due to the summation of many small, noise-sensitive negative eigenvalues in the expanded Hilbert space, the most negative eigenvalue of the partial transpose of $\rho_{\mathrm{MLE}}$~\cite{Eisert01011999} provides a stable lower bound on the total negativity, and consequently, the logarithmic negativity. Although it is not a proper measure of total entanglement, our numerical evaluations confirm its utility as a noise-robust entanglement indicator. In particular, we observed that the most negative eigenvalue is insensitive to the chosen cut-off [Table~\ref{tab:negeig}], successfully isolating the dominant physical quantum correlations from the experimental noise.  Given these observations and the results presented in Figs.~\ref{fig:cut-off_robustness} and~\ref{fig:mle_fidelity}, setting $n_c = 6$ for the state reconstruction by MLE is a robust choice that faithfully captures the ground-truth state while actively preventing the algorithmic introduction of spurious entanglement artifacts.

\begin{table}[]
    \centering    
\caption{Comparison of Log negativity and the most negative eigenvalue of the partial transpose of $\rho_{\mathrm{MLE}}$ versus photon-number cut-off $n_c$ for the $0,0$- and $1,1$-photon-subtraction measurements.}
    \label{tab:negeig}
    \begin{tabular}{ccccc}
    \hline\hline
       $n_c$ & $E_\mathcal{N}(\rho^M_{0,0})$ & $E_\mathcal{N}(\rho^M_{1,1})$ & $|\min(\lambda^M_{0,0})|$& $|\min(\lambda^M_{1,1})|$ \\
       \hline
       4 & 0.42 & 0.47 & 0.1110 & 0.1193  \\
       5 & 0.46 & 0.50 & 0.1104 & 0.1198 \\
       6 & 0.49 & 0.52 & 0.1120 & 0.1197  \\
       7 & 0.52 & 0.55 & 0.1135 & 0.1193  \\
       8 & 0.54 & 0.57 & 0.1134 & 0.1198  \\
       9 & 0.57 & 0.61 & 0.1127 & 0.1196  \\
       10 & 0.60 & 0.63 & 0.1120 & 0.1197  \\
       11 & 0.63 & 0.66 & 0.1121 & 0.1200 \\
       12 & 0.66 & 0.69 & 0.1117 & 0.1199  \\
       13 & 0.68 & 0.72 & 0.1118 & 0.1203 \\
       \hline\hline
       \end{tabular}
\end{table}

\section{Photon-subtraction state derivation}
\begin{figure}
  \centering
  \includegraphics[scale=1]{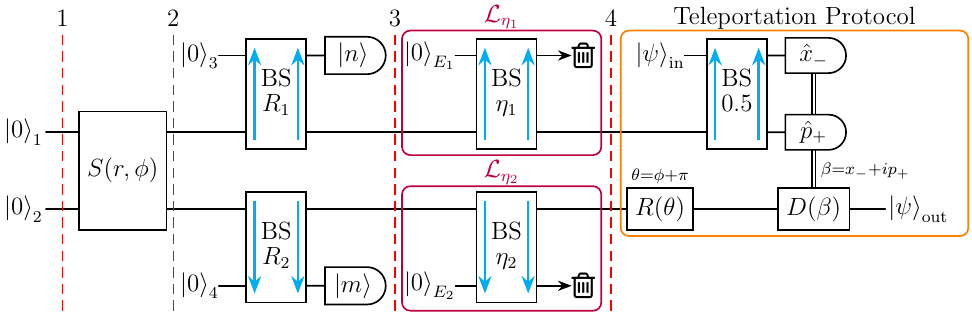}
  \caption{Circuit representation of the protocol with potential photon loss before and after subtraction. 
  Without loss, the result is \cref{eq:PSTMS}. Including loss from the last set of ancillas, the result is \cref{eq:rho_loss}.
  The arrows in beamsplitter (BS) point to the ancillary qumodes, indicating the direction of subtraction and loss.
  Two-mode squeezing operator $S(r,\phi) = {\exp}[r(e^{-i\phi}\hat{a}_1\hat{a}_2 - e^{i\phi}\hat{a}_1^\dagger\hat{a}_2^\dagger)]$.
  The Braunstein--Kimble unconditional quantum teleportation protocol  in the orange box is included for conceptually understanding the teleportation fidelity calculation, but it is beyond the scope of experiment in this work. The full circuit is a simplified theoretical model of the diagram in \cref{fig:NGTelsimpsetup} in the main text.
  The optical phase shifter (phase-space rotation gate) $R(\theta) = {\exp}(-i\theta \hat{a}_2^\dagger\hat{a}_2)$ with $\theta = \phi + \pi$, and the displacement operator $D(\beta) = {\exp}(\beta \hat{a}_2^\dagger - \beta^*\hat{a}_2)$.
  $R(\theta = \phi + \pi)$ gate is necessary to rotate the effective phase of the resource state to $\pi$ in order to match Alice's choice of continuous variable Bell basis measurements and Bob's feed-forward displacement in the teleportation protocol. Although we apply the phase shift to mode 2 on Bob's side in our calculation, experimentally it may be more advantageous to equivalently (if the input state's characteristic function has isotropic magnitude $|\chi(\alpha)|$ in phase space) apply the phase shift to mode 1 on Alice's side instead because this mode will be measured and discarded.
  }
  \label{fig:circuit}
\end{figure}

A normalized two-mode squeezed vacuum state can be expressed using the unitary squeezing operator $S_2(r,\phi)$ with a squeezing parameter $s = re^{i\phi}$, where $r \geq 0$ and $\phi \in [0, 2\pi)$, as follows:
\begin{align}
    \ket{\text{TMSV}}
 &= S_2(r,\phi) \ket{0}_{1} \otimes \ket{0}_2
 \equiv \exp[ r(e^{-i\phi} a_1 a_2 - e^{i\phi}a_1^\dagger a_2^\dagger) ] \ket{00}
  = \frac{1}{\cosh r} \exp(-a_1^\dagger a_2^\dagger e^{i\phi}\tanh r) \ket{00}.
  \label{eq:S2_r_phi}
\end{align}
We suppress the mode indices with the shorthand notation $\ket{0}_1\otimes\ket{0}_2 \equiv \ket{00}$ when there is no confusion. A Baker-Campbell-Hausdorff (BCH) type of disentangling formula is used in the final step above. For simplicity, assume the phase $\phi = 0$ and denote $\tanh r = t_r$ and $\cosh r = c_r = \sqrt{1-t_r^2}$. Applying Taylor series expansion to the above and using $(a^{\dagger})^n \ket{0} = \sqrt{n!} \ket{n}$, we find
\begin{align}
    \ket{\text{TMSV}}
 &= \sqrt{1-t_r^2} \sum_{k=0}^{\infty} \frac{(-a^{\dagger}_1 a^{\dagger}_2 t_r)^k}{k!} \ket{00}
  = \sqrt{1-t_r^2} \sum_{k=0}^{\infty} (-t_r)^k \ket{kk}.
 \label{eq:TMSV}
\end{align}

Next, consider a series of transformations with photons being subtracted on auxiliary modes $3$ and $4$, which can be modeled, as well as experimentally achieved, by a beamsplitter. A beamsplitter (BS) acting between the $i$ and $j$ modes, denoted as $\text{BS}(\theta_{ij})$, is given by $\text{BS}_{ij}(\theta, \phi) = {\exp}[\theta_{i,j}(e^{i\phi} a_i a^\dagger_j - h.c.)]$. For $\phi=\pi$,  this simplifies to $\text{BS}(\theta, \phi=\pi)= \exp{\theta_{i,j}(a^\dagger_i a_j - a_i a^\dagger_j)}$. In the Heisenberg picture, the BS's transform field operators as,
\begin{align}
    \text{BS}^\dagger(\theta_{i,j}) \mqty[a_i \\ a_j]\text{BS}(\theta_{i,j})
 &= \mqty[ \cos\theta_{i,j} & \sin\theta_{i,j} \\
          -\sin\theta_{i,j} & \cos\theta_{i,j}]
    \mqty[a_i \\ a_j], \label{eq:Bn}
\end{align}
and similarly for $(a_i^\dagger, a_j^\dagger)^{T}$.

After the TMSV of modes $(1,2)$ is entangled with modes $(3,4)$ using beamsplitters $\text{BS}(\theta_{1,3})$ and $\text{BS}(\theta_{2,4})$, the $(n,m)$ photon subtraction on modes $(3,4)$ results in the following unnormalized state:
\begin{align*}
    \ket{\psi_{n,m}}
 &= (\openone \otimes \bra{n,m}_{3,4})
    \qty[\text{BS}(\theta_{1,3})\text{BS}(\theta_{2,4}) \ket{\text{TMSV}}_{1,2} \otimes \ket{00}_{3,4}] \\
 &= \sqrt{1-t_r^2} \sum_{k=0}^{\infty} (\openone \otimes \bra{n,m}_{3,4})
    \qty[\text{BS}(\theta_{1,3})\text{BS}(\theta_{2,4})
    \frac{(-a_1^\dagger a_2^\dagger t_r)^k}{k!}
    \text{BS}^\dagger(\theta_{1,3})\text{BS}^\dagger(\theta_{2,4})
    \text{BS}(\theta_{1,3})\text{BS}(\theta_{2,4})
    \ket{\text{00}}_{1,2}\otimes \ket{00}_{3,4}],
\end{align*}
where the identity $\openone = \text{BS}^\dagger(\theta)\text{BS}(\theta)$ is inserted to make the Heisenberg picture transformation explicit.

It is convenient to introduce the experimental reflectivity $R_i = \sin^2(\theta_{i,j})$.
Heisenberg evolving the field operators from \cref{eq:TMSV} through the BS-gates and subsequently projecting with the state where ($n,m$) photons are measured results in
\begin{align}
    \ket{\psi_{n,m}}
 &= \sqrt{1-t_r^2}\sum_{k=0}^{\infty}\frac{(-t_r)^k}{k!} \mel**{n,m}{\qty(\sqrt{1-R_1}a_1^\dagger - \sqrt{R_1}a_3^\dagger)^k  \qty(\sqrt{1-R_2}a_2^\dagger - \sqrt{R_2}a_4^\dagger)^k}{00}_{3,4} \ket{00}_{1,2} \notag\\
 &= \sqrt{1-t_r^2} \sum_{k=\max(n,m)}^{\infty}\frac{(-t_r)^k}{k!}
    \mel**{n}{\binom{k}{n} \qty(\sqrt{1-R_1}a_1^\dagger)^{k-n} \qty(- \sqrt{R_1}a_3^\dagger)^{n}}{0}_{3} \notag \\
 & \mspace{90mu} \times \mel**{m}{\binom{k}{m} \qty(\sqrt{1-R_2}a_2^\dagger)^{k-m} \qty(-  \sqrt{R_2}a_4^\dagger)^{m}}{0}_{4} \ket{00}_{1,2} \\
 &= (-1)^{n+m}\sqrt{1-t_r^2}\sum_{k=\max(n,m)}^{\infty} (-t_r)^k B_{k,n}(R_1) B_{k,m}(R_2) \ket{k-n,k-m}_{1,2} \\
 &\equiv (-1)^{n+m}\sqrt{1-t_r^2} \sum_{k=\max(n,m)}^{\infty} c_{k,n,m} \ket{k-n,k-m}_{1,2},
\label{eq:PSTMS_suppl}
\end{align} 
which is the unnormalized state given in \cref{eq:PSTMS} in the main text (without the prefactor).

In the last step we used a compact notation (following the parameterization form of the probability mass function of a binomial distribution), 
\begin{align}
B_{k,n}(R) \coloneqq \sqrt{\binom{k}{n} (1-R)^{k-n} R^{n} },
\end{align}
for the beamsplitter binomial factors arising when $(k-n)$ photons are transmitted and $n$ are reflected. The normalized photon subtracted state is $\ket{\Psi(n,m)} \equiv \ket{\psi_{n,m}}/\norm{\ket{\psi_{n,m}}}$.
%
The normalization for the unnormalized state given in \cref{eq:PSTMS} is $C_{n,m} = [\sum_{k = \max(n,m)} c_{k,n,m}^2]^{1/2}$, which can be calculated as follows:
\begin{align}
    C^{2}_{n,m}
 &= \sum_{k=\max(n,m)}^{\infty} t_r^{2k}
    \binom{k}{n}\binom{k}{m} (1-R_1)^{k-n}R_1^n (1-R_2)^{k-m} R_2^m \\
 &= \frac{R_1^n R_2^m}{n! m! (1-R_1)^n(1-R_2)^m}
    \sum_{k=0}^{\infty}
    \eval[{x^n y^m \pdv[n]{x} \pdv[m]{y} x^k y^k}|_{\substack{x=1\\ y=t_r^2 (1-R_1)(1-R_2)}} \\
 &= \frac{R_1^n R_2^m t_r^{2m}(1-R_1)^{m-n}}{n! m!} 
    \eval[{\pdv[n]{x} \pdv[m]{y} \frac{1}{1-xy}}|_{\substack{x=1\\ y=t_r^2 (1-R_1)(1-R_2)}} \\
 &= \frac{R_1^n R_2^m t_r^{2m}(1-R_1)^{m-n}}{n!} 
    \eval[{\pdv[n]{x} x^{m}[1-t_r^2 (1-R_1)(1-R_2) x]^{-(m+1)}}|_{\substack{x=1}}.
\end{align}
Therefore, the probability of obtaining the $(n,m)$-photon subtracted resource state in \cref{eq:PSTMS_suppl} is $p_{n,m} = \norm{\ket{\psi_{n,m}}}^2 = (1-t_r^2) C^2_{n,m}$. For $n=m=j$ and $R_1 = R_2 = R_S$, $c_{k,j,j} = (-t_r)^{k} \binom{k}{j} (1-R_S)^{k-j}R_S^{j}$. 

Consider experimentally relevant values $(n,m) = (0,0)$ and $(1,1)$, and assume $R_1 = R_2 = R_S$. Using the above results, we have
\begin{subequations}
\begin{align}
    C^2_{0,0} &= \frac{1}{1-t_r^2 (1-R_S)^2}, \\
    C^2_{1,1} &= \frac{t_r^2 R_S^2 [1+t_r^2 (1-R_S)^2]}{[1-t_r^2 (1-R_S)^2]^3}, \\
    p_{0,0} &= \frac{1-t_r^2}{1-t_r^2 (1-R_S)^2}, \\
    p_{1,1} &= \frac{t_r^2 R_S^2 (1-t_r^2) [1+t_r^2 (1-R_S)^2]}{[1-t_r^2 (1-R_S)^2]^3}, \\
    c_{k,0,0} &= (-t_r)^k (1-R_S)^k,\quad
    c_{k,1,1}  = k(-t_r)^k (1-R_S)^{k-1}R_S, \\
    \ket{\Psi_{0,0}}
 &= \sqrt{1-\lambda^2}\sum_{k=0}^{\infty}(-\lambda)^k \ket{k,k},
 \ (\lambda \equiv t_r(1-R_S)), \label{eq:Psi_00} \\
    \ket{\Psi_{1,1}}
 &= -\sqrt{\frac{(1-\lambda^2)^3}{1 + \lambda^2}} \sum_{k=0}^{\infty} (k+1)(-\lambda)^{k} \ket{k,k}. \label{eq:Psi_11}
\end{align}
\end{subequations}
In the normalized state above we defined an effective squeezing parameter $\lambda = t_r(1-R_S) \in [0, 1]$ so that $\ket{\Psi_{1,1}}$ has a similar form as Eq.~(3) in Ref.~\cite{PhysRevA.65.062306} up to some phase differences.
$\ket{\Psi_{0,0}}$ has the same form as the input TMSV with reduced squeezing, thus reduced entanglement, and same zero stellar rank, compared to the input TMSV.
For low reflectivity ($R_S = \sin^2\theta_S \approx \theta_S^2 \ll 1$), we have $\lambda \approx \tanh r$, $p_{0,0} \approx 1$ and $p_{1,1} \approx \theta_S^4  \lambda^2(1+\lambda^2)/(1-\lambda^2)^2 \ll 1$. 
This means there is a low success probability of obtaining the photon subtracted resource state unless $\lambda \to 1$ or $r \to \infty$
(we obtained a quantitatively different $p_{1,1}$ than the Eq.~(4) in Ref.~\cite{PhysRevA.65.062306} but our $p_{0,0}$ and $p_{1,1}$ agree with Eq.~(8) in Ref.~\cite{PhysRevA.107.012418}).

The fidelity with respect to $\ket{1}\otimes\ket{1}$ used to measure stellar rank is then given by
\begin{align}
    F_{j}(\lambda) \equiv
    F\qty(\rho_{n=j,m=j},\ket{1}\otimes\ket{1})
  = \frac{c^2_{j+1,j,j}}{C^2(j,j)}
 &= \begin{cases}
    \lambda^2 (1-\lambda^2), &(j=0); \\
    4\lambda^2 (1-\lambda^2)^3 / (1+\lambda^2), &(j=1).
    \end{cases}
 \label{eq:11_fid_noloss}
\end{align}

After some simple algebraic exercise, we find $\max_{\lambda\in [0, 1]} F_0(\lambda) = F_0(\lambda_\star) = 0.25$ at $\lambda_\star = t_r(1-R_S) = 1/\sqrt{2}$, so $\ket{\Psi_{0,0}}$ has stellar rank 0+ and must be a Gaussian state; $\max_{\lambda\in [0, 1]} F_1(\lambda) = F_1(\lambda_\star) = \frac{8}{27}(316-119\sqrt{7}) \approx 0.342$ at $\lambda_\star = \sqrt{(\sqrt{7} - 2)/3} \approx 0.464$ so the non-Gaussian resource state $\ket{\Psi_{1,1}}$ has stellar rank at least 1, according to the quantum non-Gaussian testing described in \cref{sec:nonGauss_test}. To maximize the testing value for one-photon subtracted TMSV, $\lambda_\star = t_r(1-R_S)$ agrees with numerical values $(r, R_S)$ listed in \cref{sec:nonGauss_test}, which means the minimal squeezing $r_\star = \operatorname{arctanh} \lambda_\star \approx 0.50$ for $R_S=0$. To receive a positive non-Gaussian testing result (stellar rank at least 1), i.e., $F_1(\lambda) > 0.25$, we find $0.30 \lesssim \lambda \lesssim 0.63$, so we only need $r \gtrsim 0.31$ given $R_S \geq 0$. The numerical values given here assume no loss other than just photon subtraction.

The negativity $\mathcal{N}(\rho)$ and logarithmic negativity $E_\mathcal{N}(\rho)$ are defined as follows:
\begin{subequations}
\begin{align}
    \mathcal{N}(\rho)
 &= \frac{1}{2}\norm{\rho^\text{PT}}_1 - \frac{1}{2}, \\
    E_\mathcal{N}(\rho)
 &= \log_2[2\mathcal{N}(\rho) + 1] = \log_2\norm{\rho^\text{PT}}_1,
\end{align}
\end{subequations}
where the trace norm (also known as nuclear norm) is $\norm{M}_1 \equiv \Tr\sqrt{M^\dagger M}$ and $\rho^\text{PT}$ is the partial transpose. Given the Schmidt decomposition of a pure state $\ket{\psi} = \sum_j \alpha_j \ket{u_j}_A\otimes\ket{v_j}_B$, the density operator $\rho = \op{\psi} = \sum_{j,k} \alpha_j \alpha_k^* \op{u_j}{u_k}\otimes \op{v_j}{v_k}$. The partial transpose for  system $B$ is $\rho^\text{PT} = \sum_{j,k} \alpha_j \alpha_k^* \op{u_j}{u_k}\otimes \op{v_k}{v_j}$, so $\sqrt{(\rho^\text{PT})^\dagger \rho^\text{PT}} = \sum_{k,j} |\alpha_k| |\alpha_j| \op{u_k} \otimes \op{v_j}$ and $\norm{\rho^\text{PT}}_1 = (\sum_j |\alpha_j|)^2 = 1 + 2 \sum_{j<k} |\alpha_j| |\alpha_k|$. Thus, we have
\begin{subequations}
\begin{align}
    E_\mathcal{N}(\rho_{0,0})
 &= \log_2\frac{1+\lambda}{1-\lambda}, \\
    E_\mathcal{N}(\rho_{1,1})
 &= \log_2\frac{(1+\lambda)^3}{(1+\lambda^2)(1-\lambda)}
  = E_\mathcal{N}(\rho_{0,0}) + \log_2\frac{(1+\lambda)^2}{1+\lambda^2}.
\end{align}
\end{subequations}

\subsection{Lossy two-mode Squeezing}
\label{sec:TMSV_loss}

\begin{figure}
  \centering
  \includegraphics[width=0.75\linewidth]{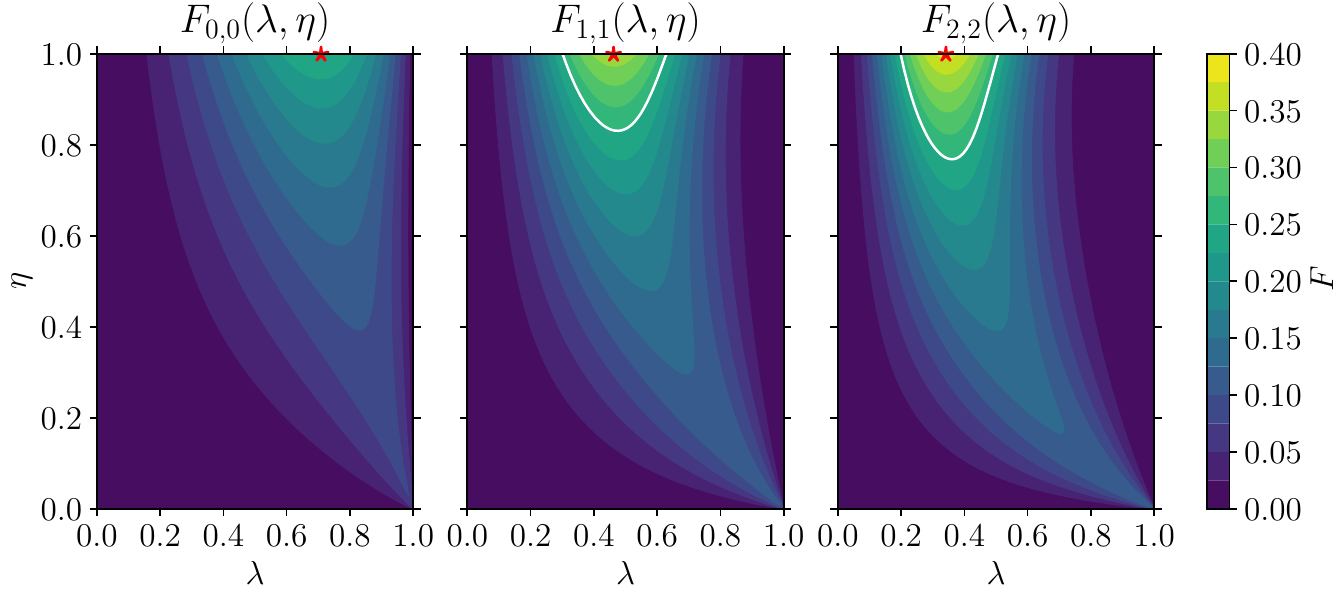}
\caption{The contour plots for the fidelity functions $F_{n,n}(\lambda,\eta)$ in \cref{eq:11_fid_withloss} as the stellar-rank witness. The fidelity is between the non-Gaussian state $\ket{1}\otimes \ket{1}$ and the resource state $(n,n)$-photon subtracted two-mode squeezed vacuum with loss. The loss is modeled by beamsplitters with the transmissivity $\eta_1 = \eta_2 = \eta$. The photon subtraction beamsplitters has the reflectivity $R_1 = R_2 = R_S$, the squeezing is $r$, and the effective squeezing parameter $\lambda = (1-R_S) \tanh r$. For all parameters, $F_{0,0} \leq 0.25$, indicating a stellar rank $0+$. To receive a stellar rank $1+$, that is $F_{1,1} > 0.25$ or $F_{2,2} > 0.25$, the parameters $(\lambda, \eta)$ must take the values inside the small pocket enclosed by the white curve in the middle and last panels. The lowest point on the white curve gives the minimal transmissivity $\eta_\text{min}$ for witnessing a stellar rank $1+$: Numerically, we found $\eta_\text{min} = 0.83$ for $(1,1)$-photon subtraction and $\eta_\text{min} = 0.77$ for $(2,2)$-photon subtraction. The red star ($\color{red} \star$) in each panel indicates the point with $\max F_{n,n}(\lambda,\eta)$. The trend shows that with increasing number of photons subtracted the maximal fidelity increases and is reached at a smaller squeezing. (However, it appears that $\lim_{n\to \infty}\max F_{n,n}(\lambda,\eta) < 0.5$ so a stellar rank $2+$ would never be witnessed).}
  \label{fig:stellar_rank_fid}
\end{figure}

In addition to perfect photon-subtraction from the squeezed input state, one also wishes to consider photon loss. Similar to the calculation above, photon loss will be modeled via a Stinespring-Sz.-Nagy dilation (see \cref{fig:circuit}). We consider a beamsplitter scattering photons into the environmental modes, which are subsequently traced over. Loss will result in a mixed state $\rho_{n,m}^\text{loss}$ that is a classical statistical mixtures over two-mode photon-subtracted states, with additional $h,l$ photons being subtracted.The pre-subtraction loss $\eta_1^\text{pre}$ can commute through the subtraction beam splitter with reflectivity $R_1$, becomes a smaller effective loss $\eta_1^\text{pre}(1-R_1)/(1-\eta_1^\text{pre}R_1)$, and be combined with the post-subtraction loss $\eta_1^\text{post}$. So, in our model we use the effective loss $\eta_1 = \eta_1^\text{post} \eta_1^\text{pre}(1-R_1)/(1-\eta_1^\text{pre}R_1)$ for channel A, while using the effective reflectivity $R_S = \eta_1^\text{pre} R_1$ for the subtraction beam splitter.

Labeling the ancillary (environment) vacuum modes as $E_1$ and $E_2$, for modes $1$ and $2$, photons are scattered via environmental beamsplitters with transmissivities $\eta_1$ and $\eta_2$:
\begin{align}
a_1^\dagger &\to
  \sqrt{\eta_1}\,a_1^\dagger + \sqrt{1-\eta_1}\,e^\dagger_1 , \notag \\
a_2^\dagger &\to
  \sqrt{\eta_2}\,a_2^\dagger + \sqrt{1-\eta_2}\,e^\dagger_2 . \notag 
\end{align}
Again, these unitary transformations act on Fock states as,
\begin{align*}
    \ket{n}_1\ket{0}_{E_1}
 &\to \sum_{h=0}^{n} B_{n,h}(1-\eta_1) \ket{n-h}_1\ket{h}_{E_1}, \\
    \ket{m}_2\ket{0}_{E_2}
 &\to \sum_{l=0}^{m} B_{m,l}(1-\eta_2) \ket{m-l}_2\ket{l}_{E_2}.
\end{align*}
Applying the environmental beamsplitters to the Fock states in Eq.~\eqref{eq:PSTMS} results in
\begin{align}
    \ket{k-n,k-m}_{1,2} 
 &\to \sum_{h=0}^{k-n}\sum_{l=0}^{k-m}
     B_{k-n,h}(1-\eta_1) B_{k-m,l}(1-\eta_2)
     \ket{k-n-h}_1\ket{k-m-l}_2
     \ket{h}_{E_1}\ket{l}_{E_2}.
\label{eq:psi_loss_global}
\end{align}
Taking the outer product of this state, and tracing out the environment modes $E_1$ and $E_2$, using the orthonormality $\Tr_{E_1 E_2} (\op{h,l}{h',l'}) = \delta_{h,h'}\delta_{l,l'}$, 
we obtain the normalized reduced state of modes \(1\) and \(2\):
\begin{align}
    \rho_{n,m}^\text{loss}
 = \frac{1}{C^{2}(n,m)} 
     &\sum_{k,k'=\max(n,m)}^{\infty}
     (-t_r)^k (-t_r)^{k'}
     B_{k,n}(R_1) B_{k',n}(R_1)
     B_{k,m}(R_2) B_{k',m}(R_2) \notag \\
 & \times
     \sum_{h=0}^{\min(k,k')-n}
     \sum_{l=0}^{\min(k,k')-m}
     B_{k-n,h}(1-\eta_1) B_{k'-n,h}(1-\eta_1)
     B_{k-m,l}(1-\eta_2) B_{k'-m,l}(1-\eta_2) \notag\\
 &\times
     \ket{k-n-h,\,k-m-l}_{12}\!
     \bra{k'-n-h,\,k'-m-l}.
\label{eq:rho_loss}
\end{align}
Note that the overall normalization factor $C^2(n,m)$ is the same one obtained in photon subtraction because the loss channel is trace preserving.

The operator $\rho_{n,m}^\text{loss}$ represents the mixed two-mode state of modes $1$ and $2$ after photon subtraction and propagation through lossy channels characterized by transmissivities $\eta_1$ and $\eta_2$. In the limit $\eta_1 = \eta_2 = 1$, all loss terms vanish and the state reduces to the pure photon-subtracted two-mode squeezed vacuum of Eq.~\eqref{eq:PSTMS}.

\subsubsection{Stellar-rank fidelity witness}
Now we calculate the stellar-rank witness, fidelity $F_{n,m} \equiv F(\rho_{n,m}^\text{loss}, \ket{1}\otimes \ket{1}) = \Tr(\rho_{n,m}^\text{loss} \op{1,1}{1,1})$. The fidelity is equal to the coefficient of matrix element $\op{1,1}{1,1}$ in \cref{eq:rho_loss}, corresponding to the three constraints $k-n-h = 1$, $k-m-l = 1$, and $k'-n-h = 1$. The fourth constraint, $k'-m-l = 1$ is not independent of the other three. This reduces the quadruple sums in \cref{eq:rho_loss} into one sum over $k$ as follows (after plugging in the three constraints $k' = k$, $h= k-n-1$ and $l=k-m-1$):
\begin{align}
    F_{n,m}
 &= \frac{1}{C^2(n,m)} \sum_{k=\max(n,m)}^{\infty} t_r^{2k} B_{k,n}^2(R_1) B_{k,m}^2(R_2)
    B_{k-n,k-n-1}^2(1-\eta_1) B_{k-m,k-m-1}^2(1-\eta_2) \notag\\
 &= \frac{(n+1)(m+1)R_1^n R_2^m \eta_1\eta_2}
    {C^2(n,m)(1-R_1)^n(1-R_2)^m (1-\eta_1)^{n+1}(1-\eta_2)^{m+1}} \notag\\
 &\quad \times \sum_{k=\max(n,m) + 1}^{\infty}
 \binom{k}{n+1} \binom{k}{m+1} [t_r^2 (1-R_1) (1-R_2) (1-\eta_1) (1-\eta_2)]^k \notag\\
 &= (1+m) \lambda^2 \eta_1 \eta_2 (1-\eta_2)^{n-m}
    \frac{\eval{\dv[n+1]{x} x^{m+1} (1-x)^{-m-2}}_{x = \lambda^2\nu^2}}{\eval{\dv[n]{x} x^m (1-x)^{-m-1}}_{x = \lambda^2}} \\
 &= (1+n)^2  \eta_1 \eta_2 \lambda^2 (1-\lambda^2)^{2n+1} (1-\lambda^2\nu^2)^{-2n-3}
    \frac{\sum_{j=0}^{n+1} \binom{n+1}{j}^2 \lambda^{2j}\nu^{2j}}{\sum_{j=0}^{n} \binom{n}{j}^2 \lambda^{2j}}
    \quad (\text{if $m = n$}).
\end{align}
In the last step, we defined $\nu = \sqrt{(1-\eta_1)(1-\eta_2)}$ and an effective squeezing parameter $\lambda = t_r \sqrt{(1-R_1)(1-R_2)}$, and used the following identity $\dv[k]{x} x^k (1-x)^{-k-1} = k!(1-x)^{-2k-1} \sum_{j=0}^{k} \binom{k}{j}^2 x^j$.

Consider experimentally relevant values $(n,m)=(0,0)$, $(1,1)$ and $(2,2)$ for the number of photons subtracted from the two modes of the resource state, and assume $R_1 = R_2 = R_S$ and $\eta_1 = \eta_2 = \eta$. The stellar-rank witness (i.e., the resource state fidelity with respect to the two-mode Fock state $\ket{1,1} = \ket{1}\otimes \ket{1}$) is given by
\begin{subequations}
 \label{eq:11_fid_withloss}
\begin{align}
    F_{0,0}(\lambda, \eta) = \ev{\rho_{0,0}^\text{loss}(\lambda,\eta)}{1,1}
 &= \frac{\eta^2 \lambda^2 \qty(1-\lambda^2)  \qty[1 + \lambda^2(1-\eta)^2]}{\qty[1 - \lambda^2(1-\eta)^2]^{3}}, \\
    F_{1,1}(\lambda, \eta) = \ev{\rho_{1,1}^\text{loss}(\lambda,\eta)}{1,1}
 &= \frac{4\eta^2 \lambda^2 \qty(1-\lambda^2)^3 \qty[1 + 4\lambda^2(1-\eta)^2 + \lambda^4(1-\eta)^4]}{\qty[1 - \lambda^2(1-\eta)^2]^{5} \qty(1+\lambda^2)}, \\
    F_{2,2}(\lambda, \eta) = \ev{\rho_{2,2}^\text{loss}(\lambda,\eta)}{1,1}
 &= \frac{9\eta^2 \lambda^2 \qty(1-\lambda^2)^5 \qty[1 + 9\lambda^2(1-\eta)^2 + 9\lambda^4(1-\eta)^4 + \lambda^6(1-\eta)^6]}{\qty[1 - \lambda^2(1-\eta)^2]^{7} \qty(1 + 4\lambda^2 + \lambda^4)}.
\end{align}
\end{subequations}
Setting $\eta = 1$ (lossless transmission), we recover the result in \cref{eq:11_fid_noloss}. The contour plots for the fidelity functions in \cref{eq:11_fid_withloss} are shown in \cref{fig:stellar_rank_fid}; see the figure caption for more details.

\subsubsection{Entanglement negativity}
For symmetric losses $\eta_1 = \eta_2 = \eta$, the entanglement negativity $\mathcal{N} = (\Tr|\rho^{T_2}| - 1)/2$ and the logarithmic negativity $E_\mathcal{N} = \log_2(2\mathcal{N}+1) = \log_2(\Tr|\rho^{T_2}|)$ can also be calculated analytically using \cref{eq:rho_loss}, as follows. We specialize to equal photon-subtraction outcomes, $n = m = l$, but allow the two photon-subtraction beamsplitters to have different reflectivities $R_1$, and $R_2$. Their dependence is contained in the effective squeezing parameter $\lambda = t_r \sqrt{(1 - R_1) (1 - R_2)}$, $0 \leq \lambda \leq 1$.

Let $\phi$ denote the phase of the two-mode squeezing operator, $\theta = \phi + \pi$, and $\ket{\Phi} = \sum_q \ket{q,q}$ the unnormalized ideal EPR state. Define the phase-matched EPR state
\begin{align}
    \ket{\Phi_\theta} 
 &= e^{i\theta \hat{n}_2} \ket{\Phi} = \sum_{q=0}^{\infty} e^{iq\theta} \ket{q,q}.
\end{align}
The corresponding phase-twisted SWAP operator is
\begin{align}
    \hat{V}_\theta
 &= \qty(\op{\Phi_\theta})^{T_2}
  = \sum_{p,q = 0}^{\infty} \qty(e^{ip\theta} \ket{pp} e^{-iq\theta}\bra{qq})^{T_2}
  = \sum_{p,q = 0}^{\infty} e^{i(p-q)\theta} \op{p}{q} \otimes (\op{p}{q})^{T}
  = \sum_{p,q = 0}^{\infty} e^{i(p-q)\theta} \op{p,q}{q,p} .
\end{align}
It is Hermitian and unitary, $\hat{V}_\theta^\dagger = \hat{V}_\theta$ and $\hat{V}_\theta^2 = \openone$. For equal losses and equal photon-subtraction outcomes, it is also the sign operator of the partially transposed resource state:
\begin{align}
   &\hat{V}_\theta (\rho_{l,l}^{\text{loss}})^{T_2} \geq 0 \label{eq:SWAP_sign_op}\\
   \implies &
   \abs{(\rho_{l,l}^{\text{loss}})^{T_2}} \equiv 
   \sqrt{[(\rho_{l,l}^{\text{loss}})^{T_2}]^\dagger (\rho_{l,l}^{\text{loss}})^{T_2}}
 = \sqrt{[\hat{V}_\theta (\rho_{l,l}^{\text{loss}})^{T_2}]^\dagger
         [\hat{V}_\theta (\rho_{l,l}^{\text{loss}})^{T_2}]}
 = \hat{V}_\theta (\rho_{l,l}^{\text{loss}})^{T_2}.
\end{align}
Therefore,
\begin{align}
    2\mathcal{N}_l + 1 
 &\equiv \norm{(\rho_{l,l}^{\text{loss}})^{T_2}}_1
  = \Tr[ \abs{(\rho_{l,l}^{\text{loss}})^{T_2}} ]
  = \Tr[ \hat{V}_\theta (\rho_{l,l}^{\text{loss}})^{T_2} ] \\
 &= \Tr[ (\hat{V}_\theta)^{T_2} \rho_{l,l}^{\text{loss}} ]
  = \ev{\rho_{l,l}^{\text{loss}}}{\Phi_\theta} \equiv L_\theta (\lambda, \{\eta_1,\eta_2\}) .
  \label{eq:negativity_EPR_overlap}
\end{align}

We omit the proof of \cref{eq:SWAP_sign_op} here because it is implied by the Theorem 2 in Ref.~\cite{Zhang2010} if we recognize that the anti-identity matrix (also called exchange matrix) $J_K$ is precisely the restriction of the two-mode SWAP operator to the $K$-photon subspace $J_K \ket{i, K-i} = \ket{K-i,i}$.

We now evaluate the right-hand side $L_\theta (\lambda, \eta)$ of \cref{eq:negativity_EPR_overlap} directly from \cref{eq:rho_loss}. For asymmetric loss with different transmissions $\eta_1 \neq \eta_2$,
\begin{align}
      L(\lambda, \eta_\text{max})
 \geq 2\mathcal{N}_l + 1
 \geq L(\lambda, \{\eta_1, \eta_2\})
 \geq L(\lambda, \eta_\text{min}),
\end{align}
where $\eta_\text{max} = \max(\eta_1, \eta_2)$ and $\eta_\text{min} = \min(\eta_1, \eta_2)$. The first inequality is due to decreasing entanglement under addition local loss from the symmetric low loss state $\tilde{\rho}_{l,l}^{\text{loss}}$ with  $\tilde{\eta_1} = \tilde{\eta}_2 = \eta_\text{max}$ to the true loss state $\rho_{l,l}^{\text{loss}}$. The second inequality is due to variational definition of the trace norm $\norm{X}_1 \geq \Tr(UX)$ for any unitary $U$.

For the phase convention used in \cref{eq:rho_loss}, $\phi = 0$ and hence
\begin{align}
    2\mathcal{N}_l + 1
 &= \frac{1}{C^2(l,l)} \sum_{h=0}^{\infty}.
\end{align}

\subsection{Quantum teleportation fidelity using heralded non-Gaussian resource state}

\subsubsection{Characteristic functions and teleportation fidelity}

The performance of continuous-variable (CV) teleportation is 
quantified by the teleportation fidelity $F$, which measures the 
overlap between the input state $\rho_{\mathrm{in}}$ and the 
teleported output state $\rho_{\mathrm{out}}$. In the phase-space 
representation, evaluating this overlap is highly efficient when 
utilizing the characteristic functions because the output state's characteristic function has a product form~\cite{Marian2006}.
We will also use this product form to derive a fidelity formula in terms of the matrix elements of the resource state in Fock basis as well as in the homodyne detection sampling form. The Fock-basis fidelity formula is similar to the one given in Ref.~\cite{Lee2017} that was derived from alternative unitary dilation description of the teleportation channel~\cite{Nha2012} instead of the measurements and feedforward description used in experiments and adopted below. The homodyne measurement sampling kernel form allows direct estimation of the fidelity from the tomography experimental data without reconstructed $\rho_\text{MLE}$. Our sampling kernel can be viewed as an extension of the result of Ref.~\cite{Fiurasek2001}, providing a practical alternative to Ref.~\cite{Nha2012} by extracting the fidelity directly from standard, independent homodyne data without the need for an additional mixing beam splitter.

For a single-mode quantum state with density matrix $\rho$, the 
characteristic function is defined as the expectation value of the 
displacement operator:
\begin{align}\label{eq:sm-chi}
    \chi(\alpha)
 &= \Tr[\rho \hat{D}(\alpha)],
\end{align}
where $\alpha$ is a complex phase-space amplitude, and the displacement 
operator is defined in terms of the creation and annihilation operators 
as $\hat{D}(\alpha) = \exp(\alpha \hat{a}^\dagger - \alpha^* \hat{a})$ (see \cite{Brask2021} for a quick reference).
Then, by Fourier--Weyl relation, the density operator is given by
\begin{align}
    \rho = \frac{1}{\pi}\int d^2\alpha\, \chi(\alpha) \hat{D}(-\alpha),
    \label{eq:sm-rho}
\end{align}
which can be obtained using orthogonality of Weyl displacement operator
\begin{align}
    \Tr(\hat{D}_\alpha \hat{D}_{-\beta})
 &= \pi \delta^2(\alpha - \beta)
 \label{eq:D_orth}
\end{align}
and \cref{eq:sm-chi}.

For a bipartite resource state $\rho_{12}$ shared between Alice (mode $1$) and Bob (mode $2$), the joint characteristic function is 
\begin{align}\label{eq:tm-chi}
    \chi_{\text{res}}(\alpha_1, \alpha_2)
 &= \Tr[\rho_{12} \hat{D}_1(\alpha_1) \otimes \hat{D}_2(\alpha_2)], 
\end{align}
where $\hat{D}_1 (\alpha_1) = {\exp}(\alpha_1 \hat{a}_1^\dagger - \alpha_1^* \hat{a}_1)$ and 
$\hat{D}_2(\alpha_2) = {\exp}(\alpha_2 \hat{a}_2^\dagger - \alpha_2^* \hat{a}_2)$.

While the characteristic functions in Eqs.~\eqref{eq:sm-chi} and~\eqref{eq:tm-chi} 
are expressed as functions of the complex amplitude $\alpha$, they are
frequently formulated using the real phase-space variables $\tau$ and
$\sigma$. To explicitly connect these two representations, it is 
sufficient to demonstrate the equivalence for a single mode, as the 
bipartite case follows simply by applying the identity  
locally to each mode. We define the position and momentum quadrature 
operators as $\hat{x} = (\hat{a} + \hat{a}^\dagger)/\sqrt{2}$ and $\hat{p} = -i(\hat{a} - \hat{a}^\dagger)/\sqrt{2}$, satisfying the commutation relation $[\hat{x}, \hat{p}] = i$ (set $\hbar = 1$) due to $[\hat{a}, \hat{a}^\dagger] \equiv \hat{a} \hat{a}^\dagger - \hat{a}^\dagger \hat{a} = 1$. 
By setting $\alpha = (\tau + i\sigma)/\sqrt{2}$, the exponent of the displacement operator becomes
\begin{align}
    \alpha \hat{a}^\dagger - \alpha^* \hat{a} &= \left( \frac{\tau + i\sigma}{\sqrt{2}} \right) \left( \frac{\hat{x} - i\hat{p}}{\sqrt{2}} \right) - \left( \frac{\tau - i\sigma}{\sqrt{2}} \right) \left( \frac{\hat{x} + i\hat{p}}{\sqrt{2}} \right) \nonumber \\
    &= \frac{1}{2} \left[ (\tau\hat{x} - i\tau\hat{p} + i\sigma\hat{x} + \sigma\hat{p}) - (\tau\hat{x} + i\tau\hat{p} - i\sigma\hat{x} + \sigma\hat{p}) \right] \nonumber \\
    &= i(\sigma \hat{x} - \tau \hat{p})
     = i\mqty(\hat{x} &\hat{p}) \mqty(0 & 1 \\ -1 &0) \mqty(\tau \\ \sigma)
     \equiv i \vu{x}^{T} \Omega \vb*{\xi}.
\end{align}
Thus, the displacement operator can be equivalently written as $\hat{D}(\tau, \sigma) = {\exp}[i(\sigma \hat{x} - \tau \hat{p})] = {\exp}[i\vu{x}^{T}\Omega \vb*{\xi}] \equiv D(\vb*{\xi})$, 
which directly recovers the standard quadrature-based characteristic function $\chi(\tau, \sigma) = \chi(\vb*{\xi}) = \Tr[\rho D(\vb*{\xi})]$ (Eq.~(12) in~\cite{Weedbrook2012}). 
The complex-$\alpha$ representation is chosen here as it directly facilitates the efficient numerical evaluation of the teleportation fidelity within the truncated photon-number basis, as detailed below in \cref{sec:numer_tele_fid}.

In the standard Braunstein-Kimble teleportation protocol~\cite{PhysRevLett.80.869,Braunstein2005} with unity gain ($g=1$), shown in the orange box in \cref{fig:circuit}, 
the characteristic function of the teleported output state is related to the input and the resource state via~\cite{Marian2006}
\begin{align}
    \chi_{\mathrm{out}}(\alpha) 
 &= \chi_{\mathrm{in}}(\alpha) \chi_{\text{res}}(\alpha^*, \alpha e^{i\theta}),
  \label{eq:out-chi}
\end{align}
where $\theta$ specifies an optional phase shift depending on the resource state used, see below for a self-contained derivation. 
We remark that incorporating the resource state's phase degree of freedom generalizes the standard product form~\cite{PhysRevA.107.012418} of the teleported state's characteristic function. However, this is distinct from the generalization presented in Ref.~\cite{Marian2006}, which instead considers an unbalanced beam splitter for mixing the input and resource modes prior to homodyne detection.

\paragraph{Teleportation fidelity in characteristic function form}
We now give a detailed derivation for this product form in a more straightforward way than Ref.~\cite{Marian2006}. The positive operator-valued measure (POVM) basis Alice uses is constructed from the simultaneous eigenvectors of the two commuting quadrature operators, $\hat{x}_- = (\hat{x}_0 - \hat{x}_1)/\sqrt{2}$ and $\hat{p}_+ = (\hat{p}_0 + \hat{p}_1)/\sqrt{2}$, acting on the joint system of the input mode (denoted as mode 0 here) and the mode 1 of the resource state (see \cref{fig:circuit}). The simultaneous eigenvectors are the so-called continuous-variable Bell states~\cite{Braunstein2005,Hofmann2000}
\begin{align}
    \ket{\Phi_\beta}
 &= \frac{1}{\sqrt{\pi}} \hat{D}_0(\beta) \ket{\Phi}
  = \frac{1}{\sqrt{\pi}} [\hat{D}_1(\beta)]^{T} \ket{\Phi}
  = \frac{1}{\sqrt{\pi}} \hat{D}_1(-\beta^*) \ket{\Phi},
\end{align}
where $(\hat{x}_- + i\hat{p}_+)\ket{\Phi_\beta} = \beta \ket{\Phi_\beta}$ (see below) and the ideal EPR state
\begin{subequations}
\begin{align}
    \ket{\Phi}
 &= \sum_{n=0}^\infty \ket{n}_0\otimes \ket{n}_1 
 \label{eq:EPR-Fock}\\
 &= \iint dx_0 dx_1 \qty[\sum_n \psi_n(x_0) \psi_n(x_1)] \ket{x_0}_0 \otimes\ket{x_1}_1 \notag\\
 &= \int_{-\infty}^{\infty}dx\, \ket{x}_0 \otimes\ket{x}_1
  = \int_{-\infty}^{\infty}dp\, \ket{p}_0 \otimes\ket{-p}_1 .
  \label{eq:EPR-xp}
\end{align}
\end{subequations}
In the middle step we plugged in the completeness relation of harmonic oscillator eigen-functions (real in position space) $\sum_n \psi_n(x_0) \psi_n(x_1) = \delta(x_0 - x_1)$ and the final step follows the Fourier transform to momentum space.

\emph{Remarks:} The Fourier transform of the above standard ideal EPR state is $\ket*{\tilde{\Phi}} \equiv {\exp}[-i\frac{\pi}{2}(\hat{n}_0 + \hat{n}_1)]\ket{\Phi} = \sum_{n} e^{-i\pi n} \ket{n}_0 \otimes \ket{n}_1 = \sum_{n}  (-1)^n \ket{n}_0 \otimes \ket{n}_1 = \int dx \, \ket{x}_0 \otimes\ket{-x}_1 = \int dp\, \ket{p}_0 \otimes\ket{p}_1$. The position space expression can be derived from $\sum_n (-1)^n \psi_n(x_0) \psi_n(x_1) = \sum_n \psi_n(-x_0) \psi_n(x_1) = \delta(x_0 + x_1)$, where $(-1)^n \psi_n(x_0) = \psi_n(-x_0)$ is due to the symmetry of Hermite polynomials. $\ket{\Phi}$ is infinitely squeezed in $x_0 - x_1$ and $p_0 + p_1$ directions, while $\ket*{\tilde{\Phi}}$ is infinitely squeezed in $x_0 + x_1$ and $p_0 - p_1$ directions. The choice of the feedback in the teleportation protocol shown in \cref{fig:circuit} assumes the resource state is squeezed in the $x_1 - x_2$ and $p_1 + p_2$ directions, so correlated in position space and anti-correlated in momentum space.

The POVM set $\{\Pi_\beta = \op{\Psi_\beta}\}$ satisfies the completeness relation $\int d^2\beta\, \Pi_\beta = \openone$, which can be shown using the displacement twirling identity $\pi^{-1}\int d^2\beta\, \hat{D}(\beta) \hat{O} \hat{D}^\dagger(\beta) = {\Tr}(\hat{O})\openone$ (see, e.g., Supplemental Material of Ref.~\cite{Oh2024}). Evidently, the Bell state POVM is a Weyl-covariant phase-space POVM corresponding to $\hat{O} = \op{\Psi}$ (density operator of the ideal EPR state) in the twirling identity (partial trace of mode being displaced), while heterodyne measurement $\{\op{\beta}/\pi\}$ is a Weyl-covariant phase-space POVM corresponding to $\hat{O} = \op{0}$.

Now we can show $\ket{\Phi_\beta}$ is an eigenvector for both $\hat{x}_-$ and $\hat{p}_+$. Denote $\beta = (x+ip)/\sqrt{2}$. Because $\hat{D}(\beta) = e^{\beta \hat{a}^\dagger - \beta^* \hat{a}} = e^{i(p\hat{x}-x\hat{p})} = e^{-ixp/2} e^{ip\hat{x}} e^{-ix\hat{p}}$ due to BCH formula, and $(\hat{x}_0 - \hat{x}_1)\ket{\Psi} = 0$ and $(\hat{p}_0 + \hat{p}_1)\ket{\Psi} = 0$ (EPR nullifier relation) due to \cref{eq:EPR-xp}, we have
\begin{subequations}
\begin{align}
    \hat{x}_- \ket{\Phi_\beta}
 &= \frac{1}{\sqrt{2\pi}} \hat{D}_0(\beta)\qty[ \hat{D}_0^\dagger(\beta)(\hat{x}_0 - \hat{x}_1)\hat{D}_0(\beta)]\ket{\Phi}
  = \frac{1}{\sqrt{2\pi}} \hat{D}_0(\beta) (\hat{x}_0 + x - \hat{x}_1)\ket{\Phi} \notag\\
 &= (\Re\beta) \ket{\Phi_\beta}, \\
    \hat{p}_+ \ket{\Phi_\beta}
 &= \frac{1}{\sqrt{2\pi}} \hat{D}_0(\beta)\qty[ \hat{D}_0^\dagger(\beta)(\hat{p}_0 + \hat{p}_1)\hat{D}_0(\beta)]\ket{\Phi}
  = \frac{1}{\sqrt{2\pi}} \hat{D}_0(\beta) (\hat{p}_0 + p + \hat{p}_1)\ket{\Phi} \notag\\
 &= (\Im\beta) \ket{\Phi_\beta}.
\end{align}
\end{subequations}

Conditioned on Alice's measurement outcome $\beta$ corresponding to the Bell state POVM element $\Pi_\beta$, the output state before Bob's correction
\begin{align}
    \tilde{\rho}_\text{out}(\beta)
 &= \frac{1}{p(\beta)} \ev{\rho_0\otimes \rho_{12}}{\Phi_\beta}
  = \frac{1}{p(\beta)} \Tr_{01}[(\op{\Phi_\beta} \otimes \openone_2) (\rho_0\otimes \rho_{12})], \\
\intertext{and the probability distribution}
    p(\beta)
 &= \Tr_{2}(\ev{\rho_0\otimes \rho_{12}}{\Phi_\beta})
  = \Tr[(\op{\Phi_\beta} \otimes \openone_2) (\rho_0\otimes \rho_{12})].
\end{align}
Thus, after Bob's correction using the displacement amplitude $\beta$ sent by Alice via classical communication channel, the unconditioned teleportation output
\begin{align}
    \rho_\text{out}
 &= \int d^2\beta\, p(\beta) \hat{D}_2(\beta)
    \qty[R_2(\theta)
    \tilde{\rho}_\text{out}(\beta)
    R_2^\dagger(\theta)]
    \hat{D}_2^\dagger(\beta).
\end{align}
Here, the phase shifter $R_2(\theta) = e^{-i\theta\hat{a}_2^\dagger \hat{a}_2}$, $\theta = \phi + \pi$, is necessary to rotate the effective phase of the resource state to $\pi$ if the two-mode squeezing operator phase $\phi \neq \pi$ (see \cref{fig:circuit}). The characteristic function of the teleported output state is
\begin{subequations}
\begin{align}
    \chi_\text{out}(\alpha)
 &= \Tr(\rho_\text{out} \hat{D}_2(\alpha)) \\
 &= \int d^2\beta\,
    \Tr{\qty[\op{\Phi_\beta} \otimes \qty(
         R_2^\dagger(\theta)
         \hat{D}_2^\dagger(\beta) \hat{D}_2(\alpha) \hat{D}_2(\beta)
         R_2(\theta) 
       )] (\rho_0\otimes \rho_{12})} \\
 &= \int d^2\beta\,
    \Tr{\qty[\op{\Phi_\beta} \otimes \qty(e^{\alpha\beta^* - \alpha^*\beta}
         R_2^\dagger(\theta)
         \hat{D}_2(\alpha)
         R_2(\theta)
       )] (\rho_0\otimes \rho_{12})} \\
 &= \Tr{\qty[\int d^2\beta\,e^{\alpha\beta^* - \alpha^*\beta}\op{\Phi_\beta} \otimes \hat{D}_2(\alpha e^{i\theta})] (\rho_0\otimes \rho_{12})} \\
 &= \Tr{\qty[\hat{D}_0(\alpha) \otimes \hat{D}_1 (\alpha^*) \otimes \hat{D}_2(\alpha e^{i\theta})] (\rho_0\otimes \rho_{12})} \\
 &= \Tr[\hat{D}_0(\alpha) \rho_0)]
    \Tr[\qty(\hat{D}_1 (\alpha^*) \otimes \hat{D}_2(\alpha e^{i\theta})) \rho_{12}]
  = \chi_\text{in}(\alpha) \chi_\text{res}(\alpha^*,\alpha e^{i\theta}).
\end{align}
\end{subequations}
In the above, we used the following identities:
\begin{align*}
    \hat{D}^\dagger(\beta)\hat{D}(\alpha) \hat{D}(\beta) 
 &= e^{\alpha \beta^* - \alpha^* \beta} \hat{D}(\alpha).
\\
    \hat{D}_0(\alpha) \hat{D}_1 (\alpha^*) \ket{\Phi_\beta}
 &= \hat{D}_0(\alpha) \hat{D}_1 (\alpha^*) \hat{D}_0(\beta) \ket{\Phi_0}
  = \hat{D}_0(\alpha) \hat{D}_0(\beta) \hat{D}_1 (\alpha^* )\ket{\Phi_0} \\
 &= \hat{D}_0(\alpha) \hat{D}_0(\beta) [\hat{D}_0 (\alpha^*)]^{T} \ket{\Phi_0}
  = \hat{D}_0^\dagger(-\alpha) \hat{D}_0(\beta) \hat{D}_0 (-\alpha) \ket{\Phi_0}\\
 &= e^{-\beta \alpha^* + \beta^* \alpha} \hat{D}_0(\beta) \ket{\Phi_0}
  = e^{\alpha \beta^* - \alpha^* \beta} \ket{\Phi_\beta}.
\\
    \int d^2\beta\, \op{\Phi_\beta}
 &= \openone .
\end{align*}

The final teleportation fidelity is calculated by integrating the product of the input and output characteristic functions over the entire complex phase space as in \cref{eq:out-fid}, which can be derived as follows using \cref{eq:sm-rho} for both input and output modes and the orthogonality \cref{eq:D_orth}:
\begin{align}
    F
 &= \Tr(\rho_\text{in} \rho_\text{out})\notag \\
 &= \frac{1}{\pi^2} \Tr \iint d^2\alpha_1 d^2\alpha_2\,
    \chi_\text{in}(\alpha_1) \hat{D}(-\alpha_1)
    \chi_\text{out}(\alpha_2) \hat{D}(-\alpha_2) \notag\\
 &= \frac{1}{\pi^2} \iint d^2\alpha_1 d^2\alpha_2\,
    \chi_\text{in}(\alpha_1) \chi_\text{out}(\alpha_2)
    \Tr[\hat{D}(-\alpha_1)\hat{D}(-\alpha_2)] \notag\\
 &= \frac{1}{\pi} \iint d^2\alpha_1 d^2\alpha_2\,
    \chi_\text{in}(\alpha_1) \chi_\text{out}(\alpha_2) \delta^2(\alpha_1 + \alpha_2) \notag \\
\implies
    F
 &= \frac{1}{\pi} \int d^2\alpha \, \chi_{\mathrm{in}}(\alpha) \chi_{\mathrm{out}}(-\alpha)  \label{eq:out-fid} \\
 &= \frac{1}{\pi} \int d^2\alpha \, |\chi_{\mathrm{in}}(\alpha)|^2 \chi_{\text{res}}(\alpha^*, e^{i\theta} \alpha). \label{eq:in-res-fid}
\end{align}
$\chi(-\alpha) = \chi^*(\alpha)$ is used in \cref{eq:in-res-fid} and $\theta = \phi + \pi$ where $\phi$ is the phase of the two-mode squeezing operator to generate the resource state at the initial step. If the input state's characteristic function is isotropic $\chi_\text{in}(\alpha) = \chi_\text{in}(|\alpha|)$ (such as coherent state, Fock state, etc.), it is easy to see the phase shift can be placed on the first mode $\chi_{\text{res}}(e^{i\theta}\alpha^*, \alpha )$ and the above gives the same fidelity. Strictly speaking, the above formula is the Hilbert--Schmidt overlap and only equal to Uhlmann--Jozsa fidelity if at least one of the input or output state is a pure state. (For example, if the input is mixed and the resource is the ideal EPR state, the obtained output would be the same as the input and thus the fidelity is equal to one, but in general the formula gives the purity of the input not the fidelity when the ideal EPR resource state is used.)
The \cref{eq:out-chi,eq:out-fid} (with $\phi = \pi$ and $e^{i\theta} = e^{i(\phi + \pi)} = 1$) correspond to Eqs.\ (10) and (11) in Ref.~\cite{PhysRevA.107.012418}, but in the above we have given a self-contained derivation of these two equations.

\paragraph{Teleportation fidelity in Fock-basis density matrix form}
We consider a coherence state $\ket{\alpha_0}$ as the input state. The characteristic function $\chi_\text{in} (\alpha) = {\Tr}[\op{\alpha_0} \hat{D}(\alpha)] = e^{-|\alpha|^2/2 + \alpha \alpha_0^* - \alpha^*\alpha_0}$, so the magnitude $|\chi_\text{in} (\alpha)| = e^{-|\alpha|^2/2}$ is independent of $\alpha_0$ or the phase of $\alpha$ and thus isotropic in the phase space. Using \cref{eq:in-res-fid}, we have
\begin{align*}
    F_\theta 
 &= \frac{1}{\pi} \int d^2\alpha\, |\chi_\text{in}(\alpha)|^2
    \chi_\text{res}(\alpha^*, e^{i\theta} \alpha) \\
 &= \frac{1}{\pi} \int d^2\alpha\, |\chi_\text{in}(\alpha)|^2
    \Tr[\rho_\text{res} \hat{D}_1(\alpha^*)\otimes \hat{D}_2(e^{i\theta} \alpha)] \\
 &\equiv \Tr(\rho_\text{res} \hat{\Omega}_\theta),
\end{align*}
where the teleportation fidelity operator
\begin{align}
   \hat{\Omega}_\theta 
 &= \frac{1}{\pi} \int d^2\alpha\, |\chi_\text{in}(\alpha)|^2
    \hat{D}_1(\alpha^*)\otimes \hat{D}_2(e^{i\theta} \alpha)
    \label{eq:fid_op_general}\\
 &= \frac{1}{\pi} \int d^2\alpha\, e^{-|\alpha|^2}
    [\hat{D}_1(-\alpha)]^{T}\otimes \hat{D}_2(e^{i\theta} \alpha).
\end{align}
In the above, $[\hat{D}_1(-\alpha)]^{T} = [{\exp}(-\alpha \hat{a}_1^\dagger + \alpha^* \hat{a}_1)]^{T} = {\exp}(-\alpha \hat{a}_1 + \alpha^* \hat{a}_1^\dagger) = \hat{D}_1(\alpha^*)$ due to $\hat{a}^T = \hat{a}^\dagger$, which is because the $\hat{a}$ and $\hat{a}^\dagger$ are real matrices in Fock basis.

Because the vacuum density operator can be expressed by its characteristic function $\chi_{\ket{0}}(\alpha) = e^{-|\alpha|^2/2}$ using Fourier-Weyl transform $\op{0} = \pi^{-1} \int d^2\alpha\, e^{-|\alpha|^2/2}\hat{D}(-\alpha) = \pi^{-1} \int d^2\alpha\, e^{-|\alpha|^2/2}\hat{D}(\alpha)$, we have the following partially transposed fidelity operator
\begin{align}
    \hat{\Omega}_\theta^{T_1}
 &= \frac{1}{\pi} \int d^2\alpha\, e^{-|\alpha|^2}
    \hat{D}_1(-\alpha)\otimes \hat{D}_2(e^{i\theta} \alpha) \notag\\
 &= \frac{1}{\pi} \int d^2\alpha\, e^{-|\alpha|^2}
    \exp[-\alpha\hat{a}_1^\dagger + \alpha^*\hat{a}_1 + 
         \alpha (e^{-i\theta}\hat{a}_2)^\dagger - \alpha^* (e^{-i\theta}\hat{a}_2)] \notag \\
 &= \frac{1}{\pi} \int d^2\alpha\, e^{-|\alpha|^2} I_{1'}\otimes \hat{D}_{2'}(\sqrt{2}\alpha) \notag\\
 &= \frac{1}{2\pi} \int d^2\alpha' \, e^{-|\alpha'|^2/2} I_{1'}\otimes \hat{D}_{2'}(\alpha')  \notag \\
 &= \frac{1}{2} I_{1'}\otimes {\ket{0}_{2'}}{\bra{0}}.
    \label{eq:fid_op}
\end{align}
In the above, we have defined two canonical modes $\hat{a}_{1'} = (e^{-i\theta}\hat{a}_2 + \hat{a}_1)/\sqrt{2}$ and $\hat{a}_{2'} = (e^{-i\theta}\hat{a}_2 - \hat{a}_1)/\sqrt{2}$, which are unitary transformed from $\hat{a}_1$ and $\hat{a}_2$. Note that under this transformation, the global vacuum state is the same $\ket{0}_{1'}\otimes \ket{0}_{2'} = \ket{0}_{1}\otimes \ket{0}_{2}$, but not the local vacuum state, e.g., $\ket{2}_{1'}\otimes \ket{0}_{2'} = \frac{1}{\sqrt{2}} (\hat{a}_{1'}^{\dagger})^2 \ket{0}_{1'}\otimes \ket{0}_{2'} = \frac{1}{2\sqrt{2}}(e^{i\theta}\hat{a}_2^\dagger + \hat{a}_1^\dagger)^2 (\ket{0}_{1}\otimes \ket{0}_{2}) = \frac{1}{2}(\ket{2}_{1}\otimes \ket{0}_{2} + e^{i\theta}\sqrt{2}\ket{1}_{1}\otimes \ket{1}_{2} + e^{2i\theta}\ket{0}_{1}\otimes \ket{2}_{2})$. In general,
\begin{align}
    \ket{N}_{1'}\otimes \ket{0}_{2'}
 &= \frac{(\hat{a}_{1'}^\dagger)^N}{\sqrt{N!}} \ket{0}_{1'}\otimes \ket{0}_{2'} \notag\\
 &= \frac{(\hat{a}_{1}^\dagger + e^{i\theta} \hat{a}_{2}^\dagger)^N}{\sqrt{N!2^N}}
    \ket{0}_{1}\otimes \ket{0}_{2} \notag \\
 &= \frac{1}{\sqrt{N!2^N}} \sum_{m=0}^{N} \binom{N}{m}
    (\hat{a}_{1}^\dagger)^{m}  (e^{i\theta} \hat{a}_{2}^\dagger)^{N-m}
    \ket{0}_{1}\otimes \ket{0}_{2} \notag \\
 &= 2^{-N/2}\sum_{m=0}^{N} \sqrt{\binom{N}{m}}
    e^{i\theta(N-m)}\ket{m}_1\otimes \ket{N-m}_2 .
    \label{eq:Fock-basis_BS_transform}
\end{align}

Using the expansion of identity $I_{1'} = \sum_{N=0}^\infty {\ket{N}_{1'}}{\bra{N}}$ and \cref{eq:fid_op,eq:Fock-basis_BS_transform}, we have
\begin{align}
    \hat{\Omega}_\theta^{T_1}
 &= \frac{1}{2} \sum_{N=0}^\infty 
    {\ket{N}_{1'}}{\bra{N}} \otimes
    {\ket{0}_{2'}}{\bra{0}} \notag\\
 &= \frac{1}{2} \sum_{N=0}^\infty 
    (\ket{N}_{1'} \otimes \ket{0}_{2'})
    (\bra{N}_{1'} \otimes \bra{0}_{2'}) \notag\\
 &= \sum_{N=0}^\infty \frac{1}{2^{N+1}} \sum_{m,p=0}^{N} 
    \sqrt{\binom{N}{m} \binom{N}{p}} e^{i\theta(p-m)} 
    (\ket{m}_1\otimes \ket{N-m}_2)
    (\bra{p}_1\otimes \bra{N-p}_2), \\
\implies
    \hat{\Omega}_\theta
 &= \sum_{N=0}^\infty \frac{1}{2^{N+1}} \sum_{m,p=0}^{N} 
    \sqrt{\binom{N}{m} \binom{N}{p}} e^{i\theta(p-m)} 
    (\ket{p}_1\otimes \ket{N-m}_2)
    (\bra{m}_1\otimes \bra{N-p}_2).
\end{align}

Finally, given the Fock basis representation of the resource state $({\bra{n}_1\otimes \bra{m}_2})\rho_\text{res} ({\ket{n'}_1\otimes \ket{m'}_2}) = \rho_{n,m;\, n',m'}$, the teleportation fidelity can be evaluated as follows:
\begin{align}
    F_\theta = \Tr(\rho_\text{res} \hat{\Omega}_\theta)
 &= \sum_{N=0}^\infty \frac{1}{2^{N+1}} \sum_{m,p=0}^{N} 
    \sqrt{\binom{N}{m} \binom{N}{p}} e^{i\theta(p-m)}
    \rho_{m,N-p;\, p,N-m}.
    \label{eq:fid_Fock-basis}
\end{align}
Here, $\theta = \phi + \pi$, where $\phi$ is the phase of the two-mode squeezing operator used in the resource state generation before the photon subtraction. In the standard Braunstein--Kimble teleportation setting, $\phi = \pi$ so $e^{i\theta(p-m)} = 1$. For reconstructed density matrix using a photon-number cut-off $n_c$ per mode, we can evaluate the coherence state teleportation fidelity using \cref{eq:fid_Fock-basis} by truncating the outer sum over $N$ at $2n_c$ and restricting the inner sums over $m$ and $p$ to the range $[\max(0, N - n_c), \min(N, n_c)]$ (alternatively the same range $[0, N]$ can be used by padding zeros for matrix elements out of the bound of the photon-number cut-off). We verified that given the same density matrix, the numerical method detailed in \cref{sec:numer_tele_fid} yields perfectly consistent  results, confirming the validity of this 
algebraic evaluation.

\paragraph{Teleportation fidelity in homodyne detection sampling form}
The homodyne measurement basis states are the eigenstates of the quadrature operator for each mode:
\begin{align}
    \hat{x}_{j, \vartheta}
 &= \frac{1}{\sqrt{2}}\left(\hat{a}_j e^{-i\vartheta} + \hat{a}_j^\dagger e^{i\vartheta}\right)
  = \hat{x}_j \cos\vartheta + \hat{p}_j\sin\vartheta
  = \mqty(\hat{x}_j &\hat{p}_j) \mqty(0 &1 \\ -1 &0)
    \mqty(\cos(\vartheta+\frac{\pi}{2}) \\
          \sin(\vartheta+\frac{\pi}{2}) ),
\end{align}
where $\hat{x}_j = \hat{x}_{j, 0} = \frac{1}{\sqrt{2}}(\hat{a}_j + \hat{a}_j^\dagger)$ and $\hat{p}_j = \hat{x}_{j, \frac{\pi}{2}} = \frac{-i}{\sqrt{2}}(\hat{a}_j - \hat{a}_j^\dagger)$. For $\alpha = (\tau + i\sigma)/\sqrt{2} = k e^{i\vartheta}/\sqrt{2}$, we have $\alpha \hat{a}_j^\dagger - \alpha^* \hat{a}_j = ik\hat{x}_j(\vartheta - \frac{\pi}{2}) = i\vu{x}_j^T \Omega \vb*{\xi}_j$, where $\vu{x}_j^T = (\hat{x}_j, \hat{p}_j)$, $\Omega = \spmqty{0 &1\\ -1&0}$, $\vb*{\xi}_j^T = (k\cos\vartheta, k\sin\vartheta)$. Using \cref{eq:fid_op_general} and plugging in $\alpha = ke^{i\varphi}/\sqrt{2}$ and $d^2 \alpha = k\, dk\, d\varphi/2$, we have
\begin{align}
    \hat{\Omega}_\theta
 &= \frac{1}{2\pi} \int_{0}^{2\pi} d\varphi \int_{0}^{\infty} k\,dk\,
    \abs{\chi_\text{in}\qty(\frac{ke^{i\varphi}}{\sqrt{2}})}^2
    \exp(ik\hat{x}_{1, -\varphi - \frac{\pi}{2}} + ik\hat{x}_{2, \varphi + \theta - \frac{\pi}{2}}) \notag \\
 &\equiv \frac{1}{2\pi} \int_{0}^{2\pi} d\varphi \int_{0}^{\infty} k\,dk\,
    \abs{\chi_\text{in}\qty(\frac{ke^{i\varphi}}{\sqrt{2}})}^2
    \exp(ik\hat{S}_{\varphi,\theta}),
\end{align}
where $\hat{S}_{\varphi,\theta} = \hat{x}_{1, -\varphi - \frac{\pi}{2}} + \hat{x}_{2, \varphi + \theta - \frac{\pi}{2}} = \hat{x}_{1, -\varphi - \frac{\pi}{2}} + \hat{x}_{2, \varphi + \frac{\pi}{2} + \phi}$. Here, $\phi$ is the phase of the two-mode squeezing operator. For coherent state input, we can change the integral variable $\varphi + \pi/2 \to \varphi$ and use $\hat{S}_{\varphi+\pi,\theta} = - \hat{S}_{\varphi,\theta}$ to simplify the above:
\begin{align}
    \hat{\Omega}_\theta
 &= \frac{1}{2\pi} \int_{0}^{2\pi} d\varphi \int_{0}^{\infty} k\,dk\, e^{-k^2/2} \cos(k\hat{S}_{\varphi - \frac{\pi}{2}, \theta})
 \equiv \frac{1}{2\pi} \int_{0}^{2\pi} d\varphi\,
    K\qty(\hat{S}_{\varphi - \frac{\pi}{2}, \theta}). \\
    K(x)
 &= \int_{0}^{\infty} k e^{-k^2/2} \cos(kx) \,dk
  = -\int_{0}^{\infty} \cos(kx)\, d\qty(e^{-k^2/2}) \notag \\
 &= 1 - x\int_{0}^{\infty} e^{-k^2/2}\sin(kx)\,dk
  = 1 - \sqrt{2}x\int_{0}^{\infty} e^{-r^2}\sin(\sqrt{2}rx)\,dr \notag\\
 &= 1 - \sqrt{2}x F_\text{D}\qty(x/\sqrt{2}).
\end{align}
where the kernel function $K(x) = \int_0^{\infty} k e^{-k^2/2}\cos(kx)\, dk = 1 - \sqrt{2}xF_D(x/\sqrt{2})$ and the Dawson function $F_D(x) \equiv e^{-x^2}\int_0^{x} e^{r^2}\, dr$. It can be shown that $F_D(x) = I_\text{sin}(x)$, where $I_\text{sin}(x) \equiv \int_0^{\infty} e^{-r^2} \sin(2xr) \, dr$, by the fact that they satisfy the same differential equation $y' = 1 - 2xy$ and $y(0) = 0$ for $y = F_D(x) = I_\text{sin}(x)$.

The teleportation fidelity is given by
\begin{align}
    F_{\theta = \phi + \pi}
 &= \Tr(\rho_\text{res}\hat{\Omega}_\theta)
  = \frac{1}{2\pi} \int_{0}^{2\pi} d\varphi \ev{K(\hat{S}_{\varphi-\frac{\pi}{2}, \theta})} \notag \\
 &= \frac{1}{2\pi} \int_{0}^{2\pi} d\varphi \,
    \ev{K(\hat{x}_{1,-\varphi} + \hat{x}_{2,\varphi + \phi})}
  = \frac{1}{\pi} \int_{0}^{\pi} d\varphi \,
    \ev{K(\hat{x}_{1,-\varphi} + \hat{x}_{2,\varphi + \phi})} .
    \label{eq:fid_homodyne}
\end{align}
The final step is due to $\hat{x}_{j,\vartheta + \pi} = -\hat{x}_{j,\vartheta}$ and $K(-x) = K(x)$.

\textbf{Evaluation from homodyne tomography data.}
For homodyne settings lying exactly on the phase-matching line, $\vartheta_1=-\varphi$ and $\vartheta_2=\varphi+\phi$, \cref{eq:fid_homodyne} can be evaluated directly from the measured quadrature outcomes. For a uniform set of $L$ angles $\varphi_l \in [0,\pi)$, with $N_l$ repeated measurements at each angle, the corresponding fidelity estimator is
\begin{align}
    \widehat{F}
 &= \frac{1}{L} \sum_{l=1}^{L} \frac{1}{N_l} \sum_{j=1}^{N_l} K\qty(x_{1,(l,j)}+x_{2,(l,j)}), \label{eq:fid_discrete_sampling}
\end{align}
where $K(x) = 1 - \sqrt{2}xF_D(x/\sqrt{2})$ is the kernel function given above.

In the tomography data, the local-oscillator phases do not lie exactly on the phase-matching line $\vartheta_1 + \vartheta_2 = \phi$ unless the experiment is specifically designed to scan the phase grid $(\vartheta_1, \vartheta_2) = (-\varphi_l, \varphi_l + \phi)$ for a fixed $\phi$ and $1\leq l \leq L$. To address this, we propose the following procedure to estimate the teleportation fidelity using homodyne tomography data without first reconstructing the density matrix.

For each recorded shot $i$, using the identity $\hat{x}_{\vartheta - \pi} = -\hat{x}_\vartheta$, each homodyne record is mapped to an equivalent phase convention by defining $m_i = \operatorname{round}[(\vartheta_{1,i}+\vartheta_{2,i}-\phi)/\pi]$, $\tilde{\vartheta}_{2,i} = \vartheta_{2,i} - m_i\pi$, and $\tilde{x}_{2,i} = (-1)^{m_i}x_{2,i}$. This folds the phase sum onto the branch nearest the phase-matching line without altering the physical measurement outcome. The phase mismatch and angular coordinate of the shot are then $\delta_i = \vartheta_{1,i} + \tilde{\vartheta}_{2,i} -\phi$, $-\frac{\pi}{2}\leq \delta_i < \frac{\pi}{2}$, and $\varphi_i =(\tilde{\vartheta}_{2,i} - \vartheta_{1,i} -\phi)/2 \pmod{\pi}$. Thus, $\delta_i=0$ corresponds exactly to the settings required in \cref{eq:fid_homodyne}. For each shot we calculate $s_i = x_{1,i} + \tilde{x}_{2,i}$ and $y_i = K(s_i)$.

This 1D kernel approach is specifically designed to estimate the teleportation fidelity from independent homodyne measurements. An alternative homodyne measurement scheme is given in Ref.~\cite{Nha2012}, which evaluates a 2D exponential kernel. While that approach yields better sample efficiency, it strictly requires adding a physical beam splitter to mix the resource modes prior to the simultaneous measurement of commuting quadratures, and is only advantageous if the loss of the additional beam splitter is negligible.

\begin{figure*}[tbh]
  \centerline{
  \includegraphics[width=0.7\textwidth]{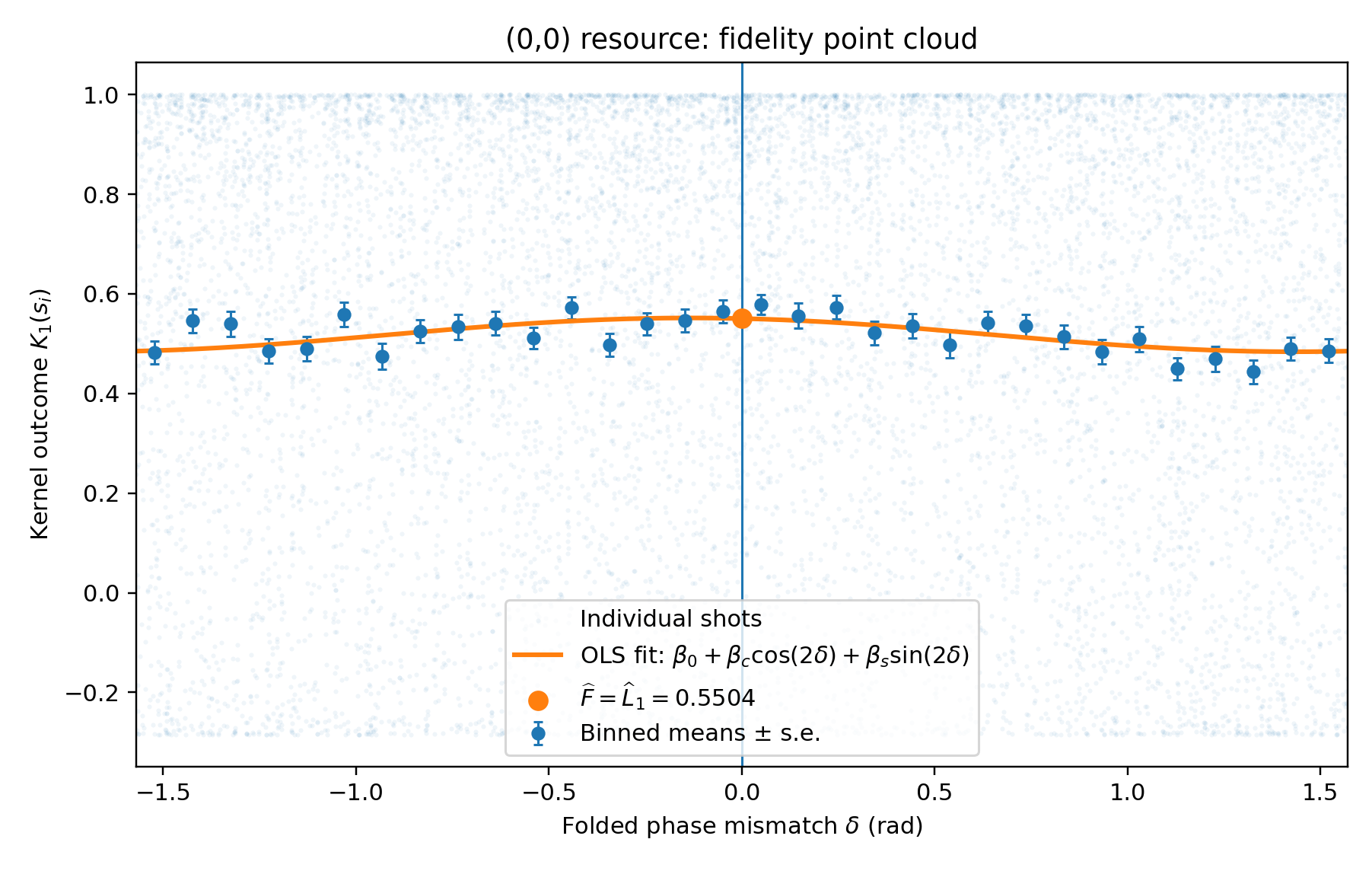}}
  \centerline{
  \includegraphics[width=0.7\textwidth]{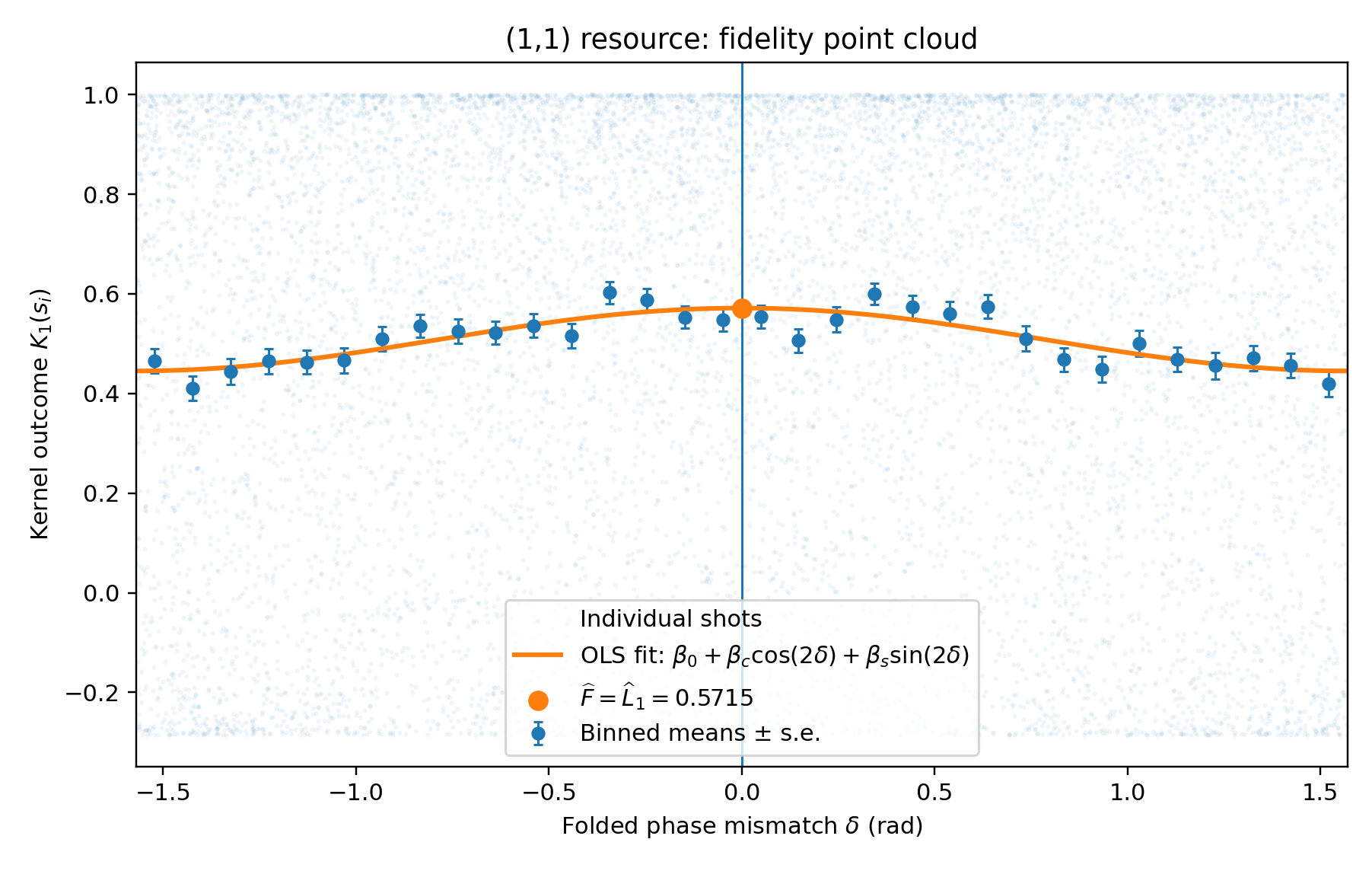}}
\caption{Kernel function $K(s)$ evaluated using tomography data. The $y$-axis represents the $K(s)$ value, where $s = x_1 + \tilde{x}_2$ (the sum of the amplitudes of the two quadrature variables) and the $x$-axis represents $\delta = \vartheta_1 + \tilde{\vartheta_2}$ (the sum of the phases of the two corresponding quadrature variables).
The orange dot marks the estimated value for the expected teleportation fidelity for the resource state prepared from (0,0) subtraction (top panel) and (1,1) subtraction (bottom panel). The background blue point cloud represents kernel function values from the full homodyne data set; the orange curve is an ordinary least squares (OLS) fit to the full data set; and the blue dots with error bars are binned mean values of the kernel function, plotted as a guide to the eye.}
  \label{fig:teleFidelity_tomoData}
\end{figure*}

To avoid selecting an arbitrary finite window around $\delta=0$, we estimate the conditional kernel mean using the empirical periodic form $g(\delta) =  \mathbb{E}[K(s)|\delta] \simeq \widehat{\beta}_{0} + \widehat{\beta}_{c}\cos(2\delta) + \widehat{\beta}_{s}\sin(2\delta)$. 
%
This model assumes that any dependence on the angular coordinate $\varphi$ is negligible for the present data. The $\beta$ coefficients are obtained by minimizing the squared shot-level residuals $\sum_i [y_i - \beta_0 - \beta_c \cos(2\delta_i) - \beta_s \sin(2\delta_i)]^2$ using ordinary least squares (see \cref{fig:teleFidelity_tomoData}).
The phase-matched fidelity is therefore $\widehat{F} = \widehat{g}(0) = \widehat{\beta}_{0} + \widehat{\beta}_{c}$. For the resource phase $\phi=0$, the Fourier-regression analysis gave the estimated fidelities $0.5504$ for $(0,0)$-photon subtraction and $0.5715$ for $(1,1)$-photon subtraction. These values are in agreement with the fidelities $0.5539$ and $0.5635$ obtained independently from the maximum-likelihood reconstructed density matrices.

\subsubsection{Analytical evaluation of teleportation fidelity}
For Gaussian states and non-Gaussian resource from photon subtraction, analytical results can be obtained, so we will derive these results here and then describe numerical method applicable to arbitrary input and resource states, including those obtained from tomography experiment measurements, in \cref{sec:numer_tele_fid}. Our analytical results will also serve as test cases to our numerical evaluation code. We will consider the expected fidelity teleporting a coherence state using our resource state generated from photon-subtracted (PS) TMSV followed by additional pure loss channels.

A pure loss channel $\mathcal{L}_\eta(\rho)$ can be accounted for in a very simple way with the characteristic function $\chi(\alpha) = {\Tr}[\rho\hat{D}(\alpha)]$:
\begin{align}
    \chi^\text{loss}(\alpha) 
 &= \Tr[\mathcal{L}_\eta(\rho) \hat{D}(\alpha)]
  = \Tr[\rho \mathcal{L}_\eta^\dagger\qty(\hat{D}(\alpha))],
\end{align}
where $\mathcal{L}_\eta^\dagger$ is the dual of the channel in Heisenberg picture~\cite{Holevo2001}. The trace preservation of $\mathcal{L}$ is equivalent to the unitality of the dual $\mathcal{L}^\dagger(\openone) = \openone$. It is easy to find
\begin{align}
    \mathcal{L}_\eta^\dagger\qty(\hat{D}(\alpha))
 &= D(\sqrt{\eta} \alpha) \exp(-\frac{1-\eta}{2}|\alpha|^2).
\end{align}
This follows directly from the beam-splitter dilation shown in \cref{fig:circuit} (purple boxes):
\begin{align*}
    \mathcal{L}_\eta(\rho) 
 &= \Tr_E[ U_\eta (\rho \otimes \ket{0} {\! {}^{}_E \!} \bra{0}) U_\eta^\dagger] \\
    \implies
    \mathcal{L}_\eta^\dagger\qty(\hat{D}(\alpha)) 
 &= {}^{}_E\! {\ev{U_\eta^\dagger \hat{D}(\alpha) U_\eta}{0}} {}^{}_E \\
 &= {}^{}_E\! {\ev{\hat{D}(\sqrt{\eta}\alpha) \hat{D}_E(\sqrt{1-\eta}\alpha)}{0}} {}^{}_E \\
 &= \hat{D}(\sqrt{\eta}\alpha) \exp(-\frac{1-\eta}{2}|\alpha|^2).
\end{align*}
Here, we used the fact the beam-splitter unitary $U_\eta^\dagger (\alpha\hat{a}^\dagger - \alpha^* \hat{a}) U_\eta = \sqrt{\eta}(\alpha\hat{a}^\dagger - \alpha^* \hat{a}) + \sqrt{1-\eta}(\alpha\hat{a}_E^\dagger - \alpha^* \hat{a}_E)$.

Therefore, for a single mode through a pure loss channel, the characteristic function
\begin{align}
    \chi^\text{loss}(\alpha) 
 &= \Tr[\rho \hat{D}(\sqrt{\eta}\alpha)] \exp(-\frac{1-\eta}{2}|\alpha|^2)
  = \chi(\sqrt{\eta} \alpha) \exp(-\frac{1-\eta}{2}|\alpha|^2).
\end{align}
Similarly, for two modes, the resource state has the following characteristic function after going through respective loss channels:
\begin{align}
    \chi_\text{res}^\text{loss}(\alpha_1, \alpha_2)
 &= \chi_\text{res}(\sqrt{\eta_1}\alpha_1, \sqrt{\eta_2}\alpha_2) 
    \exp(-\frac{1-\eta_1}{2}|\alpha_1|^2 - \frac{1-\eta_2}{2}|\alpha_2|^2).
\end{align}

Then, a similar result to Eq.~(12) in~\cite{PhysRevA.107.012418} can be obtained if we include the loss. But for the special case $m_1 = m_2 = 0$ and $n_1 = n_2 = n$, without using the full machinery developed in~\cite{PhysRevA.107.012418},  the coherence-state teleportation fidelity using (0,0)- and (1,1)-photon subtracted non-Gaussian resource state can be explicitly calculated and the result is given by
\begin{subequations}
\begin{align}
    F^{\text{NG},0}_\text{coh} 
 &= \frac{1-\lambda^2}{D}, \label{eq:tel_F_NG0_coh}\\
    F^{\text{NG},1}_\text{coh} 
 &= \frac{(1-\lambda^2)^3}{(1+\lambda^2)D^3}\qty[D^2 + (2g\lambda + 3h\lambda^2)D + 2\lambda^2(g+h\lambda)^2], \label{eq:tel_F_NG1_coh}
\end{align}
\end{subequations}
where $D = 2 - 2g\lambda -h\lambda^2$, $g = \sqrt{\eta_1 \eta_2}$, $h = 2 - \eta_1 - \eta_2$, and the effective squeezing parameter $\lambda = (\tanh r) (1-R_S)$. For $r = 0.3$, $R_S = 0.14$, $\eta_1 = 0.55$ and $\eta_2 = 0.5$, we find $F^{\text{NG},0}_\text{coh} \approx 0.5587$ and $F^{\text{NG},1}_\text{coh} \approx 0.5953$. Compared to these analytical results, the results from our numerical evaluation using a finite Fock space cut-off are $F^{\text{NG},0}_\text{coh} \approx 0.5586$ and $F^{\text{NG},1}_\text{coh} \approx 0.5952$.

We now derive \cref{eq:tel_F_NG0_coh,eq:tel_F_NG1_coh} using \cref{eq:in-res-fid}. The input state is a coherence state $\ket{\alpha_0}$ with characteristic function
\begin{align}
    \chi_\text{in}(\alpha)
 &= \Tr[\op{\alpha_0} \hat{D}(\alpha)]
  = \ev{e^{\alpha\hat{a}^\dagger - \alpha^* \hat{a}}}{\alpha_0} \notag\\
 &= \ev{e^{-|\alpha|^2/2}e^{\alpha\hat{a}^\dagger} e^{-\alpha^* \hat{a}}}{\alpha_0}
  = e^{-|\alpha|^2/2} e^{\alpha\alpha_0^*} e^{-\alpha^* \alpha_0} \notag\\
 &= e^{-|\alpha|^2/2 + \alpha\alpha_0^* - \alpha^* \alpha_0}
  = \chi_{\ket{0}}(\alpha) \exp(2i \Im\alpha\alpha_0^*),
\end{align}
where $\chi_{\ket{0}}(\alpha) = e^{-|\alpha|^2/2}$ is the characteristic function of the vacuum state. Therefore, $|\chi_\text{in}(\alpha)|^2 = |\chi_{\ket{0}}(\alpha)|^2 = e^{-|\alpha|^2}$.

The characteristic function for the resource state before loss can be calculated from resource states in Fock basis $\ket{\Psi_{0,0}}$ in \cref{eq:Psi_00} and $\ket{\Psi_{1,1}}$ in \cref{eq:Psi_11} after substituting $\lambda \to -\lambda$ because the squeezing operator \cref{eq:S2_EPR} for the EPR resource state in standard teleportation protocol corresponds to $\phi = \pi$ in \cref{eq:S2_r_phi} instead of $\phi = 0$ used in \cref{eq:Psi_00,eq:Psi_11} (this means the resource state in the analytical evaluation of the teleportation fidelity will include the phase shifter effect; see \cref{fig:circuit}). \cref{eq:Psi_00} has the same form as TMSV \cref{eq:TMSV}, so we can rewrite
\begin{subequations}
\begin{align}
    \rho_{\text{res}}^{(0,0)}
 & = \op{\Psi_{0,0}}
   = \hat{S}(\bar{r})\op{00} \hat{S}^\dagger(\bar{r}), \\
    \hat{S}(\bar{r})
 &= \exp[-\bar{r}\qty(\hat{a}_1 \hat{a}_2 - \hat{a}_1^\dagger \hat{a}_2^\dagger)], \label{eq:S2_EPR}\\
    \tanh(\bar{r})
 &= \lambda = \tanh(r) (1-R_S).
\end{align}
\end{subequations}

Because the dual of squeezing unitary (channel) $\hat{S}$ acts on the Weyl displacement operator $\hat{D}(\alpha_1, \alpha_2) = \hat{D}_1(\alpha) \otimes \hat{D}_2(\alpha_2)$
\begin{align}
    \hat{S}^\dagger(\bar{r}) \hat{D}(\alpha_1, \alpha_2) \hat{S}(\bar{r})
 &= \hat{D}(\alpha_1\cosh \bar{r} - \alpha_2^*\sinh \bar{r},
            \alpha_2\cosh \bar{r} - \alpha_1^*\sinh \bar{r}),
\end{align}
the characteristic function of the resource state can be obtained from the vacuum characteristic function $\chi_{\ket{00}}(\alpha_1,\alpha_2) = \ev{\hat{D}(\alpha_1,\alpha_2)}{00} = {\exp}(-|\alpha_1|^2/2 -|\alpha_2|^2/2)$ as follows
\begin{align}
     \chi_{\text{res}}^{(0,0)}(\alpha_1, \alpha_2)
  &= \Tr [\rho_{\text{res}}^{(0,0)} \hat{D}(\alpha_1,\alpha_2)]
   = \chi_{\ket{00}}(\alpha_1\cosh \bar{r} - \alpha_2^*\sinh \bar{r}, \alpha_2\cosh \bar{r} - \alpha_1^*\sinh \bar{r}) \notag \\
  &= \exp[-\frac{1}{2}\cosh(2\bar{r}) (|\alpha_1|^2 + |\alpha_2|^2) + \sinh(2\bar{r}) \Re(\alpha_1 \alpha_2)].
\end{align}

Similarly, we can calculate $\chi_{\text{res}}^{(1,1)}$ as follows.
\begin{align*}
 \because \ket{\Psi_{1,1}}
 &\propto \hat{a}_1 \hat{a}_2 \ket{\Psi_{0,0}} 
  = \hat{S}(\bar{r}) \qty[ \hat{S}^\dagger(\bar{r})\hat{a}_1 \hat{a}_2 \hat{S}(\bar{r})] \ket{00} \\
 &= \hat{S}(\bar{r}) \qty[
     (\hat{a}_1 \cosh \bar{r} + \hat{a}_2^\dagger \sinh \bar{r})
     (\hat{a}_2 \cosh \bar{r} + \hat{a}_1^\dagger \sinh \bar{r})
    ] \ket{00} \\
 &= \hat{S}(\bar{r})[\cosh \bar{r}\sinh \bar{r} \ket{00}  + \sinh^2 \bar{r}\ket{11}], \\
 \therefore \ket{\Psi_{1,1}}
 &= \hat{S}(\bar{r}) \frac{\ket{00} + \lambda\ket{11}}{\sqrt{1 + \lambda^2}}
  \equiv \hat{S}(\bar{r}) \ket{\phi},
\end{align*}
where $\ket{\phi} = (\ket{00} + \lambda\ket{11})/\sqrt{1+\lambda^2}$ has the characteristic function
\begin{align}
    \chi_{\ket{\phi}}(\alpha_1, \alpha_2)
 &= \frac{\chi_{\ket{00}}(\alpha_1, \alpha_2)}{1+\lambda^2}
    \qty[1 + \lambda^2 (1-|\alpha_1|^2) (1-|\alpha_2|^2) +2\lambda\Re(\alpha_1 \alpha_2)].
\end{align}
Thus,
\begin{align}
    \chi_\text{res}^{(1,1)}(\alpha_1, \alpha_2)
 &= \chi_{\ket{\phi}}(\alpha_1\cosh \bar{r} - \alpha_2^*\sinh \bar{r}, \alpha_2\cosh \bar{r} - \alpha_1^*\sinh \bar{r}).
\end{align}

After putting everything together, we find
\begin{align}
    \chi_\text{res}^{(0,0),\text{loss}}(\alpha^*, \alpha)
 &= \chi_\text{res}^{(0,0)}(\sqrt{\eta_1}\alpha^*, \sqrt{\eta_2}\alpha) e^{-(2 - \eta_1 -\eta_2)|\alpha|^2/2}
  = \exp(-c_0|\alpha|^2). \\
    \chi_\text{res}^{(1,1),\text{loss}}(\alpha^*, \alpha)
 &= \exp(-c_0|\alpha|^2)
    \qty(1 + c_1|\alpha|^2 + c_2|\alpha|^4), \\
 &\begin{cases}
 \displaystyle
 c_0 = \frac{1}{1-\lambda^2}
       \qty[\lambda^2(\eta_1 + \eta_2) - 2\sqrt{\eta_1 \eta_2}\lambda + 1-\lambda^2], \\
 \displaystyle
 c_1 = \frac{\lambda}{1-\lambda^4}
       \qty[2(1+3\lambda^2)\sqrt{\eta_1 \eta_2} - \lambda(3+\lambda^2)(\eta_1 + \eta_2)], \\
 \displaystyle
 c_2 = \frac{\lambda^2}{(1+\lambda^2)(1-\lambda^2)^2}
      \qty[(1+\lambda^2)\sqrt{\eta_1 \eta_2} - \lambda(\eta_1 + \eta_2)]^2.
 \end{cases}
\end{align}

Now we calculate the fidelity using \cref{eq:in-res-fid} (with $\phi = \pi$ and $e^{i\theta} = e^{i(\phi + \pi)} = 1$):
\begin{align}
    F_\text{coh}^{\text{NG},0}
 &= \frac{1}{\pi} \int d^2\alpha\, |\chi_\text{in}(\alpha)|^2
    \chi_\text{res}^{(0,0),\text{loss}}(\alpha^*, \alpha) \notag \\
 &= \frac{1}{\pi} \int d^2\alpha\, e^{-(c_0+1)|\alpha|^2} \notag \\
 &= \frac{1}{\pi}\int_0^{2\pi} d\theta\, \int_0^{\infty} rdr\, e^{-(c_0+1)r^2} \notag\\
 &= \frac{1}{c_0 + 1} \\
 &= \frac{1-\lambda^2}{D},
\end{align}
where $D = 2 - 2g\lambda - h\lambda^2$, $g = \sqrt{\eta_1 \eta_2}$, and $h = 2 - \eta_1 - \eta_2$. This is \cref{eq:tel_F_NG0_coh}. For (1,1)-photon subtracted resource state:
\begin{align}
    F_\text{coh}^{\text{NG},1}
 &= \frac{1}{\pi} \int d^2\alpha\, |\chi_\text{in}(\alpha)|^2
    \chi_\text{res}^{(1,1),\text{loss}}(\alpha^*, \alpha) \notag \\
 &= \frac{1}{\pi} \int d^2\alpha\, e^{-(c_0+1)|\alpha|^2}
    \qty(1 + c_1|\alpha|^2 + c_2|\alpha|^4)\notag \\
 &= \frac{1}{\pi}\int_0^{2\pi} d\theta\, \int_0^{\infty} rdr\, e^{-(c_0+1)r^2}
    (1+c_1 r^2 + c_2 r^4) \notag\\
 &= \frac{1}{c_0 + 1} + \frac{c_1}{(c_0+1)^2} + \frac{2c_2}{(c_0+1)^3} \\
 &= \frac{(1-\lambda^2)^3}{(1+\lambda^2)D^3}
    \qty[D^2 + (2g\lambda + 3h\lambda^2)D + 2\lambda^2(g+h\lambda)^2],
\end{align}
which is \cref{eq:tel_F_NG1_coh}.

For the symmetric loss $\eta_1 = \eta_2 = \eta$, the simplified forms are
\begin{align}
    F_\text{coh}^{\text{NG},0}
 &= \frac{1+\lambda}{2[1+(1-\eta)\lambda]} 
  = \frac{1}{2-\eta + \eta e^{-2r} + \eta R_S(1-e^{-4r})/[2 - R_S(1-e^{-2r})]}. 
  \label{eq:tel_F0_symloss} \\
    F_\text{coh}^{\text{NG},1}
 &= \frac{(1+\lambda)^3[2-2\eta\lambda + (\eta^2 - 2\eta + 2)\lambda^2]}{4(1+\lambda^2)[1 + (1-\eta)\lambda]^3}. \label{eq:tel_F1_symloss}
\end{align}
By setting $\eta_1 = \eta_2 = 1$, the lossless case 
\begin{align}
    F_\text{coh}^{\text{NG},0}
 &= \frac{1 + \lambda}{2}
  = \frac{1 + \tanh \bar{r}}{2}
  = \frac{1}{1 + \exp(-2\bar{r})}. \label{eq:tel_F0_noloss}\\
    F_\text{coh}^{\text{NG},1}
 &= \frac{(1+\lambda)^3(2 -2\lambda +\lambda^2)}{4(1+\lambda^2)}, \label{eq:tel_F1_noloss}
\end{align} 
where $\lambda = \tanh(r) (1-R_S) \equiv \tanh(\bar{r})$. By the substitution $\lambda \to \tanh r$ (or equivalently $R_S\to 0$) in \cref{eq:tel_F0_symloss}, we obtain the teleportation fidelity using a standard TMSV resource state with the squeezing parameter $r$ and the symmetric loss $\eta$ but without any photon subtraction construction:
\begin{align}
    F_\text{TMSV}^\text{loss}
 &= \frac{1 + \tanh r}{2[1 + (1-\eta)\tanh r]}
  = \frac{1}{2-\eta + \eta e^{-2r}}.
  \label{eq:tel_F_TMSV_loss}
\end{align}

\subsubsection{Numerical evaluation of teleportation fidelity}
\label{sec:numer_tele_fid}
Below we describe the numerical framework for evaluating the teleportation fidelity of experimentally achievable photon-subtracted two-mode squeezed vacuum (TMSV) states, including standard TMSV states as a baseline. We detail the phase-space integration strategy, the dynamic truncation of the Hilbert space, the comparison between our numerical and analytical results, and the fidelity evaluation of experimental states reconstructed via maximum likelihood estimation.

\paragraph{Numerical-Integration Strategy}
To compute the teleportation fidelity numerically, the continuous phase space 
must be discretized, and the infinite-dimensional Hilbert space must be safely 
truncated. Specifically, the evaluation procedure proceeds in two steps: first, 
we represent the density matrix in a truncated photon-number basis and expand 
the characteristic function using a unified algebraic formulation of the 
displacement operator; second, we evaluate the fidelity via a discretized polar 
integration. Here, both the photon-number cut-off threshold and the phase-space integration domain are dynamically scaled according to the specific two-mode resource state used.

To formalize the first step, we construct the state representation within a finite-dimensional subspace. For a chosen photon-number
cut-off $n_c$, the single-mode photon-number space is spanned by the basis 
$\{|0\rangle, |1\rangle, \dots, |n_c\rangle\}$, resulting in a Hilbert-space dimension 
of $(n_c + 1)$. A single-mode density matrix is thus represented as:
\begin{align}
    \rho = \sum_{m,n=0}^{n_c} \rho_{n,m} |n\rangle\langle m|.
\end{align}
By substituting this representation into the definition from Eq.~\eqref{eq:sm-chi}, 
the single-mode characteristic function can be expressed as a weighted sum over the 
matrix elements of the displacement operator:
\begin{align}
    \chi(\alpha) = \sum_{m,n=0}^{n_c} \rho_{n,m} \langle m | \hat{D}(\alpha) | n \rangle = \sum_{m,n=0}^{n_c} \rho_{n,m} D_{mn}(\alpha), 
\end{align}
where $D_{mn}(\alpha) \equiv \langle m | \hat{D}(\alpha) | n \rangle$. To compute 
$D_{mn}(\alpha)$ accurately without encountering numerical singularities, we use a 
unified expansion of the displacement operator in the normal ordered form $\hat{D}(\alpha) = e^{-|\alpha|^2/2} e^{\alpha \hat{a}^\dagger} e^{-\alpha^* \hat{a}}$ derived via the Baker-Campbell-Hausdorff 
formula. The matrix element $D_{mn}(\alpha)$ evaluates to be 
\begin{align}
    D_{mn}(\alpha) = e^{-|\alpha|^2/2} \sqrt{m!n!} \sum_{j=0}^{\min(m,n)} \frac{\alpha^{m-j} (-\alpha^*)^{n-j}}{j!(m-j)!(n-j)!}.
\end{align}
Similarly, for a two-mode resource state, the density matrix ${\rho}_{12}$ can be 
expanded in the joint photon-number basis $\{|n,m\rangle \equiv |n\rangle_1 \otimes |m\rangle_2\}$ up to the cut-off $n_c$:
\begin{align}
    {\rho}_{12} =\sum_{n,n'=0}^{n_c} \sum_{m,m'=0}^{n_c} \rho_{nm, n'm'} |n,m\rangle\langle n',m'|.
\end{align}
Applying the definition from Eq.~\eqref{eq:tm-chi}, the joint characteristic function becomes
\begin{align}
    \chi_{\text{res}}(\alpha_1, \alpha_2) &= \sum_{n,n'=0}^{n_c} \sum_{m,m'=0}^{n_c} \rho_{nm, n'm'} \langle n', m' | \hat{D}(\alpha_1) \otimes \hat{D}(\alpha_2) | n, m \rangle \nonumber \\
    &= \sum_{n,n'=0}^{n_c} \sum_{m,m'=0}^{n_c} \rho_{nm, n'm'} \langle n' | \hat{D}(\alpha_1) | n \rangle \langle m' | \hat{D}(\alpha_2) | m \rangle \nonumber \\
    &= \sum_{n,n'=0}^{n_c} \sum_{m,m'=0}^{n_c} \rho_{nm, n'm'} D_{n'n}(\alpha_1) D_{m'm}(\alpha_2).
\end{align}

With the characteristic functions computed in the truncated photon-number
basis, we proceed to the second step: evaluating the teleportation fidelity via 
numerical integration over the phase space. By defining polar coordinates
$(k, \theta)$ in the quadrature plane such that $\tau + i\sigma = k e^{i\theta}$, 
the complex amplitude becomes $\alpha = k e^{i\theta}/\sqrt{2}$. The 
differential area element transforms as $d^2\alpha = \frac{1}{2} d\tau d\sigma = \frac{k}{2} \, dk \, d\theta$. 
Consequently, the fidelity formula transforms into a double integral:
\begin{align}
    F = \frac{1}{2\pi} \int_0^{2\pi} \int_0^{k_{\mathrm{max}}} \chi_{\mathrm{in}}(k, \theta) \chi_{\mathrm{out}}(-k, \theta) \, k \, dk \, d\theta.
\end{align}
To accurately capture the full probability mass of the state, the upper
integration boundary $k_{\mathrm{max}}$ must be scaled dynamically
to enclose the effective phase-space support of the input and resource 
states. Because photon subtraction physically increases the phase-space 
spread of the resulting two-mode squeezed state, this boundary can be 
safely heuristically bounded by $k_{\mathrm{max}} \approx 3.5 e^r + 0.5(s_1 + s_2)$, where $r$ is the squeezing parameter and $s_1$ and $s_2$
are the numbers of photons subtracted from mode 1 and 2, respectively. 
In our numerical evaluation, the continuous integral is calculated by 
discretizing the amplitude $k$ and phase $\theta$ into a high-resolution
grid (specifically, $150 \times 150$ grid points in our numerical evaluation) up to this $k_{\mathrm{max}}$ boundary, and executing
two-dimensional numerical quadrature.
Correspondingly, the photon-number cut-off $n_c$ must be chosen large 
enough to prevent boundary truncation errors. A strictly safe upper limit
tracks with the integration boundary as 
$n_c = \lceil k_{\mathrm{max}}^2 / 2 \rceil + 3$. In practice, however, 
the physical states possess a bounded mean photon number 
with rapidly vanishing probability tails at high photon numbers.
As a result, exact numerical convergence of the fidelity is often achieved 
at a smaller $n_c$ than this conservative upper bound requires, further 
accelerating the evaluation process.

\begin{figure}[htbp]
     \centering
     \includegraphics[width=0.6\linewidth]{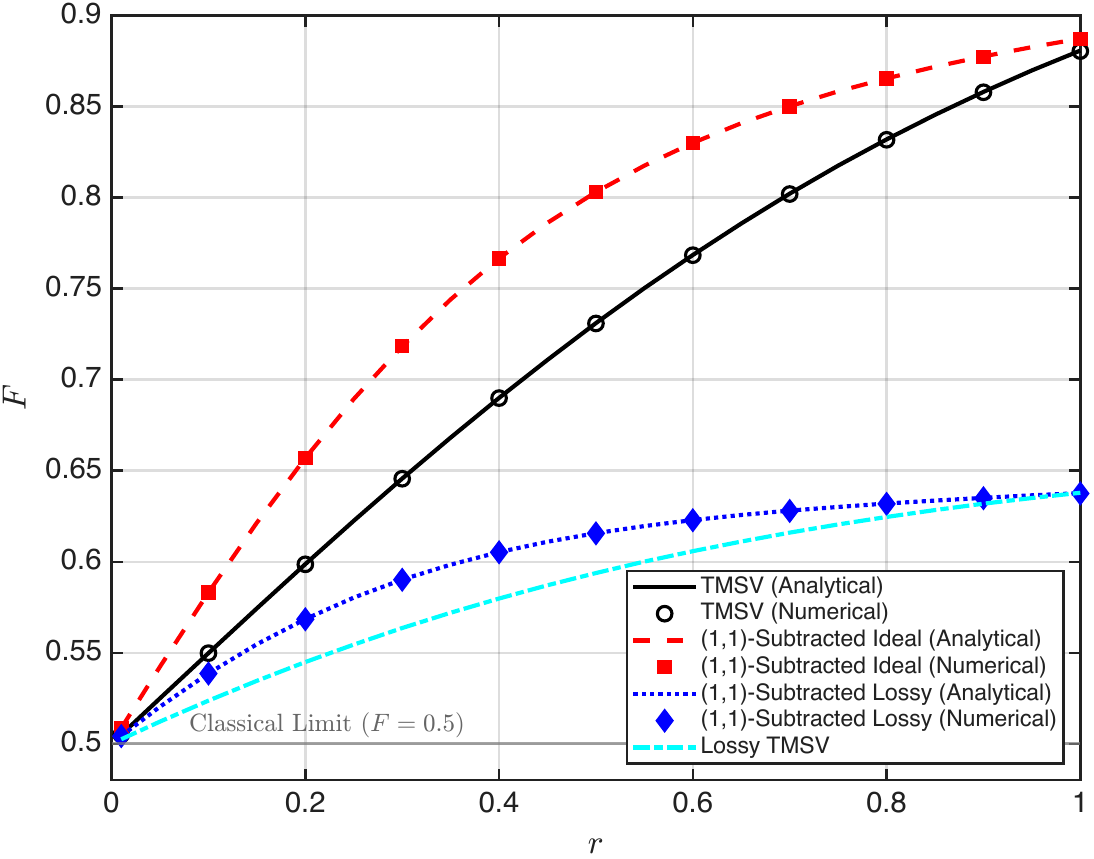}
     \caption{Coherent-state teleportation fidelity $F$ as a function of the squeezing parameter $r$. The plot compares analytical (curves) and numerical (symbols) results for three resource states: a standard TMSV state (solid black curve and circles), an ideal $(1,1)$-photon-subtracted TMSV state without photon loss (dashed red curve and squares), and a $(1,1)$-photon-subtracted TMSV state subjected to photon loss with transmission efficiency $\eta = 0.5$ (dotted blue curve and diamonds). Photon subtraction at each optical mode is performed via photon detection at the reflected port of a beam splitter with reflectivity $R = 0.14$. Additionally, the plot shows the results for a standard TMSV state subjected to the same photon loss (dash-dotted cyan curve); numerical markers for this case are omitted for clarity, as they perfectly overlap with the analytical curve.}
     \label{fig:fidelity}
\end{figure}

\paragraph{Numerical Validation and Resource State Comparison} 
Our numerical framework was first validated against the analytical formula 
for the teleportation fidelity using a standard TMSV resource state with 
squeezing parameter $r$ [Eq.\ (13) of Ref.~\cite{PhysRevA.107.012418}]:
\begin{equation}
    F_{\mathrm{TMSV}} = \frac{1}{1 + e^{-2r}}.
\end{equation}
The numerical integration matches the theoretical limit perfectly across all 
tested squeezing regimes ($r \in [0, 1.0]$), confirming that the numerical 
calculation of teleportation fidelity is accurate and independent of the 
specific coherent input amplitude. See \cref{fig:fidelity}.

To analytically verify this independence regardless of the initial complex 
amplitude $\alpha_0$, we examine the fidelity integrand. For an input coherent state 
$|\alpha_0\rangle$, the single-mode characteristic function is $\chi_{\mathrm{in}}
(\alpha) = \exp(-\frac{1}{2}|\alpha|^2 + \alpha\alpha_0^* - \alpha^*\alpha_0)$. 
The Braunstein-Kimble protocol yields an output state with characteristic function 
$\chi_{\mathrm{out}}(\alpha) = \chi_{\mathrm{in}}(\alpha) \chi_{\mathrm{res}}(\alpha^*, \alpha)$. Consequently, the overlap integral for the fidelity is 
determined by the product 
$\chi_{\mathrm{in}}(\alpha) \chi_{\mathrm{out}}(-\alpha) = \chi_{\mathrm{in}}(\alpha) \chi_{\mathrm{in}}(-\alpha) \chi_{\mathrm{res}}(-\alpha^*, -\alpha)$. 
When taking the product of the input characteristic functions, the 
amplitude-dependent phase terms perfectly cancel:
\begin{align}
    \chi_{\text{in}}(\alpha) \chi_{\text{in}}(-\alpha)
 &= \exp\left(-\frac{1}{2}|\alpha|^2  + \alpha\alpha_0^* - \alpha^*\alpha_0\right) 
    \exp\left(-\frac{1}{2}|-\alpha|^2 - \alpha\alpha_0^* + \alpha^*\alpha_0\right)
  = \exp(-|\alpha|^2).
\end{align}
Because the input amplitude $\alpha_0$ completely vanishes from the final integrand, 
the resulting teleportation fidelity depends exclusively on the correlation properties
of the resource state $\chi_{\mathrm{res}}$. Hence, the teleportation performance of 
an arbitrary two-mode resource state is strictly independent of the amplitude of the 
input coherent state. For this reason, we will not specify the amplitude of the input coherent state in the following evaluations.  

When evaluating non-Gaussian resource states, the numerical results demonstrate a 
distinct advantage for symmetrically photon-subtracted states. We numerically
confirmed that symmetrically subtracting one photon from each mode (that is, 
$(1,1)$-photon subtraction) enhances quantum entanglement, yielding a higher 
teleportation fidelity than a standard TMSV state at low-to-moderate squeezing 
levels ($r \leq 1.0$). 

However, this non-Gaussian advantage is highly fragile to photon loss. To 
demonstrate this, we simulated realistic experimental conditions. In 
particular, we considered the case where photon subtraction at each mode is 
performed via photon detection at the reflected port of a beam splitter with 
reflectivity $R = 0.14$, after which each optical mode is transmitted with a 
$50$\% transmission efficiency ($\eta = 0.5$). Our numerical results show that 
the presence of photon loss severely suppresses the teleportation fidelity, 
dragging the performance curve down toward the classical teleportation boundary 
of $F = 0.5$.
However, the fidelity remains above the classical value $0.5$ if $\eta > 0$ and is upper bounded by $1/(2-\eta)$ (for $r\to \infty$ and $R_S=0$) --- an identical limit for both $(0,0)$- and $(1,1)$-photon subtracted TMSV with loss.
Nevertheless, if a benchmark TMSV resource state suffers from the same level of loss degradation, the non-Gaussian advantage of the photon-subtracted state remains even after taking into account the reduction of the effective squeezing due to photon-subtraction tapping via the beam splitter. As shown by the cyan dash-dotted line in \cref{fig:fidelity}, which 
plots \cref{eq:tel_F_TMSV_loss} for standard TMSV under the same loss but without photon-subtraction tapping, the standard state strictly underperforms the photon-subtracted state. Altogether, these numerical results, illustrating both the non-Gaussian 
advantage and its degradation under photon loss, are summarized in Fig.~\ref{fig:fidelity}. The figure also confirms the exact consistency between our numerical and analytical results. 

\paragraph{Expected Teleportation Performance with Reconstructed Experimental States}
The finalized numerical pipeline was subsequently applied to physically reconstructed resource states generated via maximum-likelihood estimation (MLE) as detailed in Sect.~\ref{sect:mle}. Two experimental density matrices were evaluated: a baseline state with $(0,0)$-photon subtraction, and a non-Gaussian state with $(1,1)$-photon subtraction.

Crucially, the raw states produced by the experiment are geometrically \textit{anti-correlated} rather than correlated. Because the standard Braunstein-Kimble protocol relies on standard quadrature correlations to cancel quantum noise via classical feed-forward, directly inputting the reconstructed density matrices results in noise amplification. 

To bridge the gap between the experimental resource and the theoretical protocol, a local mathematical correction is applied to the reconstructed density matrices $\rho_{\mathrm{MLE}}$ prior to computing the characteristic function. By applying a $\pi$ phase shift to mode 2 using the unitary operator $\hat{U} = \hat{I}_1 \otimes \exp(i \pi \hat{n}_2)$ with $\hat{n}_2=\hat{a}_2^\dagger \hat{a}_2$, the anti-correlated states $\rho_{\mathrm{MLE}}$ are perfectly rotated into standard correlated states:
\begin{equation}
    \rho_{\mathrm{correlated}} = \hat{U} \rho_{\mathrm{MLE}} \hat{U}^\dagger.
\end{equation}

Once this unitary transformation is applied, the resulting density matrices $\rho_{\mathrm{correlated}}$ can be seamlessly integrated into the standard characteristic function formulas, allowing for the numerical calculation of the expected teleportation fidelity with experimental resources. Using the reconstructed density matrix conditional on $(0,0)$-photon subtraction, the expected fidelity for teleporting any coherent state is $0.5539$. Remarkably, when utilizing the reconstructed density matrix conditional on $(1,1)$-photon subtraction, this expected fidelity increases to $0.5635$, suggesting the presence of a non-Gaussian advantage in our experimental realization subject to optical loss. This fidelity increase using photon subtraction $0.5635$ about breaks even with the estimated teleportation fidelity with $(0,0)$-photon subtraction when the beamsplitter is removed ($0.5670$). But we expect that with improved photon-subtraction filtering reducing some of the thermal noise coupled into the measurements that we could get closer to the theoretical teleportation fidelity using the model of our experimental parameters ($0.5952$) which would be more than break even improvement in teleportation fidelity.


\section{Full experimental density matrices}
\begin{figure*}
    \centerline{\includegraphics[width=1\textwidth]{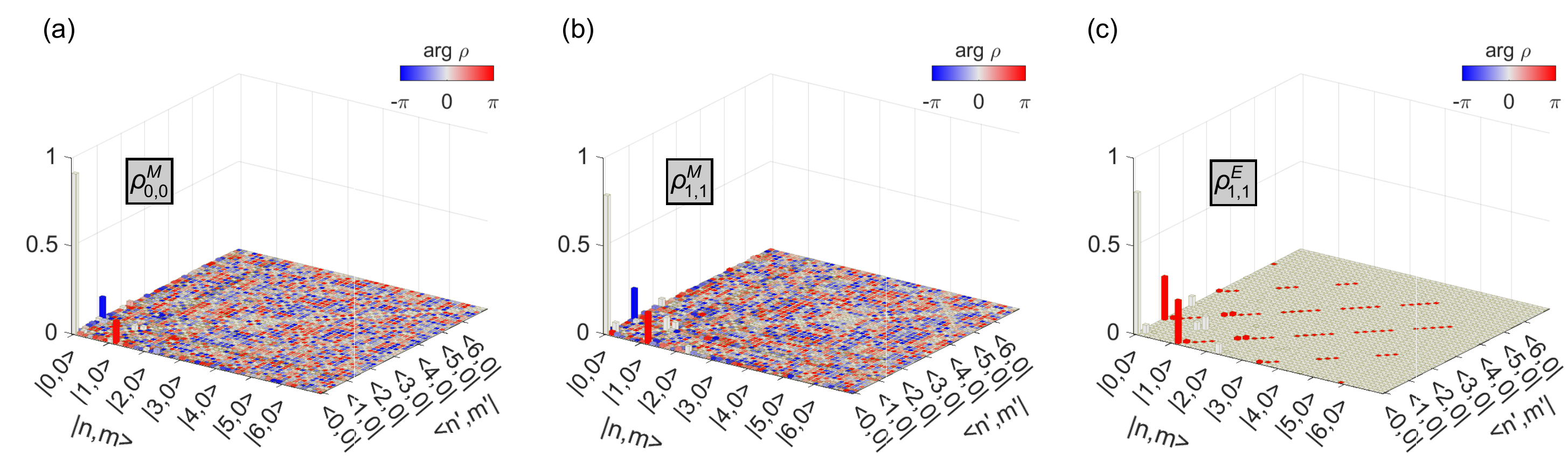}}
    \caption{Density matrices from tomographic reconstruction of measured data for (a) zero photons subtracted ($\rho^M_{0,0}$) (b) 1 photon subtracted per mode ($\rho^M_{1,1}$), as well as (c) expected theoretical state for 1 photon subtracted per mode ($\rho^E_{1,1}$). For these measurement reconstructions, the photon-number cut-off per mode is 6.}
    \label{fig:DMs6969full}
\end{figure*}

Fig.~\ref{fig:DMs6969full} shows the full (real and imaginary components) density matrices from Fig.~\ref{fig:DMs6969} of the main text.

The dominant matrix elements in \cref{fig:DMs6969full}, the $\mel{0,0}{\rho}{0,0}$ population and the $\mel{1,1}{\rho}{0,0} = \mel{0,0}{\rho}{1,1}^*$ coherence, can be expressed in closed analytical form using \cref{eq:rho_loss} and the set of parameters [$R_1 = R_2 = R_S$, $\lambda =  (\tanh r)(1 - R_S)$, $\kappa = \lambda^2 (1 - \eta_A) (1 - \eta_B)$] that are derived from analysis on measurements independent of tomography data. By evaluating the resulting infinite geometric series, we obtain:
\begin{subequations}
\begin{alignat}{2}
    \mel{0,0}{\rho_{0,0}}{0,0}
 &= \frac{1-\lambda^2}{1-\kappa},
 &\quad
    \mel{1,1}{\rho_{0,0}}{0,0}
 &= -\frac{\sqrt{\eta_A \eta_B} \lambda (1-\lambda^2)}{(1-\kappa)^2}. \\
    \mel{0,0}{\rho_{1,1}}{0,0}
 &= \frac{(1-\lambda^2)^3 (1+\kappa)}{(1+\lambda^2)(1-\kappa)^3},
 &\quad
    \mel{1,1}{\rho_{1,1}}{0,0}
 &= -\frac{2\sqrt{\eta_A \eta_B} \lambda (1-\lambda^2)^3 (1+2\kappa)}{(1+\lambda^2)(1-\kappa)^4}.
\end{alignat}
\end{subequations}

Substituting the experimentally determined values $R_1 = R_2 = R_S = 0.14$, $r = 0.3$, and $\eta_A = 0.55$, $\eta_B = 0.5$ into these analytical formulas yields
$\mel{0,0}{\rho_{0,0}}{0,0} \approx  0.9507$,
$\mel{1,1}{\rho_{0,0}}{0,0} \approx -0.1267$,
$\mel{0,0}{\rho_{1,1}}{0,0} \approx  0.8198$, and
$\mel{1,1}{\rho_{1,1}}{0,0} \approx -0.2215$.
These analytical expectations are in excellent agreement with both the experimentally reconstructed density matrices in \cref{fig:DMs6969full}(a,b) and the numerically calculated expected density matrix in panel (c).

\section{Detailed experimental description}
\subsection{Pump laser and squeezed light source}
\begin{figure*}
    \centerline{\includegraphics[width=1\textwidth]{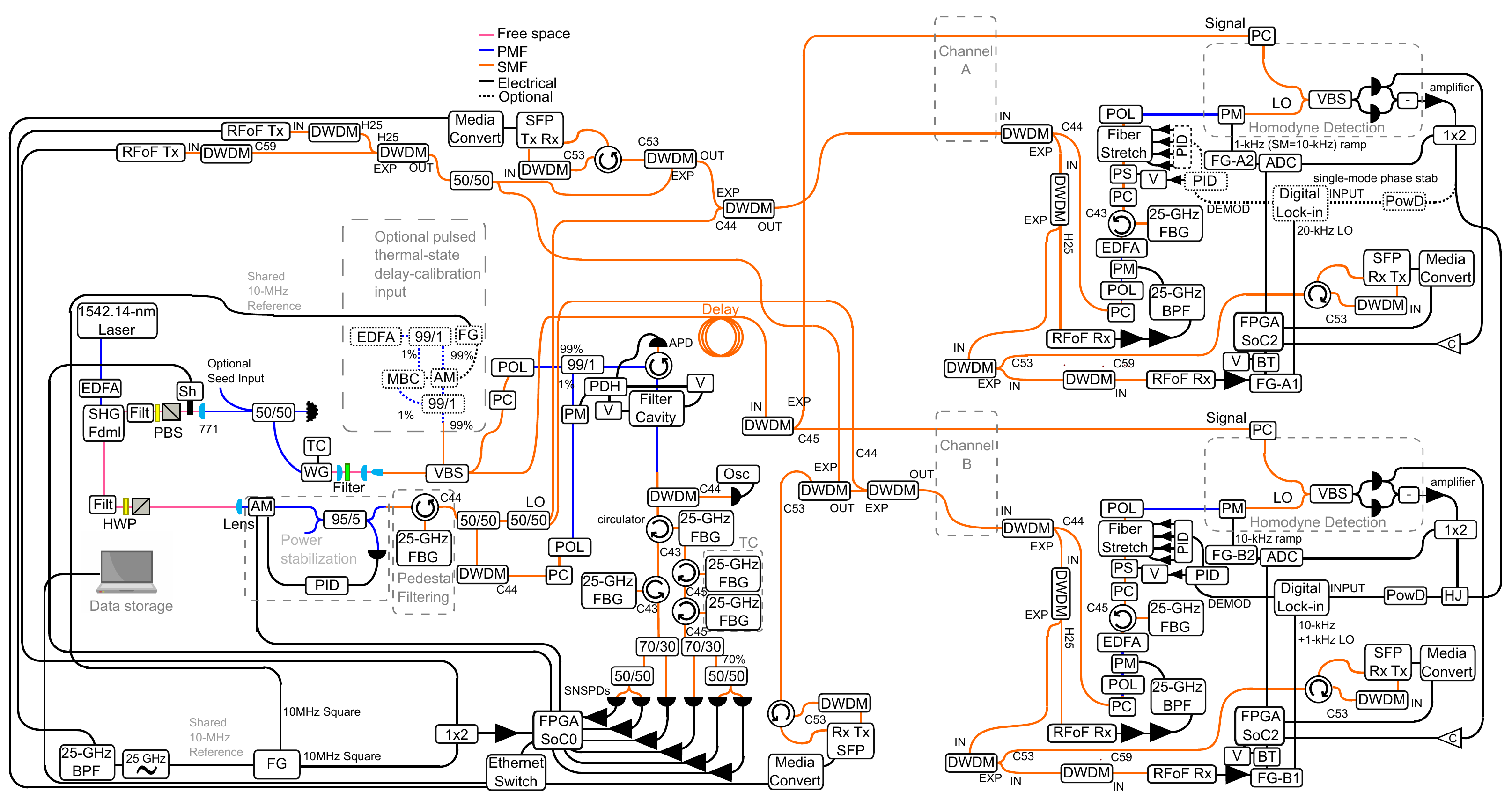}}
    \caption{Full experimental setup. Definitions (in alphabetical order). ADC: analog-to-digital converter. AM: amplitude modulator. APD: avalanche photodiode. Atten: attenuator. BPF: band-pass filter. BT: bias tee. C: comparator. CLK: clock. DEMUX: demultiplex. DWDM: dense-wavelength division multiplexer. EDFA: erbium-doped fiber amplifier. FBG: fiber Bragg grating. Fdml: fundamental. FG: function generator. Filt: filter. FPGA-SoC: field programmable gate array - system on chip. HJ: hybrid junction. HWP: half-wave plate. LO: local oscillator. MBC: modulator-bias controller. MUX: multiplex. Osc: oscilloscope. PC: polarization controller. PDH: Pound-Drever-Hall controller. PID: proportional-integral-derivative controller. PM: phase modulator. POL: polarizer. PowD: power detector. PS: phase shifter. RFoF: radio-frequency over fiber. Rx: receiver. SFP: small-form-factor pluggable transceiver. Sh: shutter. SHG: second-harmonic generation. TC: temperature controller. Tx: transmitter. V: voltage source. VBS: variable beamsplitter. WG: waveguide. 100-GHz channel center wavelength H25: 1556.96~nm.  C43: 1542.95~nm. C44: 1542.14~nm. C45: 1541.35~nm. C53: 1535.04~nm. C59: 1530.33~nm.}
    \label{fig:fullexpsetup}
\end{figure*}

 The full experimental setup is shown in Fig.~\ref{fig:fullexpsetup}. We use an amplified and doubled continuous-wave fiber laser (NKT Photonics Koheras Adjustik X15 and Harmonik 1W) emitting two beams; the fundamental wavelength at $\lambda_f=1542.14$~nm, aligned with International Telecommunications Union (ITU) grid 100-GHz Channel C44, and the second-harmonic at 771.07~nm  are filtered and independently fiber coupled~\cite{ChapmanDeployedTMSV2023}. The second-harmonic is used to pump a temperature-controlled (Thorlabs TEC200C) input-fiber-coupled type-0 periodically poled reverse-proton-exchange lithium niobate waveguide (WG) with free-space output (AdvR, Inc.). Compared to a ridge waveguide, the non-linearity is higher in reverse proton-exchange waveguides leading to the possibility of more efficient pumping with lower power. The WG is phase-matched at an elevated temperature (about 113$^{\circ}$C) to reduce the effects of photo-refractive damage, which is more likely to be an issue with reverse-proton-exchange waveguides. Initially, we were able to pump with less power than with a comparable ridge waveguide~\cite{ChapmanDeployedTMSV2023} for a similar $r$ value of the squeezed light. We could use about 75~mW to produce $r=0.5$ compared 180~mW with a ridge wavegiude. However,   we saw a reduction in SPDC output power over the course of using the device but the squeezing power was relatively stable during a given measurement. We found that leaving the device at an  elevated phase matching temperature for several days after degradation was observed (on the order of hours of use) regained some of the lost squeezed-light optical power but not completely. This degradation over time lead us to increase the pump power to about 240~mW to have approximately  $r=0.3$ for our measurements.
 
 The WG spontaneous parametric downconversion output is broadband squeezed vacuum ($>$7~THz), which is subsequently coupled into single-mode fiber using relay optics with about 80\% coupling efficiency. The squeezed vacuum is then directed to a variable fiber beamsplitter (Newport F-CPL-1550-N-FA) where some of the light is reflected into another mode for filtering and single-photon detection. We use R=14\% to balance the trade-off between heralding rate and state quality. The transmitted modes are sent to a dense-wavelength division multiplexer (AC Photonics) which is configured to demultiplex a 100-GHz channel on the ITU grid, namely C45 (centered on 1541.35~nm) from the rest of the spectrum. Each mode (C45 and the remaining spectrum) is then directed towards a separate homodyne detector. 

 In total, the loss that each path experiences (excluding the photon-subtraction reflectivity) is estimated to be 0.3~dB (waveguide propagation loss), 0.2~dB (free-space lenses and filter), 1~dB (fiber coupling), 0.1~dB (photon-subtraction beamsplitter loss, 0~dB (spliced short delay fiber), 0.3~dB (demultiplex DWDM), 0-dB (spliced short fiber to homodyne detectors), 0.5~dB (homodyne VBS and photodiodes), 0.25~dB (electronics noise of homodyne detector and HDS ADCs) which comes out to about 0.55 for mode-A with some minor variation between the two sides leading mode-B to be about 0.5 based on measured differences.

\subsection{LO generation and homodyne detector}
\label{SM:LOandHD}
The homodyne detection and local oscillator (LO) generation largely follow the methods described in Ref.~\cite{ChapmanDeployedTMSV2023} with several improvements. The variable beamsplitters both use single-mode fiber and are passively temperature stabilized~\cite{longtermsq24} in foam leading to highly stable coupling ratio on the order of a day. Moreover, we use high-quantum efficiency diodes (Laser Components IGHQEX0500-1550) that are fiber pigtailed (AC Photonics) and spliced to the variable beamsplitter (Newport F-CPL-1550-N-FA) for a total loss from homodyne beamsplitter input to detection of about 10\%. To generate the LO, we use the optical phase reference at $\lambda_f$ and use a filtered sideband as the C45 or C43 LO. In Ref.~\cite{ChapmanDeployedTMSV2023}, we distributed a 10-MHz RF reference to lock a RF oscillator at 25~GHz to produce 25-GHz sidebands on the optical phase reference. To overcome previous issues in Ref.~\cite{ChapmanDeployedTMSV2023}, we instead use a single 25-GHz oscillator (SignalCore SC5521A) and distribute it to each homodyne detector using RF-over-fiber transceivers (RFoptic RFoF30TFL-N0-11/RFoF30RFL-N0-11), which we call the sideband reference. The received signal goes through a two-stage amplification (Fairview Microwave FMAM3260 and SPA-265-33-01-K), totaling over 70 dB of gain, and filtering (Mini-Circuits ZVBP-25875-K+) before going to the phase modulator (EOSpace) for sideband generation.

To ensure the samples from the homodyne detector are real-time quadrature samples according to methods developed in Refs.~\cite{PhysRevLett.116.233602,RTPSquadOE2017}, the detector bandwidth is chosen to match the temporal mode of the photon-subtracted light; in our case, for broadband squeezed light filtered by a single filtering cavity, this corresponds to needing a low-pass filter equal to the cavity bandwidth~\cite{PhysRevLett.116.233602,PhysRevA.52.3126}. Our filtering cavity has a HWHM of 9~MHz. By modifying the homodyne detectors described in Ref.~\cite{ChapmanDeployedTMSV2023} so the feedback resistor is 200~k$\Omega$, given the increased capacitance of the photodiodes we used, the detector bandwidth becomes 8.8~MHz.

\subsection{Homodyne detection server}
The homodyne detector's output is directed to the newly developed homodyne detection server (HDS). The HDS's contains two low-noise 14-bit analog-to-digital converters (ADC, Analog Devices AD9254) that samples, at 100~MHz, the homodyne detector output and a complimentary voltage reference signal for the phase drive. The ADCs are co-located (Terasic ADCSoC) with a field-programmable gate array (FPGA) system on chip (FPGA-SoC, Intel Cyclone V) that has a dual-core hard processor system (HPS) integrated with the FPGA having various interconnect bridges for coordination and communication between the FPGA and HPS including shared access to 1-GB of random-access memory (SDRAM). As the FPGA receives samples from the ADC, they are collected into large groups of 8 and buffered. Then the two signed 14-bit samples (one from each ADC) are padded to make a single 32-bit word, they are time-stamped and  stored into the shared SDRAM. The FPGA is given priority access to write to the RAM but first-in-first-out (FIFO) buffers are used to ensure no data is lost if the SDRAM is busy.

The FPGA needs to use the physical address for storage in SDRAM whereas linux applications in the HPS normally operate using virtual addresses. To circumvent this disconnect, when the homodyne server linux application starts, it allocates enough RAM for 67500 continuous pages (4096 bytes or 1024 32-bit words). After allocation, the page frame numbers, corresponding to the physical address (after bit-shifting by 12 bits) of the first location of each 4096-byte page of memory, are gathered for the allocated memory. The buffer is initialized with  test values and each page is checked for complete allocation. If any page is found to be incompletely allocated, the allocation is attempted again. This allocation is usually successful within 1-3 tries if the buffer size is not too big or too small depending on the total SDRAM size (1~GB in our case). The 67500 page frame numbers are stored in the FPGA's on-chip RAM, filling this small RAM. This enables the FPGA to address a buffer of 69120000 32-bit words (276.48 MB) within the non-contiguous memory allocated by the homodyne server application that will process the buffered data to service client requests. Due to the known order of the allocated page frame numbers by both the FPGA and the server application, knowing the timetag for a certain sample, is sufficient to calculate its physical address in memory so only the 32-bit words, corresponding to an ADC sample from each ADC, are stored consecutively in SDRAM while the timtags are not. It has been rigorously tested that the FPGA and the linux application can correctly place and find a given sample by using a test mode where the FPGA writes a 32-bit timetag into the corresponding memory location.

The 100-MHz ADC sample rate means that every 0.6912~s the buffer will fill up, at which point it starts writing at timetag 0 again. The FPGA uses a 32-bit register to count the number of times the buffer fills up (memory-overflow number). Through a shared 32-bit memory-mapped interface between the FPGA and the HPS, the server application can query the current memory-overflow number (besides other status values and it can write values to configure several FPGA parameters). This can be used to check for data validity since requests to the server are required to provide the memory-overflow number corresponding to the timetag(s) requested.

The homodyne server application is based on a socket server with an application programming interface (API) for configuration and basic query (e.g., current memory-overflow number and timetag) similar to the standard commands for programmable instruments (SCPI). Although not currently fully SCPI-compliant, it easily could be with more development. For data queries, a binary array of 32-bit words is expected. The received array starts with a 32-bit keyword and then the associated memory-overflow number followed by the timetags corresponding to the requested samples. Once the keyword and memory-overflow number are received, subsequent binary queries are assumed to correspond to that same memory-overflow number until the keyword and memory-overflow number are appended the beginning of a query indicating that a new request is starting. This enables flexibility on the server side to handle queries where the client's lower-level drivers may break the request up into multiple messages to the server if the request contains a large enough number of timetags, e.g., more than about 16000 in our case. The server application makes several checks to the FPGA to ensure data reliability (no FIFO overflows, no ADC samples out of range, phase-locked loops are locked) and checks the client-provided memory-overflow number against the local memory-overflow number and the local current time-stamp. It is assumed that buffer queries will be broken into two parts: a query for data in the first half of the buffer or the second half of the buffer. After the FPGA has filled up the first half of the buffer, the HDS can safely service requests for that half without concern of data overwrite since the FPGA is filling up the other half of the buffer at that time, and vice versa. By breaking the buffer into parts, the buffer can be read out safely while also using it as rolling buffer that is continuously refilled.

When a timetag is processed by the HDS, there are multiple ways for it to be processed. The basic method is for a given timetag, the server will return the corresponding buffered word containing the sample from each ADC. If the integration-window parameter is greater than 1, the samples within the integration window are integrated together, separately for each ADC, to provide an integrated sample 32-bit word. In the case of real-time quadrature detection of photon-subtracted states, the integration-window was set to 1 since each sample corresponded to an integrated quadrature sample per Sec.~\ref{SM:LOandHD}. But for general detection of squeezed light or other states, the integration window can be increased to measure continuous-wave quadrature samples at frequencies lower than 100~MHz.

To facilitate homodyne tomography, there is also mode where the slope of the phase-drive ADC is checked before saving the sample for a given timetag. Assuming a sawtooth ramp waveform, the slope is checked to make sure the timetag corresponds to a sample within the long ramp and not the short reset. At 100-MHz sample rate and about 10-kHz ramp frequency, we found this results in the rejection of about 0.1\% of the random samples queried. When a sample is rejected, a known placeholder value is inserted in place of the homodyne sample so the client knows what happened. This placeholder value is chosen so the 14-bit two's compliment ADC output (padded to 16 bits) cannot produce it.

Moreover, the server also has a threshold-counting mode for calibration purposes (discussed more fully in Sec.~\ref{SM:delcalib}) that the client can enable via the configuration API. In the threshold-counting mode, a bi-polar threshold value and signal slope, which the client can set via the configuration API, are used as the server goes through every timetag in the buffer to collect all the timetags corresponding to a threshold crossing and provide those to the client. This is useful for cross-correlation analysis but is limited to data from a single memory-overflow number since it takes the server much longer than a single memory-overflow number to scan the entire buffer. In practice, we have found that around 10000-20000 threshold crossings provides sufficient signal-to-noise ratio (SNR $>10$) for a cross-correlation measurement.

\subsection{Phase stabilization and tomography phase drive}
\label{SM:Phstbdr}
\begin{figure*}
    \centerline{\includegraphics[width=1\textwidth]{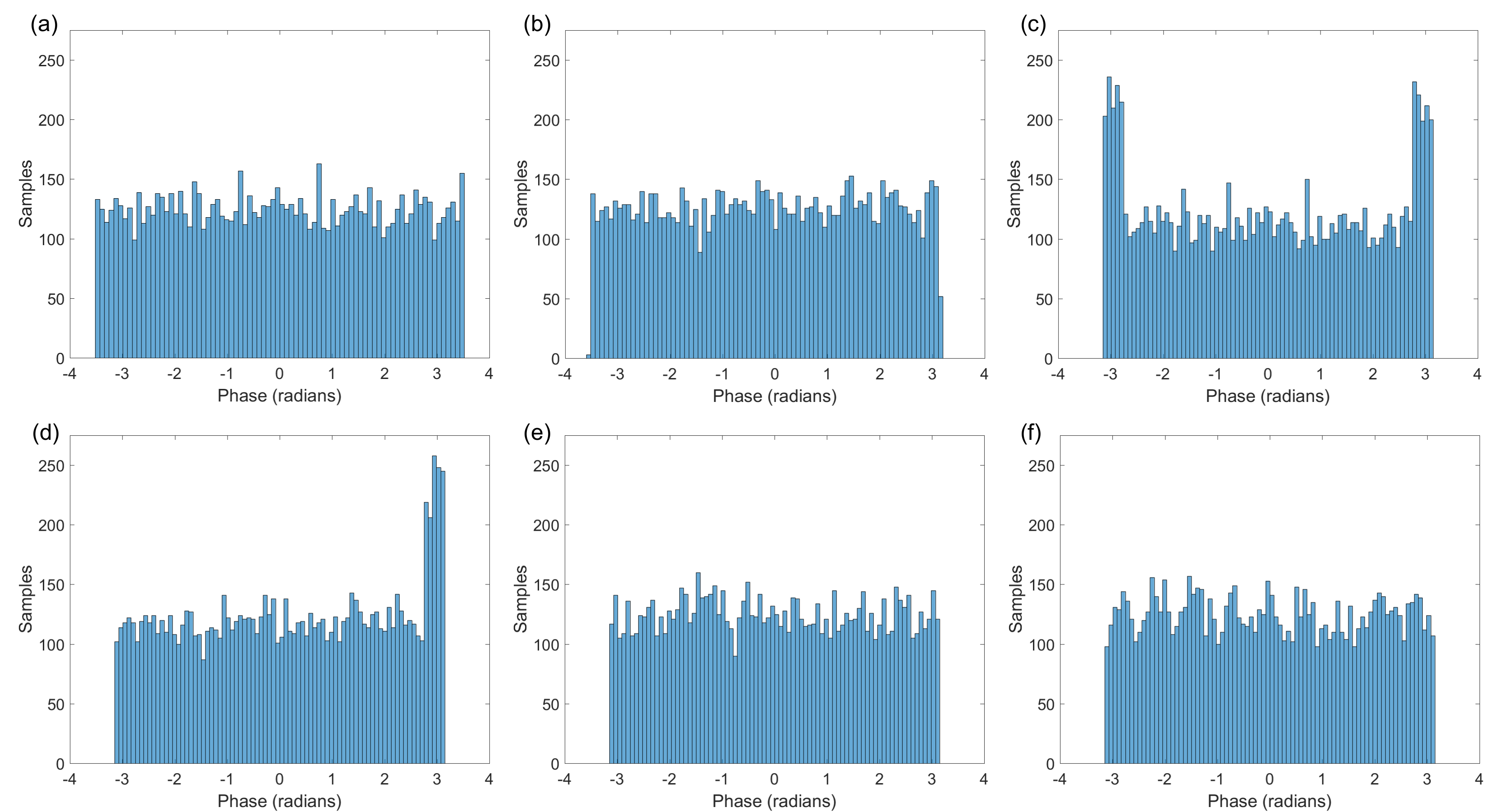}}
    \caption{Histogram of LO phase settings measured during representative tomography for state displayed in Fig.~\ref{fig:DMs6969}(b). LO phase settings on (a) side A/C43, (b) side B/C45,(a) side A/C43 with $2\pi$ modulo centered at 0, (b) side B/C45 with $2\pi$ modulo centered at 0, (c) Joint added phase with $2\pi$ modulo centered at 0. (c) Join subtracted phase with $2\pi$ modulo centered at 0. The bins are 0.1-radians wide.}
    \label{fig:jointphasehist}
\end{figure*}

To implement two-mode homodyne tomography, we use a function generator (Rigol DG2102) to drive a phase modulator on each LO to ramp the phase by approximately $2\pi$ using a sawtooth wavefrom. The reference clock for each function generator is synchronized to a common reference clock as described in SM Sec.~\ref{SM:ctrlandsync}. On side A(B), the ramp frequency is 1(10)~kHz. This is implemented at different rates so the joint phase is also varied. In Fig.~\ref{fig:jointphasehist}, we show a representative histogram of the marginal and joint phases measured during a two-mode tomography. 

To enable high-quality homodyne tomography, the LO phases need to be stable besides the phase drive but in reality the LO phase varies due to environmental changes (mainly temperature variation). To compensate for that variation, each homodyne detector output signal is split using a RF power divider (Mini-circuits Z99SC-62-S+) so half goes to the HDS and half is sampled for phase stabilization. The halves for phase stabilization are combined electrically on a hybrid junction (Pulsar Microwave JF-01-412) which enables direct measurement of two-mode squeezing. This signal is then directed towards a power detector (Linear Technology LT5537)  which gives a voltage output corresponding to the power of the input. Using lock-in detection (Liquid Instruments Moku:Go), the combined phase drive is demodulated from the squeezing signal so the residual phase drift can be extracted. The lock-in's local oscillator is a combined phase drive signal created by generating a 10-kHz sinewave amplitude modulated by a 1-kHz sinewave using a function generator (Rigol DG2102) phase locked to the phase drive signal with a shared 10-MHz reference. This residual phase drift error signal is used to control the phase of the mode-B optical homodyne local oscillator to match the drift of the mode-A homodyne local oscillator---effectively making the drift common mode. This is sufficient for two-mode tomography of our expected states since we expect the investigated marginal states to be phase insensitive. For this phase stabilization, two-stage proportional-integral-derivative control system (Liquid Instruments Moku:Go) is employed using a fast piezo-based $>55\pi$ phase shifter (Luna Innovations/General Photonics FPS-003-35-SS-FC/APC) as well as a slower fiber stretcher (Luna Innovations/General Photonics FST-001-B-04-15-SM-FC/APC) with larger range (about 3~mm or $>800\pi$) to enable fast long-term phase stabilization with 1-kHz bandwidth. The fast phase shifter and slow fiber stretcher take output signals from the controller which are amplified by piezo drivers (Thorlabs KPZ101 and MDT694B, respectively). The control loop on the fast phase shifter attempts to zero the demodulated error signal from the lock-in with a control bandwidth of about 1~kHz, while the control loop on the fiber stretcher has a much lower bandwidth (300~mHz) and that control loop adjusts the fiber stretcher to maintain the average output to the fast phase shifter near the middle of its range. To keep the fiber stretcher in range for a long-term lock, it helped that the laboratory the experiment took place in has temperature stability of about 1$^{\circ}$C and there were also vinyl curtains hung around the optical table to enclose the table as one big ``box.'' Both of these contributed significantly to maintaining a long-term phase stability such that the fiber stretcher range was sufficient.

We remark that although this method requires the homodyne detection of each mode to be co-located, even still, we employ a method which enables phase stabilization for two-mode tomography that adds no dark counts to the photon subtraction and requires no additional timing coordination or down-time, as compared to the conventional method of using an optical chopper and bright seed.

To enable truly independent phase stabilization of each LO, we also developed and tested a method using a dim seeded sideband at 20-MHz. This method and our testing are described in SM Sec.~\ref{SM:NGTelExp}.

That independent method was not used due to some unresolved technical issues with seed power instability specific to our system causing inter-signal interference leading to seed instability and coherent state contamination of the squeezing. This seemed to be mainly from reflections off the variable beamsplitter(s). We also noted seed power instability in the reverse-proton-exchange waveguide,  compared to the ridge waveguide we have (used in Ref.~\cite{ChapmanDeployedTMSV2023}), likely due to photo-refractive damage.

\subsection{Photon subtraction optics and detection}
On the reflected mode of the photon-subtraction variable beamsplitter, filtering is required for the photon-subtraction events to correspond to mode detected by the homodyne detectors. This filtering is mainly done with a  9-MHz half-width-at-half-maximum (HWHM) 25-GHz free-spectral range (FSR) fiber-coupled optical cavity (Stable Laser Systems). The cavity piezos are used to lock a cavity resonance to $\lambda_f$ using a Pound-Drever-Hall (PDH) control system (Liquid Instruments Moku:Go) on 20-MHz sidebands produced by a phase modulator (EOSpace). A fast/short piezo and slow/long piezo are provided in the cavity enabling 2-stage long-term locking ability at a certain FSR (which is adjustable by about 1~GHz). The PDH-controller fast output is amplified (Thorlabs BPA100) then provided to the fast/short piezo. The PDH-controller slow output is amplified (Thorlabs MDT694B) with a 110~V offset internally provided by the piezo amplifier to approximately adjust the free-spectral range to be close to 25~GHz.

To ensure spectral alignment of two-mode non-degenerate photon subtraction, we obtained a cavity with a tunable free-spectral range. This tunable free-spectral range contributed to significant mechanical vibrational sensitivity in the cavity. To provide passive isolation from acoustic vibrations, we surrounded the cavity in a multi-layer acoustically damping enclosure we created from mass-loaded vinyl and bonded acoustical cotton (Acoustical Surfaces). To isolate the cavity from contact vibrations through the table from the floor, we placed the acoustic enclosure, with the cavity inside, on an optical breadboard (Thorlabs B2424F) which was actively isolated (Thorlabs PWA090) from the optical table (Newport RS4000) beneath it. The optical table was also actively isolated (Newport S-2000) from the floor. Moreover, we noticed a significant reduction of the environmental vibrations when we moved this experiment from an older crowded noisy lab on the second floor of a general research building to our new ground-floor lab space purpose-built for quantum-information-science research. The cavity on the actively stabilized breadboard had the added benefit that minor work could be done on the optical table while maintaining the cavity lock.

In the PDH controller, for the fast loop, we found it optimal to just use the integrator to enable high-gain control at lower frequencies while not exciting mechanical resonances or responding to noise near the limit of our control loop bandwidth which helped to improve the stabilization. After the passive and active stabilization, the cavity stabilization is limited to about 0.4~MHz for the 9-MHz HWHM cavity due to the limited bandwidth of the faster (short) piezo for controlling the cavity phase (about 10~kHz). But this is not a fundamental limitation of our overall demonstration. A different configuration using a tunable pump laser stabilized to the cavity or using a non-FSR-tunable cavity can be used in future work, depending on RF adjustment of the LO alone for spectral-mode alignment.

The targeted two-mode squeezing is at +/- 100 GHz from the $\lambda_f$ (which align with ITU grid channels C43 = mode A and C45 = mode B). The slow piezo voltage is chosen to get close to a 25~GHz FSR. To more precisely align the cavity resonances with the local oscillators, we temporarily redirect the LOs to go through the cavity and fine-tune the 25-GHz RF oscillator used in LO generation to maximize the LO transmission through the cavity. This needed to be re-calibrated every time the cavity was unlocked for more than a few minutes otherwise the cavity FSR may differ from RF frequency by about 10-100~MHz due to cavity drift.

We found the maximum cavity transmission for each LO wavelength has a different RF frequency of about 2~MHz (which translates to an 8~MHz frequency difference due to the LO generation process) likely due to dispersion in the LO generation system. We split the difference so the maximum transmission of each mode was only reduced by a few percent but this will somewhat affect heralding and coherence leading to reduced entanglement. We would expect this mismatch to cause the state to become mixed with incoherent thermal noise. The log negativity of $E_\mathcal{N}(\rho^E_{0,0})=0.34$ and $E_\mathcal{N}(\rho^E_{1,1})=0.55$ which we compare to $E_\mathcal{N}(\rho^M_{0,0})=0.49\pm 0.03$ and $E_\mathcal{N}(\rho^M_{1,1})=0.52\pm 0.03$. We would expect $E_\mathcal{N}(\rho^E_{0,0})$ and $E_\mathcal{N}(\rho^E_{1,1})$ to be less than $E_\mathcal{N}(\rho^M_{0,0})$ and $E_\mathcal{N}(\rho^M_{1,1})$ due to the effects of sample size, as discussed in the main text, since  $E_\mathcal{N}(\rho^E_{i,i})$ is effectively for an infinite sample size being a theoretically derived density matrix but note that $E_\mathcal{N}(\rho^M_{1,1})<E_\mathcal{N}(\rho^E_{1,1})$. From the insets in Fig.~\ref{fig:DMs6969}, we see only a modest improvement in the squeezing (about $1.25-0.95 = 0.3$~dB) but a much larger increase in the anti-squeezing (about $2.92-1.76=1.16$~dB) compared to  $\rho^M_{0,0}$. These observed features of small increased entanglement and squeezing with larger increases in anti-squeezing are consistent with this description of this filtering causing the heralded state to be mixed with incoherent thermal noise. This frequency mismatch could be reduced by using a larger bandwidth cavity or fixed by having different filtering cavities for each mode and optimizing the transmission independently, by using a cavity-based squeezed-light source, or by matching the dispersion of the cavity and LO generation (using electro-optically generated sidebands). 

 After the cavity, a 100-GHz dense-wavelength-division-multiplexing filter is used to remove most of the $\lambda_f$ stabilization light transmitted through the cavity. This light is detected by a photodiode (Thorlabs DET08CFC) and monitored by an oscilloscope channel in the cavity-lock controller.  The rest of the light is transmitted through a 25-GHz fiber Bragg gratings (O/E Land) with circulators (O/E Land and AC Photonics), which are used to select the two desired resonances from the cavity transmission spectrum. The C45 FBGs needed to be heated to shift the transmission peak about 10 GHz due to manufacturing variability. The FBGs are heated using a thermo-electric cooler (heater) and temperature controller (Thorlabs TEC200C). Each resonance (one at C43 and the other at C45) is then sent to a 3-output beamsplitter tree (AC Photonics) connected to superconducting-nanowire single-photon detectors (Quantum Opus) to enable limited photon-number resolution with low-jitter and fast reset time. The single-photon detectors are measured to have about 85-90\% efficiency with measured dark counts in our system of about 50-100 counts per second. The delays for each optical and electronic detector path were adjusted using coaxial cables so that they were matched up arriving at the PSO FPGA. The signals are amplified (Mini-Circuits ZX60-100VH+) then attenuated (Mini-circuits VAT) so the pulse amplitude are compatible with the FPGA general-purpose input/output (GPIO) pins. With this filtering system, we measured about 1000 (2500) background counts per second for C43 (C45) modes due to the room background and (mostly) C44 cavity stabilization laser (about 5~$\mu$W transmitted through cavity) corresponding to a DWDM and FBG filtering isolation of about 104~dB. This high-rejection filtering combined with using high-sensitivity 20-MHz sideband detection (via amplified avalanche photodiodes~\cite{ChapmanDeployedTMSV2023}) enables continuous cavity locking with negligible added photon-subtraction noise since accidental coincidence probability from order 1000 noise counts per second for a 10-ns coincidence window is about 0.02 cps, small compared to our photon-subtracted coincidence rate of about 20 cps.

\subsection{Photon-subtraction system orchestrator}
The photon-subtraction detector outputs are collected by a newly developed photon-subtraction system orchestrator (PSO). The orchestrator is implemented on a board (Terasic DE10-Nano) containing a FPGA-SoC (Intel Cyclone V) very similar to the FPGA-SoC used in the HDS. Using the shared 10-MHz reference, the FPGA establishes 100-MHz and 300-MHz clock domains. To register a single-photon detection event, the relevant input pins are monitored for rising edges by flip-flops in the 300-MHz clock domain. The outputs of rising-edge detection registers for all detectors are then pipelined through a processing stage. First there is an addition of all events on each side (A and B) assuming there are two modes with a detector tree on each. Currently, the pipelined addition is setup for 3 detectors on each side but more detectors could easily be added by using more GPIO pins and increasing the pipelined addition stage at the expense of increased latency. The total counts for each side are put into a pipelined series of 9 registers to encompass three 10-MHz time bins using nine 300-MHz sub bins. Then these nine sub bins are pipelined through nine delay stages while a pipelined sub-bin weighted average is calculated using the total counts (adding both sides together). To avoid needing to divide in the FPGA, the weighted average numerator is calculated simultaneously with the denominator (sum) and create multiple copies of the sum after multiplying it by various bin numbers. Once the numerator and mulitplied denominator are calculated, they are compared to see where the numerator is greater than the multiplied denominator for some bins but less than others. If the weighted average bin ends up being 4, 5, or 6 (in the middle of the 9 sub bins) an event-recorded flag is raised. If not, then the system passes over these events (or there are no events) until they shift into the 4, 5, or 6 sub bins. When the event recorded flag is raised, two things happen simultaneously. (1) The data is loaded into a three-bin buffer to provide buffering for clock domain crossing into the 100-MHz clock domain. (2) A pulse trigger output is issued to be high for two 300-MHz bins to provide low-jitter real-time feedforward output capability.

To characterize the heralding trigger, we split the amplified SNSPD signal right before the orchestrator FPGA input. From there we used an oscilloscope (Agilent MSO-X 4104A) to measure the delay of the signal entering the FPGA and the heralding signal coming out of the FPGA. The measured timing latency $\pm$ jitter (standard deviation of 2000 samples) for the generation of this heralding signal is $91\pm1$~ns using 300~MHz clock-rate in the FPGA. The latency of the PSO producing the heralding signal could be further reduced if less detectors are in the beamsplitter tree, or higher-performance circuitry with higher clock-rate is used so this is not a hard lower limit.

In the 100-MHz clock domain, there are several stages of synchronizer flip-flops, then the data is processed to provide sub-bins with total sum for each side in one 64-bit word. Due to the assumption of 6 detectors, sub bins are 3-bits each and the sum is 5 bits for a total of 2x(9x3-bits+5bits)= 64 bits. At this stage, the detection event is timetagged on the 100-MHz time-stamp clock that is synchronized with the HDS. At this point, the events are checked whether they should be saved or ignored. Based on a flag in the shared memory-mapped interface, the orchestrator application can tell the FPGA to hold any event or just a coincident event (of any photon number) in the same 100-MHz time bin. If a singles (one-sided event) or coincidences (two-sided event) are recorded (based on the flag), data is held in a buffer if it is not within the (phase-stabilization) seed rejection filter window (adjusted for seed delay), which is discussed more fully in SM Sec.~\ref{SM:NGTelExp}. The data is held for the duration of the hold time to see if another event comes in too quickly. If that happened, it would distort the event type and make it a partial double photon subtraction at large delay which is technically a different state~\cite{PhysRevA.77.062315}. In practice, we find our photon-subtracted rates were low enough that the filter is not needed;  so we set the hold time to about 3 100-MHz bins, which is less than our detector dead time (4 bins). We also found for the thermal-state calibration that it was important to reduce this hold time to about 3 bins (the detector dead time) because the filter does indeed work and was filtering out all events from the thermal pulses thus blocking the cross-correlation analysis.

If no other events are detected in this hold time then the data is flagged to be saved. At this point, the detection signature (counts per sub bin per side and total counts per side) is then prepared with the timetag, memory-overflow number and loaded into a FIFO buffer with the current saved-data counter value which corresponds to its address in the SDRAM buffer. The SDRAM buffer is set up the same way as in the HDS. As there is data in the FIFO buffer and the SDRAM is available to be written to, the data in the FIFO buffer is transferred to the SDRAM buffer using the saved-data counter to calculate the SDRAM address instead of the timetag as the HDS does.

During the event checking stage, there is also the ability to save some zero-photon subtracted data, periodically taking advantage of the timetags in between photon subtraction events as if it were any other detection event. The reason for this data is to make a reference measurement on the two-mode squeezed vacuum using the same detection system at the same time as the photon-subtraction data is being gathered. By gathering the data in between photon subtraction events, we enable concurrent data collection but this does not completely trace over the photon-subtraction mode. To completely trace over the photon-subtracted mode, at least sometimes photon subtraction events should be included according to the transmission of the photon subtraction path relative to the total photon subtraction reflectivity but this transmission is such a small probability (about 0.1\%) that it is negligible. For improved photon-subtraction efficiency, a random sampling of the homodyne samples should be measured to completely trace over the photon-subtracted mode. Within the FPGA, there is some zero-detection rate logic to provide a rate at which zero-detection rate data is attempted to be saved as long as there is no other detection event during the hold time. The attempted rate is set by the orchestrator application and provided to the FPGA using the shared memory-mapped interface.

The PSO linux application begins similar to the HDS by allocating memory and share page addresses with the FPGA. Then socket connections are opened with each HDS. These connections are facilitated by the PSO's ethernet connection to an ethernet switch where each HDS is connected to the same switch via media converters using 10G fiber transceivers that are wavelength multiplexed with other control signals. After the socket connections are opened, the PSO application can be controlled by a SCPI-like command-line interface. Using the commands available, various parameters on the PSO and each HDS can be configured, tests can be run, and various data acquisition routines can be initiated. By using process threading and pointers to shared memory, the command-line interface is available during data acquisition enabling live reconfiguration of the system during data acquisition.

The main data acquisition routine queries the user for file info (name and data description) and some relevant data acquisition parameters (zero-detection rate, delays, etc.). Then the application ensures the servers and the master are synced up by directing the FPGA to send out the homodyne-trigger pulse and checking the synchronization of the memory-overflow numbers of the servers. At this point, two threads are started on the dual-core processor. A data-collection thread and a data-save thread.

In the data-collection thread, the application directs the FPGA to signal the shutter (Thorlabs SHB1) to close for a current shot-noise calibration to be measured. Afterwards, the application directs the FPGA to signal the shutter to open and the user is given an opportunity to re-enable phase stabilization before measurements proceed. The data-collection thread contains a while loop where periodically (10~ms), the PSO's FPGA time-stamp counter is queried by the application to see when it crosses over half way at which point the local buffer of detection events is processed until the timetag for half way is reached (in our case, 69120000/2=34560000). FPGA-collected events are triaged by the application into different parts of various memory buffers based on their detection signature. All the time tags with appropriate delays are written to a buffer which is then transmitted to each HDS for processing. The received ADC samples (homodyne-detector voltage sample and phase-drive voltage) are then processed and saved with their corresponding photon-detection signatures assuming the order quadrature samples are received is the order the timetags were sent which has been verified through testing. Sample counters for each type of detection signature are incremented for each event added. When the counter for a certain dataset reaches a certain threshold, e.g., 10k or 100k, then the dataset counter for that detection signature is incremented. The data buffer is large enough to store 4 full datasets for each detector signature simultaneously.

In another thread that is running on the other processor of the dual-core processor, a data-save function is running which loops through checking the dataset counters for the different detection signatures. When this function registers that a dataset counter has been incremented, after making sure that data has not been overwritten by data collection faster than data has been saved, the function will save the new full dataset(s) to a binary file(s) in non-volatile memory. For each entry in a data file, the detection signature (counts per sub bin per side and total counts per side) , memory-overflow number, time-stamp counter, and all four signed ADC values (two from each server) are saved.

After the data collection thread is done processing the detected events from the beginning of the time-stamp counter to halfway, it periodically checks the time-stamp counter to see when it overflows back to 0. At which point, the data collection described above is done again except the stopping condition is when the memory-overflow number is different. It is unknown how many samples will be collected in total (and the FPGA is not set up to provide that to the application though it could be) so that cannot be used but we do want to only look at samples collected by the previous (since the recent most overflow) memory-overflow number.

For the shot-noise data collection, the process is very similar except instead of looking in the SDRAM for new events to gather timetags to query the servers, a pseudo-random function is used to generate timetags to query the server.

We profiled the speed of PSO/HDS combined system by increasing the zero-detection rate (ZDR) by factors of 2 up to $2^{17}=131072$ which was the highest rate (per memory-overflow time, multiply by 1/0.6912 to convert to per second) that worked reliably. From a linear fit (with $R^2=0.9998$) of the execution time for the first half of the buffer vs ZDR/2, we extrapolate that the system could reliably continuously process events up to about 250k detection events per second.

This system is designed to run continuously for a very long time, limited by the 29-bit memory-overflow number (about 12 years) and somewhat by the finite non-volatile storage of the micro SD card though using an secure-shell connection, files can be offloaded to larger storage devices and deleted from the SD card to recover space over time. To date, the longest data acquisition completed was on the order of seven hours. The data-set and sample counters for each detection event type are listed in Table~\ref{tab:PSOdatastats}. Besides showing the system's ability to continuously run for hours, this also shows the ability to capture some higher-order photon-number events using multiple detectors of the tree.

\begin{table*}[]
    \centering  
       \caption{Photon-subtraction orchestrator dataset and sample counters. The location corresponds to the detected counts per mode [mode A, mode B]. Then is the final dataset counter followed by the sample counter at the time data acquisition was halted, separated by a ``$|$.'' This data was acquired using zero-detection rate = $2^7$ and with only coincidences being saved. A new dataset was saved every 5000 samples.}
    \label{tab:PSOdatastats}
    \begin{tabular}{|c|c|c|c|}
    \hline
       [0,0] = 918 $|$ 11676  &  [0,1] = 0$|$0  &  [0,2] = 0$|$0  &  [0,3] = 0$|$0\\
       \hline
       [1,0] = 0 $|$ 0  &  [1,1] = 541$|$ 9047 &  [1,2] = 0$|$3112  &  [1,3] = 0$|$2\\
       \hline
       [2,0] = 0 $|$ 0  &  [2,1] = 0$|$ 2240 &  [2,2] = 0$|$6  &  [2,3] = 0$|$0\\
       \hline
       [3,0] = 0 $|$ 0  &  [3,1] = 0$|$ 0 &  [3,2] = 0$|$0  &  [3,3] = 0$|$0\\
       \hline
    \end{tabular}
\end{table*}

\subsection{Analog control and synchronization signals}
\label{SM:ctrlandsync}
To ensure the tomography phase-drive, sampling and time-stamp clocks are synchronized between the orchestrator and each HDS, we use 2.5-GHz radio-frequency-over-fiber (RFoF) transceivers (RFoptic) for clock distribution. With this method, the clocks of the HDS and PSO are synchronized with a jitter of about 500~ps. Moreover, to synchronously start each time-stamp clock, the orchestrator issues a homodyne-trigger pulse (about 10~ns wide) which modulates the amplitude of the phase reference beam (at $\lambda_f$) so that it is distributed to each homodyne detector when the phase reference is converted into the LO. This pulse is recovered by a current monitor on a single photodiode of the homodyne detector ( the monitor is reduced from stock gain by about 10x) which is fed to a comparator (Analog Devices ADCMP600) to create a fast rise-time digital pulse to issue a start signal to the corresponding HDS. Due to the slow bandwidth of the current monitor (about 1~MHz), the delay between the time-stamp counters is dependent on the comparator threshold and the DC bias of the amplitude modulator used to make the homodyne-trigger pulse on the phase reference. In practice, the DC bias is fairly stable (even though it is controlled by a feedback loop to stabilize the phase reference power) but can result in homodyne-to-photon-subtraction event delays varying by about 1-2 bins if not accounted for. 

The clock synchronization RFoF signal is wavelength multiplexed with the sideband reference (a 25-GHz RFoF signal), an optical phase reference (sampled from $\lambda_f$), and a conventional fiber transceiver bi-directional signal using 100-GHz dense-wavelength-division-multiplexing (DWDM) components (AC Photonics) which are all sent to each homodyne detection system. All of these signals and the resource state modes are contained within the optical C-band. As shown in Ref.~\cite{ChapmanDeployedTMSV2023}, these multiplexed signals can also be multiplexed with the resource-state output modes to enable truly distributed use of this resource state with only some additional insertion loss but not any significant added noise even though the spacing between the bright phase reference and dim resource state modes is only one 100-GHz channel apart. Before preparing to take photon-subtraction data with the lowest loss our system could permit, we did have multiplex and demultiplex filters to having the resource-state modes coexisting with the multiplexed classical signals and were able to take two-mode state tomographies in that configuration. As such, we do have the required pedestal filtering~\cite{ChapmanDeployedTMSV2023} on each classical transmitter to remove noise photons at the wavelengths of the resource state as seen by the extra DWDM component in front of each transmitter.

\subsection{Photon-subtraction/quadrature-sample delay calibration}
\label{SM:delcalib}
To calibrate the delay between the time-stamped photon-subtraction detection events to the time-stamped homodyne detection samples at the servers, we inject a pulsed ( 50-ns at 100~kHz) thermal state into the unused port of the photon-subtraction beamsplitter (see Fig.~\ref{fig:fullexpsetup}). To create the pulsed thermal state, the amplified spontaneous emission of a polarization-maintaining erbium-doped fiber optical amplifier (Pritel SCG-40) is modulated by an amplitude modulator (EOSpace) controlled by a bias controller (Oz Optics MBC-PDBC-3A) with 99/1 sampling (AC Photonics before and after the modulator, to control pulse extinction ratio, and pulsed by an arbitrary waveform generator (Tek AWG 710B).

After data collection for approximately one buffer-overflow period, the systems were stopped to avoid data overwrite. Then time-stamped photon-subtraction detections were collected from the PSO's data buffer. We found it important to reduce the PSO FPGA event hold time to 3 bins (less than the detector dead time) and use a single detector and do this measurement for one detection mode at a time since there is bunching in the bright thermal state leading to the multiple-detection filter otherwise rejecting the events of interest. For this calibration, the HDSs used in a threshold-crossing counter mode and every occurrence of a threshold crossing had its time-stamp recorded. All threshold crossings from both servers were then reported back to the orchestrator which used cross-correlation (CC) analysis between the detectors for a given mode and the threshold crossings from that mode's HDS to find the time-stamp at the peak CC. Using this method, we captured about 10000 events for each input to the CC and found clear peaks between the photon-subtraction for a given mode and the homodyne detector on that mode with an SNR of about 40 and a width of about 6 bins full-width-at-half-maximum.

As verification, we also conducted a delay sweep, centered on the peak-CC delay, of single-mode photon-subtraction tomographic measurements on mode A. The truly optimal delay maximizes the variance of the single-mode photon-subtracted state~\cite{RTPSquadOE2017} which we find to be 1 bin off the thermal-state peak-CC delay but that can vary a bit based on the thermal pulse height on the homodyne detector and the DC offset on the amplitude modulator used to send the homodyne-trigger pulse.

Finally, in the two-mode case, the delay space is richer because of the 2-D landscape (Table~\ref{tab:ENdiffdels}). Using the thermal state calibrated delay as a starting point, we acquired tomographic data at several different delays adaptively (for time efficiency), where based on the state tomography data  we would not go to the next delay if it was clear from the initial data analysis of the variances (e.g., Fig.~\ref{fig:DMs6969} insets) that the current delay combination was already moving away from the optimal delay combination (with largest two-mode squeezing and anti-squeezing). Even though this adaptive delay scan mostly searches just one side of the peak, given the temporal width of the state we are confident we have found the peak.

\begin{table}[]
    \centering    
    \caption{Log negativity~\cite{PhysRevA.65.032314}  $E_\mathcal{N}(\rho)$ versus delay combination. Top section: two-mode delay combinations D($i,j$) measured. Upper-middle section: $E_\mathcal{N}(\rho^M_{0,0})$ as a function of the two-mode delay combination. Lower-middle section: $E_\mathcal{N}(\rho^M_{1,1})$ as a function of the two-mode delay combination. Bottom section: Log negativity improvement $E_\mathcal{N}(\rho^M_{1,1})-E_\mathcal{N}(\rho^M_{0,0})$ from photon subtraction. The data error bar (standard deviation) is about 0.03.}
    \label{tab:ENdiffdels}
    \begin{tabular}{|c|c|c|c|}
    \hline
       D(-2,-2)  &  D(-2,-1)   &  D(-2,0)  &  \\
       \hline
         &  D(-1,-1) &  D(-1,0)  &  D(-1,1)\\
       \hline
         &  D(0,-1) &  D(0,0)  &  D(0,1)\\
       \hline
        \hline
        0.47  &   0.48  &  0.48  &  \\
       \hline
         &  0.46 &  0.46  &  0.45\\
       \hline
         &  0.35 &  0.52  &  0.49\\
       \hline
       \hline
        0.45  &   0.48  &  0.47  &  \\
       \hline
         &  0.51 &  0.51  &  0.46\\
       \hline
         &  0.45 &  0.53  &  0.52\\
       \hline
       \hline
        -0.01  &   0  &  -0.02  &  \\
       \hline
         &  0.06 &  0.04  &  0.01\\
       \hline
         &  0.1 &  0.01  &  0.03\\
       \hline
       \end{tabular}
\end{table}

\subsection{Modified system for single-mode operation}

Additionally, with minor modifications, we also measure single-mode squeezed vacuum and single-mode photon-subtraction of squeezed vacuum. Due to the broadband nature of our squeezed-light source, this source simultaneously also emits single-mode squeezing at $\lambda_f$ (C44). Given the C45 DWDM demultiplexer after the squeezed-light source, the C44 single-mode squeezed light is directed towards the mode-A/C43 homodyne detector.

To phase stabilize the single-mode squeezing and enable simultaneous phase drive for single-mode homodyne tomography, we can reconfigure the aforementioned phase stabilization from two-mode to one-mode as follows. The side-A/C43 homodyne detector output is split in half as before. The half previously directed towards the hybrid junction for two-mode phase stabilization is then directed towards a power detector (Linear Technology LT5537)  which gives a voltage output corresponding to the power of the input. Using lock-in detection (Liquid Instruments Moku:Go), the phase drive is demodulated from the squeezing signal so the residual phase drift can be extracted. The phase ramp frequency on side A is changed from 1 kHz to 10-kHz; the lock-in's local oscillator is also changed to a phase drive signal created by generating a 20-kHz sinewave using a function generator (Rigol DG2102) phase locked to the phase drive signal with a shared 10-MHz reference. For single-mode squeezing, a 10-kHz $2\pi$ ramp, produces a 20-kHz sinewave of squeezing and anti-squeezing since squeezing and anti-squeezing are $\pi/2$ offset not $\pi$. This residual phase drift error signal is used to control the phase of the mode-A homodyne local oscillator which is used in the detection of single-mode squeezing. The rest of the phase stabilization system is the same as described above except the control system drives a phase shifter and fiber stretcher on side-A where the single-mode squeezing is directed to.

To acquire data in this single-mode configuration, without changing the PSO application to be dedicated for single-mode use, just requires having the optical and electronic connections to the side-B/C45 homodyne detector and HDS connected as normal. In that case, any data collected from the side-B/C45 homodyne detector is ignored.

\begin{figure*}
    \centerline{\includegraphics[width=1\textwidth]{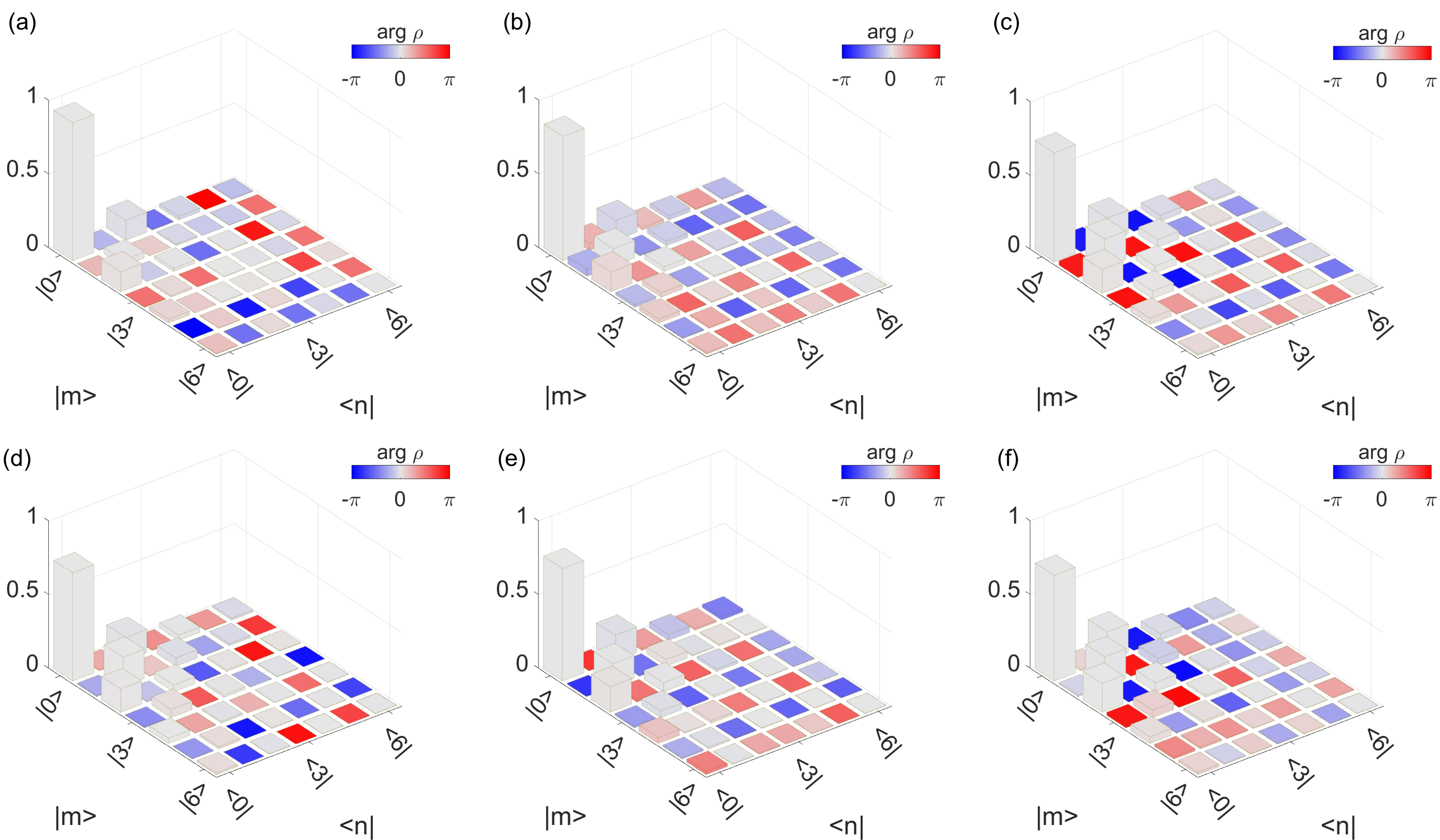}}
    \caption{Measured tomographically reconstructed density matrix of (photon-subtracted) single-mode squeezed vacuum. (a) no photons subtracted. (b) one photon subtracted using one single-photon detector (SPDs). (c) one photon subtracted using beamsplitter tree of two SPDs. (d) one photon subtracted using beamsplitter tree of three SPDs. (e) two photons subtracted using beamsplitter tree of two SPDs. (f) two photons subtracted using beamsplitter tree of three SPDs. For these measurement reconstructions, the photon-number cut-off per mode is 6.}
    \label{fig:SMdata}
\end{figure*}

Fig.~\ref{fig:SMdata} and Table~\ref{tab:SMdata} shows illustrative data of the effects of the number of detectors in a beamsplitter tree on the photon-subtracted state with $R_S=10\%$. Due to the amount of background photons we had (about 10-20\% of total singles counts which appears due to either the elevated waveguide temperature or photo-refractive crystal waveguide damage), we expect this state to be mixed with squeezed vacuum~\cite{RTPSquadOE2017}. For reference, Fig.~\ref{fig:SMdata}(a) shows the zero-photons subtracted measured state for which it is important to note the vacuum probability relative to the single-photon probability and the two-photon probability. Fig.~\ref{fig:SMdata}(b)-(d) show the measured density matrices of single-photon subtraction for increasing photon-subtraction detectors using a beamsplitter tree from one to three. All three show significantly higher single-photon probability compared to Fig.~\ref{fig:SMdata}(a). Fig.~\ref{fig:SMdata}(c)-(d) show significantly more single-photon probability compared to Fig.~\ref{fig:SMdata}(b) indicating that it is important to have at least one more detector in the tree than the number of photons being subtracted. Fig.~\ref{fig:SMdata}(e)-(f) show a similar comparison for two-photon subtraction where there are two and three detectors, respectively. Fig.~\ref{fig:SMdata}(f) having three photon-subtraction detectors in the beamsplitter tree shows reduced vacuum and increased one photon probability compared to Fig.~\ref{fig:SMdata}(e).

\begin{table}[]
    \centering    
\caption{Photon number probabilities of reconstructed density matrices in Fig.~\ref{fig:SMdata} versus detector tree size and subtracted photon number.}
    \label{tab:SMdata}
    \begin{tabular}{cccccc}
    \hline
       Fig.~\ref{fig:SMdata} panel & Detector tree size & Photons subtracted & $\ketbra{0}{0}$& $\ketbra{1}{1}$ & $\ketbra{2}{2}$\\
       \hline
       (a) & N/A & 0 & 0.94 & 0.04 & 0.02 \\
       (b) & 1 & 1 & 0.84 & 0.12 & 0.03 \\
       (c) & 2 & 1 & 0.74 & 0.19 & 0.05 \\
       (d) & 3 & 1 & 0.73 & 0.20 & 0.05 \\
       (e) & 2 & 2 & 0.76 & 0.15 & 0.06 \\
       (f) & 3 & 2 & 0.72 & 0.20 & 0.06 \\
       \end{tabular}
\end{table}

\section{Non-Gaussian Teleportation Experimental Design}
\label{SM:NGTelExp}
Here we describe our proposed design to demonstrate non-Gaussian teleportation (NGT) of a coherent state using two-mode generalized photon subtraction which coexists with conventional laser communications and other control signals enabling distributed operation between disparate locations. As we have discussed, non-Gaussian teleportation has the ability to increase the teleportation fidelity at the cost of probabilistic operation. Generalized-photon subtraction has been introduced to  provide significantly increased success probabilities by sending squeezed vacuum into the otherwise unused photon-subtraction beamsplitter port~\cite{PhysRevA.103.013710}. The two-mode version needed here has been analyzed previously in Ref.~\cite{PhysRevA.67.062320,PhysRevA.88.043818}, though in a somewhat different context. Ref.~\cite{PhysRevA.88.043818} shows that the teleportation fidelity with this resource is a monotonically increasing function of the photon-subtraction beamsplitter transmissivity. Even still, having squeezed vacuum in the second port of the photon-subtraction beamsplitters will increase the rate compared to having vacuum in those ports.

To include two-mode generalized photon subtraction, another squeezed-light source needs to be added and the two sources need to be phase stabilized with respect to one another. This phase stabilization can be done by using the single-mode squeezed light emitted by each source---that is otherwise unused. After the sources are combined, the single-mode squeezing at $\lambda_f$ is demultiplexed from the rest of the squeezing using a DWDM component and is directed towards a homodyne detector. This combination results in a state which is dependent on the phase between the squeezed-light sources and also the phase of the local oscillator. We propose to use this output to stabilize the phase between the squeezed-light sources by driving the LO phase to average it out and then adjusting the phase between the squeezed-light sources to minimize the noise variance assuming that the resultant average noise of the two sources being 90$^{\circ}$ out of phase is less than when they are in phase.

\begin{figure*}
    \centerline{\includegraphics[width=1\textwidth]{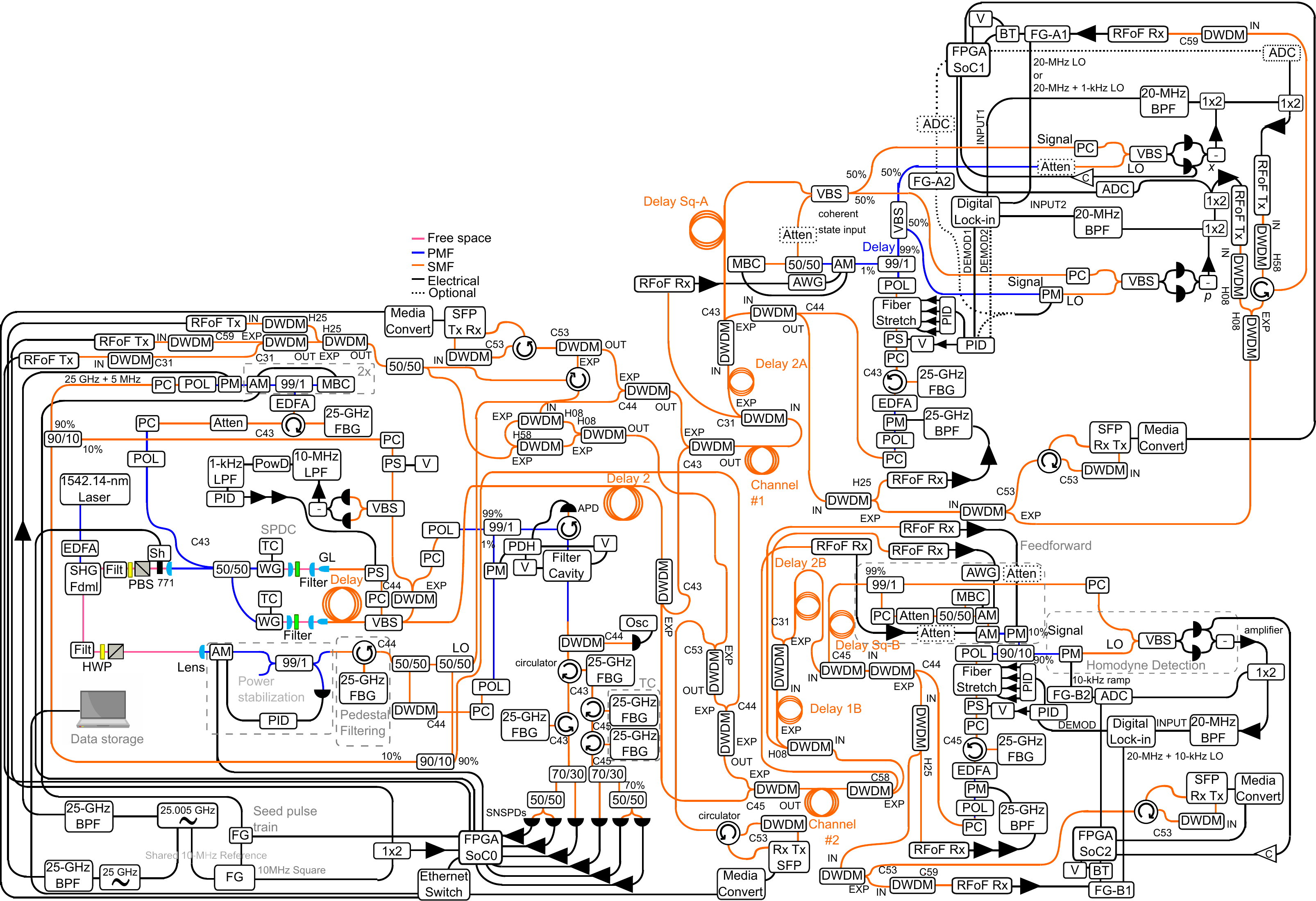}}
    \caption{Proposed experimental setup for non-Gaussian teleportation of a coherent state using two-mode generalized photon subtraction. Definitions (in alphabetical order).  ADC: analog-to-digital converter. AM: amplitude modulator. APD: avalanche photodiode. BPF: band-pass filter. BT: bias tee. C: comparator. CLK: clock. DEMUX: demultiplex. DWDM: dense-wavelength division multiplexer. EDFA: erbium-doped fiber amplifier. FBG: fiber Bragg grating. Fdml: fundamental. FG: function generator. Filt: filter. FPGA-SoC: field programmable gate array - system on chip. HJ: hybrid junction. HWP: half-wave plate. LO: local oscillator. MBC: modulator-bias controller. MUX: multiplex. Osc: oscilloscope. PC: polarization controller. PDH: Pound-Drever-Hall controller. PID: proportional-integral-derivative controller. PM: phase modulator. POL: polarizer. PowD: power detector. PS: phase shifter. RFoF: radio-frequency over fiber. Rx: receiver. SFP: small-form-factor pluggable transceiver. Sh: shutter. SHG: second-harmonic generation. TC: temperature controller. Tx: transmitter. V: voltage source. VBS: variable beamsplitter. WG: waveguide. 100-GHz channel center wavelength H08: 1570.83~nm. H25: 1556.96~nm. H58: 1530.72~nm. C31: 1552.52~nm. C43: 1542.95~nm. C44: 1542.14~nm. C45: 1541.35~nm. C53: 1535.04~nm. C59: 1530.33 nm}
    \label{fig:NGTellexpsetup}
\end{figure*}

Our proposed NGT setup uses the same general photon subtraction method whose detectors are input to a PSO. The PSO's heralding output signal is sent via RFoF to each end node. On side-A, this signal is used to trigger the creation of an temporally mode-matched input coherent state. On Side-B, this signal is used to trigger the creation of a temporally mode-matched feedforward displacement field. Moreover, the feedforward itself in this design is also transmitted using RFoF from side-A to side-B. Comparing the amplified homodyne noise power to the component noise floors, we have determined that there is sufficient SNR to use RFoF and post-amplification to produce gain-of-1 teleportation feedforward without introducing significant noise due to the RFoF conversion, transmission loss (at least on metropolitan-scale distances), and amplification.

To provide the necessary synchronization for coherence and temporal-mode matching between the various signals of the system, we have determined a variety of delays are required to be added to meet the timing constraints, as shown in Fig.~\ref{fig:NGTellexpsetup}:
\begin{itemize}
    \item Delay -- used to equalize the path length between the squeezed sources into the beamsplitter.
    \item Delay 2 -- used to equalize the path length between the quantum-light paths and the LOs.
    \item Delay 2A -- used to delay the photon-subtracted-squeezed temporal mode (PSSTM), phase and sideband references to give time for heralding trigger pulse to generate the input coherent state.
    \item Delay Sq-A -- used to delay the PSSTM so it arrives at the dual homodyne detector simultaneously with the input coherent state.
    \item Delay 1B -- used to delay the heralding trigger, PSSTM, phase and sideband references to give time for the feedforward signals to arrive.
    \item Delay 2B -- used to delay the PSSTM, phase and sideband references to give time for the heralding trigger pulse to generate the temporally mode-matched feedforward displacement light.
    \item Delay Sq-B -- used to delay the PSSTM so it arrives at the displacement beamsplitter simultaneously with the temporally-mode matched feedforward displacement light.
\end{itemize}

To enable independent phase stabilization of each mode, which enables distributed operation, we create a seed signal which creates a 20-MHz sideband easily measured and isolated by each homodyne detector. Lock-in detection can be used on each side to extract the phase drift which can be corrected using the two-stage control method described in Sec.~\ref{SM:Phstbdr}. The seed is input to the unused input port of the 50/50 splitter before the SPDC waveguides to seed and phase reference the squeezed light. Due to the waveguide being an optical parametric amplifier, an idler is produced in the conjugate mode providing a signal for each mode.

The seed signal is generated by tapping 50\% of the light going to the $\lambda_f$ power stabilization pick-off detector. From there, a seed signal is generated at C43 + 5~MHz using methods similar to our LO generation (described in Sec.~\ref{SM:LOandHD}) except a second RF oscillator (Berkeley Nucleonics Corp. Model 845-26-FILT-NM-1URM) is used, which is phase referenced to the 25-GHz main RF oscillator, to produce a 25.005-GHz signal to drive the sideband generation to create the seed at $\lambda_f$ + 100.02~GHz which is 20~MHz shifted from the LO of the homodyne detectors.

The homodyne detector signal is split and the part for phase stabilization is bandpass filtered (Mini-circuits SBP-21.4+) and amplified (Mini-circuits ZFL-1000LN+). After which it goes to a lock-in detector (Liquid Instruments Moku:Go) for 20-MHz demodulation. To facilitate a phase-referenced 20-MHz LO for the 20-MHz demodulation, A 20-MHz signal is produced by a function generator (Rigol DG2102) that is phase referenced to the shared 10-MHz reference between the 3 nodes. 

Moreover, for the dual-homodyne detector on side-A, the overall optical LO phase can be stabilized using the output of the detector without the phase modulator in the optical LO path. For teleportation, the phase between the detectors in the dual-homodyne detector needs to be stabilized. To accomplish this, the other detector's phase error signal can be maximized by adjusting the phase modulator in that LO path---this ensures a 90$^{\circ}$ phase shift between the optical LO's of each detector if the lock-in-detector LO phases are correct (which depends on propagation delay differences of each detector to the lock-in). This is the configuration used for teleportation where the output of each detector is also recorded by the HDS analog-to-digital converters. 

For two-mode homodyne tomography of the resource state, a few minor modifications are needed. The phase modulator providing the 90$^{\circ}$ phase difference for the dual-homodyne detector is repurposed to provide the tomography phase drive. The 20-MHz lock-in detection LO needs to be modified by modulating it with the tomography phase drive. These alterations are shown with dashed lines in Fig.~\ref{fig:NGTellexpsetup}. Also, the 50/50 beamsplitter on the signal path can to be removed to improve the results.

For phase stabilization of two-mode squeezing, this CW seed is sufficient and the squeezed light bandwidth can be filtered (Mini-circuits SLP-10.7+) to reject the 20-MHz seed which we have tested. But for photon-subtracted states, this CW seed needs to be pulsed. Frequency filtering of this seed which is within the photon-subtracted-state bandwidth is not compatible with the real-time operation described above. Additionally, the filtering cavity does not provide enough attenuation of the 20-MHz sideband. An amplitude modulator (EOSpace) and bias controller (Oz Optics MBC-PDBC-3A), with appropriate optical pick-off(s) around the modulator, can be used to create pulses. The electrical pulses (about  50~ns, 100~kHz) are generated from a function generator (Rigol DG2102) phase referenced to shared 10-MHz clock. The pulse parameters are chosen so the duty cycle is low to avoid disruption of data acquisition but the repetition rate is high compared to the tomography phase drive. Moreover, an additional stage of optical amplification is added (Pritel FA-30-IO) after the sidebands are filtered with the 25-GHz FBG to further amplify the seed, where, after the amplifier, the residual amplified spontaneous emission is removed using a 100-GHz bandpass filter (AC Photonics). This provides sufficient power to generate sidebands with enough power for phase stabilization based on their average value.

Using this pulsed method, will result in extra background to the photon subtraction detectors. The repetition rate is too fast for an optical chopper and the path to the detectors from the photon-subtraction beamsplitter is in fiber except the enclosed fiber-coupled cavity. One could include a fiber-based optical switch to switch out the seed light before going to the detectors but this will result in additional loss in the photon subtraction path. To avoid the extra loss, we developed a rejection filter in the PSO FPGA which takes a seed-clock signal (phase referenced to the shared 10-MHz clock) and will reject detection events which are within a certain time stamp range with respect to the seed clock based on calibrated delay and range width parameters. The calibration can be done by adjusting the delay and range width to minimize background counts from the pulse seed.

We tested this method of phase stabilization and found it to stabilize the phase on each side to $<<1^{\circ}$ depending on the SNR of the demodulated phase error signal. We also verified that the two-mode squeezing was phase stabilized to within about 0.1~dB on a swept spectrum analyzer (Agilent N9000A CXA Signal Analyzer) using the hybrid junction to directly see two-mode squeezing. In our testing, we directly distributed a 20-MHz reference which was amplified (Mini-Circuits ZKL-1R5+ and ZFL-1000LN+) and filtered (Mini-circuits SBP-21.4+) instead of deriving the 20-MHz lock-in LO from a function function generator synchronized to a 10-MHz reference but we expect this to be a trivial difference. Additionally, although we developed the seed-clock rejection in the PSO FPGA which is fairly straight forward and also compiled and synthesized without issue, it was not used or tested to date.

\end{document}